\documentclass[11pt,a4paper]{article}

\usepackage{amsmath}
\usepackage{microtype}
\usepackage{amssymb}
\usepackage{etoolbox}
\usepackage{array}
\usepackage{graphicx}
\usepackage[margin=2cm]{geometry}
\usepackage[font=small]{caption}
\usepackage{cite}

\usepackage[colorlinks=true,linkcolor=blue,citecolor=blue,urlcolor=blue,filecolor=blue]{hyperref}
\usepackage{placeins}
\newcommand{\inlfpfirst}[1]{{\let\nmfirstfracdone\relax\renewcommand{\frac}[2]{\ifx\nmfirstfracdone\relax(\nmslash{##1}{##2})\def\nmfirstfracdone{1}\else\nmslash{##1}{##2}\fi}#1}}
\AtBeginDocument{\medmuskip=4mu plus 2mu minus 1mu}

\providecommand{\nmfrac}[2]{\frac{#1}{#2}}
\providecommand{\nmInvFA}{\frac{d_F^{abcd}d_A^{abcd}}{N_R}}
\providecommand{\nmInvAA}{\frac{d_A^{abcd}d_A^{abcd}}{N_A}}
\providecommand{\nmInvFAA}{\frac{d_F^{abcd}d_A^{abef}d_A^{cdef}}{N_R}}

\newcommand{\nmDFASUthree}{\nmfrac{5}{2}}%
\newcommand{\nmDFASUN}{\frac{(N_c^2-1)(N_c^2+6)}{48}}%
\newcommand{\nmDFAASUN}{\frac{N_c^2(N_c^2-1)(N_c^2+51)}{432}}%
\newcommand{\nmDAASUN}{\frac{N_c^2(N_c^2+36)}{24}}%
\newcommand{\nmTrcSUN}{\begin{pmatrix}0 & -\frac{N_c^2-1}{N_c}\\ -\frac{1}{N_c} & \frac{2}{N_c}-\frac{N_c}{2}\end{pmatrix}}%
\newcommand{\nmBornAdjSUN}{-\frac{1}{2N_c}}%
\newcommand{\nmSigFactor}{\frac{e^{-i\pi\omega}\pm 1}{2}}%
\newcommand{\nmPsiPlusLeading}{-\nmfrac{\pi^2\omega}{2}}%
\newcommand{\nmDipoleOperator}{2a\sum_{i<j}\mathbf{T}_i\cdot\mathbf{T}_j\ln(-s_{ij}-i0)}%
\newcommand{\nmSoneData}{4C_F\zeta_2}%
\newcommand{\nmStwoData}{-8C_F^2\zeta_2}%
\newcommand{\nmPoneZetaData}{-8\zeta_2C_F^2}%
\newcommand{\nmXzeroDiffGen}{38C_FC_A^3 - 120C_F^2C_A^2 + 104C_F^3C_A - 16C_F^4 - 48\,\nmInvFA}%
\newcommand{\nmXzeroDiffABCD}{-48A + 14B + 20C + 2D}%
\newcommand{\nmXzeroDiffSUN}{\frac{(N_c^2-1)(N_c^4 + 10N_c^2 + 1)}{N_c^4}}%
\newcommand{\nmXzeroDiffSUthree}{\nmfrac{1376}{81}}%
\newcommand{\nmXzeroDiffCombination}{-\frac{S_4}{24} + \frac{C_FS_3}{6}}%
\newcommand{\nmNNLLpredSthreePart}{-\frac{C_FS_3}{3}}%
\newcommand{\nmNNLLpredCFfourPart}{-\nmfrac{56C_F^4}{3}}%
\newcommand{\nmRelSthreePart}{-\frac{C_FS_3}{6}}%
\newcommand{\nmBracketPrintedOne}{256}%
\newcommand{\nmBracketPrintedTwo}{160}%
\newcommand{\nmBracketOne}{128}%
\newcommand{\nmBracketTwo}{80}%
\newcommand{\nmPrintedXzeroQuartic}{-96}%
\newcommand{\nmBracketDiff}{48}%
\newcommand{\nmLnxFourLoopZfour}{5760}%
\newcommand{\nmLnxFourLoopZtwoSq}{2304}%
\newcommand{\nmLnxPrinted}{-(9984\zeta_2 + 2304\zeta_2^2 + 6144\zeta_2\zeta_3)}%
\newcommand{\nmLnxFourLoop}{-9984\zeta_2 + 6144\zeta_2\zeta_3 + 5760\zeta_4}%
\newcommand{\nmPthreeMinusLnSixQCD}{\nmfrac{3632}{3645}}%
\newcommand{\nmPthreeMinusLnSixPrinted}{0.99643347}%
\newcommand{\nmPoneSigDiffData}{8K\left(N^{-3} - N^{-2} + (2 - \zeta_2)N^{-1}\right)}%
\newcommand{\nmKdef}{C_F(C_A-2C_F)}%
\newcommand{\nmOneLoopPole}{\frac{2aC_F}{\omega}}%
\newcommand{\nmOneLoopSterm}{2aC_F}%
\newcommand{\nmKappa}{+1}%
\newcommand{\nmSthreeGen}{12\,\zeta_2\,C_F(C_A-2C_F)(5C_A-6C_F)}%
\newcommand{\nmSthreeVec}{(60,\,-192,\,144)}%
\newcommand{\nmSfourGen}{-48\,\zeta_2\left(19C_A^3C_F - 65C_A^2C_F^2 + 68C_AC_F^3 - 20C_F^4 - 24\,\nmInvFA\right)}%
\newcommand{\nmSfourVec}{(-912,\,3120,\,-3264,\,960,\,1152)}%
\newcommand{\nmSthreeSUN}{\frac{6(N_c^2-1)(2N_c^2 + 3)}{N_c^3}\,\zeta_2}%
\newcommand{\nmSthreeSUthree}{\nmfrac{112}{3}\,\zeta_2}%
\newcommand{\nmSfourSUN}{-\frac{12(N_c^2-1)(19N_c^2 + 5)}{N_c^4}\,\zeta_2}%
\newcommand{\nmSfourSUthree}{-\nmfrac{5632}{27}\,\zeta_2}%
\newcommand{\nmSfiveSUN}{-\frac{6(N_c^2-1)(16N_c^6 - 352N_c^4 - 244N_c^2 - 19)}{N_c^5}\,\zeta_2}%
\newcommand{\nmSfiveSUthree}{\nmfrac{305008}{81}\,\zeta_2}%
\newcommand{\nmSsixSUN}{\frac{12(N_c^2-1)(226N_c^8 - 2488N_c^6 - 2148N_c^4 - 435N_c^2 + 1)}{N_c^6}\,\zeta_2}%
\newcommand{\nmSsixSUthree}{-\nmfrac{16283776}{243}\,\zeta_2}%
\newcommand{\nmRepsIndependentDistinct}{37}%
\newcommand{\nmRepsExtCount}{57}%
\newcommand{\nmSfiveGen}{\zeta_2\left(19252C_FC_A^4 - 65344C_F^2C_A^3 + 74400C_F^3C_A^2 - 30720C_F^4C_A + 3648C_F^5 - 57600C_A\nmInvFA + 33792C_F\nmInvFA + 3072C_F\nmInvAA\right)}%
\newcommand{\nmSsixGen}{\zeta_2\left(-551600C_FC_A^5 + 1846640C_F^2C_A^4 - 2173056C_F^3C_A^3 + 1021056C_F^4C_A^2 - 165120C_F^5C_A - 768C_F^6 + 2683008C_A^2\nmInvFA - 2694144C_AC_F\nmInvFA + 552960C_F^2\nmInvFA - 296448C_AC_F\nmInvAA + 147456C_F^2\nmInvAA + 456192\,\nmInvFAA\right)}%
\newcommand{\nmReductionCoef}{-12}%
\newcommand{\nmSingletAdjoint}{-2C_F}%
\newcommand{\nmBornWeight}{C_F - \frac{C_R}{2}}%
\newcommand{\nmStwoOwn}{12\,\zeta_2\,C_F(C_A-2C_F)}%
\newcommand{\nmPonePlusZetaOwn}{-12\zeta_2C_F(C_A-2C_F)}%
\newcommand{\nmPoneSigDiffOwn}{8K\left(N^{-3} + \zeta_2N^{-1}\right)}%
\newcommand{\nmLLminusTwo}{8C_FC_A - 12C_F^2}%
\newcommand{\nmLLminusTwoSUthree}{\nmfrac{32}{3}}%
\newcommand{\nmLLminusThree}{-24C_FC_A^2 + 96C_F^2C_A - 80C_F^3}%
\newcommand{\nmLLminusThreeSUthree}{\nmfrac{928}{27}}%
\newcommand{\nmLLminusFour}{192C_FC_A^3 - 640C_F^2C_A^2 + 704C_F^3C_A - 304C_F^4}%
\newcommand{\nmLLminusFourSUthree}{\nmfrac{58112}{81}}%
\newcommand{\nmPminusQuartic}{16}%
\newcommand{\nmCtrlKappaSthreeSUthree}{\nmfrac{368}{9}\,\zeta_2}%
\newcommand{\nmLambdaVec}{(1,\,1,\,2,\,2,\,2)}%
\newcommand{\nmXkKmax}{4}%
\newcommand{\nmXkComponents}{8}%
\newcommand{\nmXkComponentsThree}{3}%
\newcommand{\nmXkComponentsFour}{5}%
\newcommand{\nmXkUncovQuarticFormula}{-16\lambda_k(\lambda_k-2)}%
\newcommand{\nmXkCovQuarticFormula}{-16\lambda_k(\lambda_k+2)}%
\newcommand{\nmXkUncovQuarticLow}{16}%
\newcommand{\nmXkUncovQuarticHigh}{0}%
\newcommand{\nmXkVthreeCovLamTwo}{(96,\,-288,\,192)}%
\newcommand{\nmXkVfourCovLamTwo}{(-2240,\,7872,\,-8832,\,3072,\,3072)}%
\newcommand{\nmXkVthreeUncLamOne}{(84,\,-256,\,176)}%
\newcommand{\nmXkVfourUncLamOne}{(112,\,-496,\,832,\,-448,\,-384)}%
\newcommand{\nmXkVthreeUncLamTwo}{(160,\,-480,\,320)}%
\newcommand{\nmXkVfourUncLamTwo}{(-448,\,1856,\,-2432,\,1024,\,0)}%
\newcommand{\nmXkDataVthreePlusXtwo}{(96,\,-288,\,192)}%
\newcommand{\nmXkDataVfourPlusXtwo}{(-2240,\,7872,\,-8832,\,3072,\,3072)}%
\newcommand{\nmXkDataVthreeMinusXzero}{(-28,\,96,\,-80)}%
\newcommand{\nmXkDataVfourMinusXzero}{(720,\,-2992,\,4032,\,-1728,\,-384)}%
\newcommand{\nmXkDataVthreeMinusXtwo}{(-96,\,288,\,-192)}%
\newcommand{\nmXkDataVfourMinusXtwo}{(1472,\,-6464,\,9088,\,-4096,\,0)}%
\newcommand{\nmFiveLoopXzeroGen}{-\nmfrac{4813}{180}C_FC_A^4 + \nmfrac{4312}{45}C_F^2C_A^3 - \nmfrac{368}{3}C_F^3C_A^2 + \nmfrac{336}{5}C_F^4C_A - \nmfrac{884}{45}C_F^5 + 80C_A\nmInvFA - \nmfrac{160}{3}C_F\nmInvFA - \nmfrac{64}{15}C_F\nmInvAA}%
\newcommand{\nmFiveLoopXzeroSUN}{-\frac{(N_c^2-1)(50N_c^8 - 212N_c^6 + 1110N_c^4 + 628N_c^2 + 221)}{360N_c^5}}%
\newcommand{\nmFiveLoopXzeroSUthree}{-\nmfrac{53857}{2187}}%
\newcommand{\nmFiveLoopXzeroRelPart}{-\nmfrac{212096}{10935}}%
\newcommand{\nmFiveLoopXzeroTowerPart}{-\nmfrac{19063}{3645}}%
\newcommand{\nmFiveLoopNsevenSUthree}{-\nmfrac{4308560}{243}}%
\newcommand{\nmFiveLoopMinusNNLLpred}{-\nmfrac{89}{243}}%
\newcommand{\nmFiveLoopLargeN}{-\nmfrac{5N_c^5}{36}}%
\newcommand{\nmTmMap}{T_m=\sum_{n=1}^{m}\binom{2(m-n)}{m-n}(2C_F)^{m-n}S_n}%
\newcommand{\nmCatalanF}{f_n=(2C_F)^n\,\mathrm{Cat}_{n-1}}%
\newcommand{\nmDtwoSUN}{-\frac{2N_c^2+3}{N_c^2}}%
\newcommand{\nmDthreeSUN}{\frac{2(19N_c^2+5)}{N_c^3}}%
\newcommand{\nmOmegaZeroDef}{\sqrt{8aC_F}}%
\newcommand{\nmLinearSub}{F_A^- = 2aN_c\,u'(\omega)/u(\omega)}%
\newcommand{\nmLinearEq}{4a^2N_c^2\,\omega\,u'' - 2aN_c(\omega^2-2aN_c)\,u' - \frac{a}{N_c}\left[\omega + 4F_1^+ + 2(N_c^2-4)F_A^+\right]u = 0}%
\newcommand{\nmAlphaSOne}{0.1}%
\newcommand{\nmAlphaSTwo}{0.2}%
\newcommand{\nmXminExp}{-8}%
\newcommand{\nmXmaxExp}{-2}%
\newcommand{\nmSizeOneXtwo}{-1.3}%
\newcommand{\nmSizeOneXeight}{-5.5}%
\newcommand{\nmSizeTwoXtwo}{-4.2}%
\newcommand{\nmSizeTwoXeight}{-14.5}%
\newcommand{\nmNLLtermCoef}{2C_A + 6C_F}%
\newcommand{\nmXadjSUN}{-\frac{N_c^2+2}{4N_c^2}}%
\newcommand{\nmPminusLLtwoKm}{4C_F(5C_F-2C_A)}%
\newcommand{\nmPplusLLtwo}{4C_F^2}%
\newcommand{\nmMoneGen}{-\frac{1}{4}(C_A-2C_F)(3C_A-4C_F)}%
\newcommand{\nmRowSum}{\frac{C_R}{2}-2C_F}%
\newcommand{\nmRowSumBorn}{b_R\left(C_A-2C_F+\frac{C_R}{2}\right)}%
\newcommand{\nmSthreeKmGen}{12\,\zeta_2\,C_F(C_A-2C_F)(7C_A-10C_F)}%
\newcommand{\nmTAXthree}{\frac{2C_A(b_A+C_F/N_A)}{C_A-C_X}}%
\newcommand{\nmSONdims}{(1,\,\frac{N(N-1)}{2},\,\frac{(N-1)(N+2)}{2})}%
\newcommand{\nmSONcasimirs}{(0,\,\frac{N-2}{2},\,\frac{N}{2})}%
\newcommand{\nmSONborn}{(\frac{N-1}{4},\,\frac{1}{4},\,-\frac{1}{4})}%
\newcommand{\nmTrcSON}{\begin{pmatrix}0 & -\frac{N-1}{2} & 0\\ -\frac{1}{N} & 0 & -\frac{N^2-4}{4N}\\ 0 & -\frac{N-2}{4} & 0\end{pmatrix}}%
\newcommand{\nmDFASON}{\frac{(N-1)(N-2)(N^2-7N+22)}{768}}%
\newcommand{\nmSthreeSON}{-\frac{3(N-1)(2N-7)}{4}\,\zeta_2}%
\newcommand{\nmSfourSON}{\frac{3(N-1)(6N^2-19N+19)}{4}\,\zeta_2}%
\newcommand{\nmSpNdims}{(1,\,\frac{(N+1)(N-2)}{2},\,\frac{N(N+1)}{2})}%
\newcommand{\nmSpNcasimirs}{(0,\,\frac{N}{2},\,\frac{N+2}{2})}%
\newcommand{\nmSpNborn}{(\frac{N+1}{4},\,\frac{1}{4},\,-\frac{1}{4})}%
\newcommand{\nmTrcSpN}{\begin{pmatrix}0 & 0 & -\frac{N+1}{2}\\ 0 & 0 & -\frac{N+2}{4}\\ -\frac{1}{N} & -\frac{N^2-4}{4N} & 0\end{pmatrix}}%
\newcommand{\nmDFASpN}{\frac{(N+1)(N+2)(N^2+7N+22)}{768}}%
\newcommand{\nmSthreeSpN}{\frac{3(N+1)(2N+7)}{4}\,\zeta_2}%
\newcommand{\nmSfourSpN}{-\frac{3(N+1)(6N^2+19N+19)}{4}\,\zeta_2}%
\newcommand{\nmSthreeBornForm}{-48\kappa\,\zeta_2\,C_F\,(b_A^-)^2-96\,\zeta_2\,C_F\,M_1}%
\newcommand{\nmSfourRevDiffGen}{S_4(b_A^-)-S_4(-b_A^-)=8\,b_A^-\,S_3}%
\newcommand{\nmPfourMinusPoleVec}{(-60672,\,56832,\,3072)}%
\newcommand{\nmInvRelationClassical}{17C_FC_A^4 - 48C_F^2C_A^3 + 36C_F^3C_A^2 - 90C_A\nmInvFA + 72C_F\nmInvFA + 6C_F\nmInvAA = 0}%
\newcommand{\nmSoneCplusShift}{-12\,\zeta_2\,c_+C_F}%
\newcommand{\nmDTPOnePlusLtwoZeta}{0}%
\newcommand{\nmDTPOnePlusLtwoRat}{2C_F^2}%
\newcommand{\nmDTPOnePlusLoneZeta}{0}%
\newcommand{\nmDTPOnePlusLoneRat}{(\nmfrac{22}{3})C_FC_A - 4C_F^2 - (\nmfrac{4}{3})C_Fn_f}%
\newcommand{\nmDTPOnePlusLzeroZeta}{-8C_F^2}%
\newcommand{\nmDTPOnePlusLzeroRat}{(\nmfrac{302}{9})C_FC_A - 4C_F^2 - (\nmfrac{44}{9})C_Fn_f}%
\newcommand{\nmDTPOneMinusLtwoZeta}{0}%
\newcommand{\nmDTPOneMinusLtwoRat}{4C_FC_A - 6C_F^2}%
\newcommand{\nmDTPOneMinusLoneZeta}{0}%
\newcommand{\nmDTPOneMinusLoneRat}{(\nmfrac{46}{3})C_FC_A - 20C_F^2 - (\nmfrac{4}{3})C_Fn_f}%
\newcommand{\nmDTPOneMinusLzeroZeta}{-8C_FC_A + 8C_F^2}%
\newcommand{\nmDTPOneMinusLzeroRat}{(\nmfrac{446}{9})C_FC_A - 36C_F^2 - (\nmfrac{44}{9})C_Fn_f}%
\newcommand{\nmDTPTwoPlusLfourZeta}{0}%
\newcommand{\nmDTPTwoPlusLfourRat}{(\nmfrac{2}{3})C_F^3}%
\newcommand{\nmDTPTwoPlusLthreeZeta}{0}%
\newcommand{\nmDTPTwoPlusLthreeRat}{(\nmfrac{22}{3})C_F^2C_A - 4C_F^3 - (\nmfrac{4}{3})C_F^2n_f}%
\newcommand{\nmDTPTwoPlusLtwoZeta}{-30C_FC_A^2 + 96C_F^2C_A - 104C_F^3}%
\newcommand{\nmDTPTwoPlusLtwoRat}{(\nmfrac{121}{9})C_FC_A^2 + (\nmfrac{472}{9})C_F^2C_A + 4C_F^3 - (\nmfrac{44}{9})C_FC_An_f - (\nmfrac{64}{9})C_F^2n_f + (\nmfrac{4}{9})C_Fn_f^2}%
\newcommand{\nmDTPTwoMinusLfourZeta}{0}%
\newcommand{\nmDTPTwoMinusLfourRat}{-C_FC_A^2 + 4C_F^2C_A - (\nmfrac{10}{3})C_F^3}%
\newcommand{\nmDTPTwoMinusLthreeZeta}{0}%
\newcommand{\nmDTPTwoMinusLthreeRat}{(\nmfrac{40}{9})C_FC_A^2 - (\nmfrac{14}{9})C_F^2C_A - 4C_F^3 - (\nmfrac{16}{9})C_FC_An_f + (\nmfrac{20}{9})C_F^2n_f}%
\newcommand{\nmDTPTwoMinusLtwoZeta}{14C_FC_A^2 - 96C_F^2C_A + 104C_F^3}%
\newcommand{\nmDTPTwoMinusLtwoRat}{81C_FC_A^2 - (\nmfrac{152}{3})C_F^2C_A - 60C_F^3 - (\nmfrac{196}{9})C_FC_An_f + (\nmfrac{80}{3})C_F^2n_f + (\nmfrac{4}{9})C_Fn_f^2}%
\newcommand{\nmDTPThreePlusLsixZeta}{0}%
\newcommand{\nmDTPThreePlusLsixRat}{(\nmfrac{1}{9})C_F^4}%
\newcommand{\nmDTPThreePlusLfiveZeta}{0}%
\newcommand{\nmDTPThreePlusLfiveRat}{(\nmfrac{22}{9})C_F^3C_A - (\nmfrac{4}{3})C_F^4 - (\nmfrac{4}{9})C_F^3n_f}%
\newcommand{\nmDTPThreePlusLfourZeta}{38C_FC_A^3 - 140C_F^2C_A^2 + 168C_F^3C_A - (\nmfrac{248}{3})C_F^4 - 48\,\nmInvFA}%
\newcommand{\nmDTPThreePlusLfourRat}{(\nmfrac{121}{9})C_F^2C_A^2 + (\nmfrac{170}{9})C_F^3C_A + (\nmfrac{16}{3})C_F^4 - (\nmfrac{44}{9})C_F^2C_An_f - (\nmfrac{20}{9})C_F^3n_f + (\nmfrac{4}{9})C_F^2n_f^2}%
\newcommand{\nmDTPThreeMinusLsixZeta}{0}%
\newcommand{\nmDTPThreeMinusLsixRat}{(\nmfrac{4}{15})C_FC_A^3 - (\nmfrac{8}{9})C_F^2C_A^2 + (\nmfrac{44}{45})C_F^3C_A - (\nmfrac{19}{45})C_F^4}%
\newcommand{\nmDTPThreeMinusLfiveZeta}{0}%
\newcommand{\nmDTPThreeMinusLfiveRat}{-(\nmfrac{8}{15})C_FC_A^3 + (\nmfrac{104}{15})C_F^2C_A^2 - (\nmfrac{586}{45})C_F^3C_A + (\nmfrac{92}{15})C_F^4 + (\nmfrac{8}{15})C_FC_A^2n_f - (\nmfrac{32}{15})C_F^2C_An_f + (\nmfrac{76}{45})C_F^3n_f}%
\newcommand{\nmDTPThreeMinusLfourZeta}{-30C_FC_A^3 + 132C_F^2C_A^2 - (\nmfrac{656}{3})C_F^3C_A + (\nmfrac{376}{3})C_F^4 + 16\,\nmInvFA}%
\newcommand{\nmDTPThreeMinusLfourRat}{-(\nmfrac{28}{3})C_FC_A^3 + (\nmfrac{883}{9})C_F^2C_A^2 - (\nmfrac{1078}{9})C_F^3C_A + (\nmfrac{56}{3})C_F^4 + (\nmfrac{26}{9})C_FC_A^2n_f - (\nmfrac{68}{3})C_F^2C_An_f + (\nmfrac{196}{9})C_F^3n_f + (\nmfrac{4}{9})C_FC_An_f^2 - (\nmfrac{4}{9})C_F^2n_f^2}%
\newcommand{\nmDTformSthree}{\zeta_2\left(4C_Fz^{(1)}_0-16C_Fc^{(1)}_2-2z^{(2)}_2\right)}%
\newcommand{\nmDTformSfour}{\zeta_2\left(8C_Fz^{(2)}_2-192C_Fc^{(2)}_4+4c^{(1)}_2z^{(1)}_0-24z^{(3)}_4\right)}%
\newcommand{\nmDTPzeroLaurent}{C_F\Big[\frac{2}{N}+1+(2-4\zeta_2)\,N+O(N^2)\Big]}%
\newcommand{\nmDTPzeroDigamma}{C_F\big[3+2/\big(N(N+1)\big)-4\psi(N+1)-4\gamma_E\big]}%
\newcommand{\nmDTMellinRule}{(-1)^m\,m!\,N^{-m-1}}%
\newcommand{\nmDTmLL}{m=2n}%
\newcommand{\nmDTmNLL}{m=2n-1}%
\newcommand{\nmDTmNNLL}{m=2n-2}%
\newcommand{\nmPzeroResidues}{(2C_F,\,2C_F,\,4C_F,\,4C_F,\,4C_F)}%
\newcommand{\nmXkSecondKmax}{5}%
\newcommand{\nmXkCovQuarticHigh}{-128}%
\newcommand{\nmLambdaGenFun}{\frac{1+x^2}{1-x}}%
\newcommand{\nmXkEquationKmax}{10}%
\newcommand{\nmAmpTwoLoopCA}{-15}%
\newcommand{\nmAmpThreeLoopCA}{12}%
\newcommand{\nmAmpThreeLoopCACF}{-58}%
\newcommand{\nmAmpKmTwoLoopCA}{-21}%
\newcommand{\nmAmpKmThreeLoopCA}{15}%
\newcommand{\nmAmpKmThreeLoopCACF}{-74}%
\newcommand{\nmAmpTwoLoopIRpart}{-18}%
\newcommand{\nmAmpTwoLoopPairPart}{3}%
\newcommand{\nmAmpThreeLoopCACFsq}{84}%
\newcommand{\nmAmpOneLoopZetaConst}{\nmfrac{13}{2}C_F}%
\newcommand{\nmAmpLLCount}{12}%
\newcommand{\nmAmpLLOddCount}{6}%
\newcommand{\nmAmpStructCount}{195}%
\newcommand{\nmAmpDataOneVec}{(0,\,\nmfrac{13}{2})}%
\newcommand{\nmAmpDataTwoVec}{(-15,\,48,\,-41)}%
\newcommand{\nmAmpDataThreeVec}{(12,\,-58,\,84,\,-\nmfrac{451}{12})}%
\newcommand{\nmAmpEqOneVec}{(-6+v_A-r_A,\,\nmfrac{21}{2}+12\ell_{+}+12\alpha+24\beta_e+v_F-r_F)}%
\newcommand{\nmAmpEqTwoVec}{(-15,\,36+p_A+2v_A-r_A,\,-15+48\ell_{+}+24\alpha+72\beta_e+p_F+2v_F-r_F)}%
\newcommand{\nmAmpEqThreeVec}{(12,\,-58,\,78+\nmfrac{2}{3}p_A+v_A-\nmfrac{1}{3}r_A,\,-\nmfrac{85}{4}+30\ell_{+}+10\alpha+40\beta_e+\nmfrac{2}{3}p_F+v_F-\nmfrac{1}{3}r_F)}%
\newcommand{\nmAmpLLOneSingEven}{\frac{N_c^2-1}{2N_c}}%
\newcommand{\nmAmpLLOneSingOdd}{\frac{N_c^2+3}{2N_c}}%
\newcommand{\nmAmpLLKmOneSing}{\frac{N_c^2-5}{2N_c}}%
\newcommand{\nmAmpLLOneAdjEven}{-\frac{2N_c^2+1}{2N_c}}%
\newcommand{\nmAmpLLOneAdjOdd}{\frac{2N_c^2+3}{2N_c}}%
\newcommand{\nmAmpLLKmOneAdj}{-\frac{2N_c^2+5}{2N_c}}%
\newcommand{\nmAmpLLTwoSingEven}{\frac{(N_c^2-1)^2}{12N_c^2}}%
\newcommand{\nmAmpLLTwoSingOdd}{\frac{N_c^4-2N_c^2-5}{6N_c^2}}%
\newcommand{\nmAmpLLKmTwoSing}{\frac{N_c^4-2N_c^2+7}{6N_c^2}}%
\newcommand{\nmAmpLLTwoAdjEven}{\frac{(2N_c^2+1)(3N_c^2+1)}{12N_c^2}}%
\newcommand{\nmAmpLLTwoAdjOdd}{-\frac{19N_c^2+5}{6N_c^2}}%
\newcommand{\nmAmpLLKmTwoAdj}{\frac{23N_c^2+7}{6N_c^2}}%
\newcommand{\nmAmpLLThreeSingEven}{\frac{(N_c^2-1)^3}{144N_c^3}}%
\newcommand{\nmAmpLLThreeSingOdd}{\frac{5N_c^6+41N_c^4+31N_c^2+19}{240N_c^3}}%
\newcommand{\nmAmpLLKmThreeSing}{\frac{5N_c^6-71N_c^4+31N_c^2-61}{240N_c^3}}%
\newcommand{\nmAmpLLThreeAdjEven}{-\frac{(2N_c^2+1)(60N_c^4+34N_c^2+5)}{720N_c^3}}%
\newcommand{\nmAmpLLThreeAdjOdd}{-\frac{16N_c^6-352N_c^4-244N_c^2-19}{240N_c^3}}%
\newcommand{\nmAmpLLKmThreeAdj}{\frac{16N_c^6-424N_c^4-324N_c^2-61}{240N_c^3}}%
\newcommand{\nmNfRun}{4}%
\newcommand{\nmRunCouplingDef}{a/(1+\beta_0a\ell)}%
\newcommand{\nmNfExcessGen}{-8C_F(C_A - 2C_F)(7C_A - 10C_F)}%
\newcommand{\nmNfExcessSUthree}{-\nmfrac{736}{27}}%
\newcommand{\nmYdef}{y=d-k_P/\zeta_2+2e}%
\newcommand{\nmPminusQuarticKm}{-48}%
\newcommand{\nmQuarticOfSFour}{1152}%
\newcommand{\nmCountSthreeRaw}{3}%
\newcommand{\nmCountSfourRaw}{5}%
\newcommand{\nmCountXzeroRaw}{4}%
\newcommand{\nmCountCovoneRaw}{8}%
\newcommand{\nmCountCovtwoRaw}{8}%
\newcommand{\nmCountPminusquarticRaw}{1}%
\newcommand{\nmCountUncquarticRaw}{1}%
\newcommand{\nmMismatchVthreeSUthree}{-\nmfrac{1984}{27}}%
\newcommand{\nmMismatchVfourSUthree}{\nmfrac{53632}{81}}%
\newcommand{\nmPzeroConstKzero}{C_F}%
\newcommand{\nmPzeroConstKone}{C_F}%
\newcommand{\nmPzeroConstKtwo}{0}%
\newcommand{\nmPzeroConstKthree}{-\nmfrac{8}{3}C_F}%
\newcommand{\nmPzeroConstKfour}{-\nmfrac{25}{6}C_F}%
\newcommand{\nmBornCorrMatchV}{6C_A-8C_F}%
\newcommand{\nmBornCorrMatchVA}{6}%
\newcommand{\nmConstSetR}{-10C_F}%
\newcommand{\nmConstFamilyCplus}{-\nmfrac{1}{3} - \beta_e}%
\newcommand{\nmConstFamilyAlpha}{-\nmfrac{1}{6} - \beta_e}%
\newcommand{\nmPminusQuarticRho}{16\rho}%
\newcommand{\nmAmpEpsZeroLLDiffCount}{8}%
\newcommand{\nmHarmonicSoneSmallN}{S_1(N)=\zeta_2N-\zeta_3N^2+\ldots}%
\newcommand{\nmMtwoGen}{\nmfrac{23}{24}C_A^3-\nmfrac{7}{2}C_A^2C_F+\nmfrac{9}{2}C_AC_F^2-2C_F^3-\frac{2}{C_F}\nmInvFA}%
\newcommand{\nmMthreeGen}{-\nmfrac{163}{144}C_A^4+\nmfrac{39}{8}C_A^3C_F-\nmfrac{33}{4}C_A^2C_F^2+\nmfrac{13}{2}C_AC_F^3-2C_F^4+\left(\frac{5C_A}{C_F}-6\right)\nmInvFA-\nmfrac{1}{3}\nmInvAA}%
\newcommand{\nmMfourGen}{\nmfrac{1103}{864}C_A^5-\nmfrac{235}{36}C_A^4C_F+\nmfrac{55}{4}C_A^3C_F^2-15C_A^2C_F^3+\nmfrac{17}{2}C_AC_F^4-2C_F^5+\left(-\frac{149C_A^2}{18C_F}+20C_A-12C_F\right)\nmInvFA+\left(\nmfrac{10}{9}C_A-\nmfrac{4}{3}C_F\right)\nmInvAA-\frac{2}{C_F}\nmInvFAA}%
\newcommand{\nmMthreeSUN}{-\frac{2N_c^6-6N_c^4+5N_c^2+2}{16N_c^4}}%
\newcommand{\nmMfourSUN}{-\frac{(N_c^2-2)(2N_c^6-4N_c^4+4N_c^2+1)}{32N_c^5}}%
\newcommand{\nmDrecFour}{d_4=4(b_A-2C_A)d_3+d_2^2+64\kappa\left(12M_3 - 17C_FM_2 + 6C_F^2M_1\right)}%
\newcommand{\nmDrecFive}{d_5=4(b_A-3C_A)d_4+2d_2d_3+64\kappa\left(297M_4 - 660C_FM_3 + 488C_F^2M_2 - 120C_F^3M_1\right)}%
\newcommand{\nmFiveLoopTestCFCAfour}{-\nmfrac{4813}{180}}%
\newcommand{\nmFiveLoopTestCAdFA}{80}%
\newcommand{\nmFiveLoopTestCFdAA}{-\nmfrac{64}{15}}%
\newcommand{\nmFiveLoopXzeroUpdateShift}{\nmfrac{17728}{3645}}%
\newcommand{\nmSourceFiveFreeParams}{7}%
\newcommand{\nmLargeNcCondVec}{(\nmfrac{1}{2},\,\nmfrac{1}{4},\,\nmfrac{1}{8},\,\nmfrac{1}{16},\,\nmfrac{1}{32},\,\nmfrac{1}{48},\,\nmfrac{1}{96},\,\nmfrac{1}{48})}%
\newcommand{\nmNthreeLLnfFreeGen}{-4\,\zeta_2\left(215C_FC_A^3 - 654C_F^2C_A^2 + 548C_F^3C_A - 72C_F^4 - 384\,\nmInvFA\right)}%
\newcommand{\nmNthreeLLnfFreeSUthree}{\nmfrac{176}{3}\,\zeta_2}%
\newcommand{\nmMisprintPrefactorPrinted}{g^2/p_i}%
\newcommand{\nmMisprintPrefactorCorrected}{b_ig^2/c_i}%
\newcommand{\nmMisprintGaussPrinted}{\exp[-(\omega/\omega_i)^2/2]}%
\newcommand{\nmMisprintGaussCorrected}{\exp[+(\omega/\omega_i)^2/4]}%
\newcommand{\nmKappaDegH}{24C_F}%
\newcommand{\nmSfiveKmHSUthree}{\nmfrac{110288}{27}\,\zeta_2}%
\newcommand{\nmRelCoefQuarter}{\nmfrac{1}{4}}%
\newcommand{\nmRelCoefFortyEighth}{\nmfrac{1}{48}}%
\newcommand{\nmRelCoefTwelfth}{\nmfrac{1}{12}}%
\newcommand{\nmRelFourCoefTwoThirds}{\nmfrac{2}{3}}%
\newcommand{\nmRelFourCoefThird}{\nmfrac{1}{3}}%
\newcommand{\nmFiveShiftFactor}{720}%
\newcommand{\nmSourceFiveFreeParamsSUthree}{1}%
\newcommand{\nmQsourceShiftSUthree}{-\nmfrac{1}{60}}%
\newcommand{\nmOneLoopAdjZetaConst}{-\nmfrac{3}{4}C_A + \nmfrac{13}{2}C_F}%
\newcommand{\nmHoneRational}{\nmfrac{85}{9}C_A - 16C_F - \nmfrac{10}{9}n_f}%
\newcommand{\nmResDiffSingletCA}{-6}%
\newcommand{\nmRungRungColor}{\frac{C_A^2}{4}}%
\newcommand{\nmLegRungColor}{\frac{C_A(C_A-2C_F)}{4}}%
\newcommand{\nmGlauberColorGroups}{9}%
\newcommand{\nmReggeJpair}{0}%
\newcommand{\nmReggeJfour}{-1}%
\newcommand{\nmHoneLrCoefCF}{4}%
\newcommand{\nmTrcTrcMtwoCoef}{-2}%
\newcommand{\nmSthreeAbelian}{144\,\zeta_2\,C_F^3}%
\newcommand{\nmSfourAbelian}{960\,\zeta_2\,C_F^4}%
\newcommand{\nmVthreeMinusAbelianDiff}{-256\,\zeta_2\,C_F^3}%
\newcommand{\nmVfourMinusAbelianDiff}{-1280\,\zeta_2\,C_F^4}%
\newcommand{\nmQuarticRelCFCAsq}{96}%
\newcommand{\nmQuarticRelCFsqCA}{24}%
\newcommand{\nmZetaThreeFourVec}{(0,\,0,\,48,\,-\nmfrac{320}{3},\,0)}%
\newcommand{\nmZetaThreeThreeVec}{(0,\,48,\,-96)}%
\newcommand{\nmZetaThreeContrastFour}{(152,\,-520,\,544,\,-160,\,-192)}%
\newcommand{\nmWordCountCovtwoRaw}{eight}%
\newcommand{\nmWordCountAmpCfreeRaw}{three}%
\newcommand{\nmWordCountConsXzero}{three}%
\newcommand{\nmWordCountConsDegOne}{three}%
\newcommand{\nmWordCountConsCoefFree}{two}%
\newcommand{\nmLLminusTwoAbelian}{-12C_F^2}%
\newcommand{\nmLLminusThreeAbelian}{-80C_F^3}%
\newcommand{\nmLLminusFourAbelian}{-304C_F^4}%
\newcommand{\nmSfiveAbelian}{3648\,\zeta_2\,C_F^5}%

\ExplSyntaxOn
\bool_new:N \l__nmsl_lead_bool
\bool_new:N \l__nmsl_sum_bool
\bool_new:N \l__nmsl_sign_bool
\bool_new:N \l__nmsl_frac_bool
\int_new:N \l__nmsl_depth_int
\int_new:N \l__nmsl_atoms_int
\int_new:N \l__nmsl_level_int
\tl_new:N \l__nmsl_prev_tl
\seq_new:N \l__nmsl_stack_seq
\str_new:N \l__nmsl_name_str
\str_new:N \l__nmsl_spec_str
\tl_new:N \l__nmsl_exp_tl
\cs_generate_variant:Nn \tl_if_head_eq_meaning:nNT { V }
\cs_generate_variant:Nn \str_case:nnF { V }
\cs_new_protected:Npn \__nmsl_measure:n #1
  {
    \bool_set_true:N \l__nmsl_lead_bool
    \bool_set_false:N \l__nmsl_sum_bool
    \bool_set_false:N \l__nmsl_sign_bool
    \bool_set_false:N \l__nmsl_frac_bool
    \int_zero:N \l__nmsl_depth_int
    \int_zero:N \l__nmsl_atoms_int
    \int_zero:N \l__nmsl_level_int
    \tl_set:Nn \l__nmsl_prev_tl { o }
    \__nmsl_scan:n {#1}
  }
\cs_new_protected:Npn \__nmsl_scan:n #1
  { \exp_args:Ne \__nmsl_scan_aux:n { \tl_trim_spaces:n {#1} } }
\cs_new_protected:Npn \__nmsl_scan_aux:n #1
  {
    \tl_if_blank:nF {#1}
      {
        \tl_if_head_is_group:nTF {#1}
          { \exp_args:Ne \__nmsl_group:n { \tl_head:n {#1} } }
          { \exp_args:Ne \__nmsl_head:n { \tl_head:n {#1} } }
        \exp_args:Ne \__nmsl_scan:n { \tl_tail:n {#1} }
      }
  }
\cs_new_protected:Npn \__nmsl_head:n #1
  {
    \str_if_eq:VnTF \l__nmsl_prev_tl { s }
      { \tl_set:Nn \l__nmsl_prev_tl { o } }
      { \__nmsl_token:N #1 }
  }
\cs_new_protected:Npn \__nmsl_atom:
  {
    \int_compare:nNnT \l__nmsl_depth_int = 0
      { \int_incr:N \l__nmsl_atoms_int }
    \bool_set_false:N \l__nmsl_lead_bool
  }
\cs_new_protected:Npn \__nmsl_group:n #1
  {
    \str_if_eq:VnT \l__nmsl_prev_tl { s } { \tl_set:Nn \l__nmsl_prev_tl { x } }
    \str_case:VnF \l__nmsl_prev_tl
      {
        { x } { \tl_set:Nn \l__nmsl_prev_tl { o } }
        { c } { }
      }
      {
        \seq_push:Nx \l__nmsl_stack_seq { \int_use:N \l__nmsl_atoms_int }
        \__nmsl_scan:n {#1}
        \seq_pop:NN \l__nmsl_stack_seq \l_tmpa_tl
        \int_set:Nn \l__nmsl_atoms_int { \l_tmpa_tl }
        \__nmsl_atom:
        \tl_set:Nn \l__nmsl_prev_tl { o }
      }
  }
\cs_new_protected:Npn \__nmsl_sign:
  {
    \int_compare:nNnT \l__nmsl_depth_int = 0
      {
        \bool_set_true:N \l__nmsl_sign_bool
        \bool_if:NF \l__nmsl_lead_bool { \bool_set_true:N \l__nmsl_sum_bool }
      }
    \bool_set_false:N \l__nmsl_lead_bool
    \tl_set:Nn \l__nmsl_prev_tl { o }
  }
\cs_new_protected:Npn \__nmsl_open:
  { \__nmsl_atom: \int_incr:N \l__nmsl_depth_int \tl_set:Nn \l__nmsl_prev_tl { o } }
\cs_new_protected:Npn \__nmsl_close:
  { \int_decr:N \l__nmsl_depth_int \tl_set:Nn \l__nmsl_prev_tl { o } }
\cs_new_protected:Npn \__nmsl_token:N #1
  {
    \bool_lazy_or:nnTF
      { \token_if_math_superscript_p:N #1 }
      { \token_if_math_subscript_p:N #1 }
      { \tl_set:Nn \l__nmsl_prev_tl { s } }
      { \token_if_cs:NTF #1 { \__nmsl_cs:N #1 } { \__nmsl_char:N #1 } }
  }
\cs_new_protected:Npn \__nmsl_char:N #1
  {
    \str_case:nnF {#1}
      {
        { + } { \__nmsl_sign: }
        { - } { \__nmsl_sign: }
        { ( } { \__nmsl_open: }
        { [ } { \__nmsl_open: }
        { ) } { \__nmsl_close: }
        { ] } { \__nmsl_close: }
      }
      {
        \str_if_in:nnTF { 0123456789 } {#1}
          {
            \str_if_eq:VnF \l__nmsl_prev_tl { d } { \__nmsl_atom: }
            \bool_set_false:N \l__nmsl_lead_bool
            \tl_set:Nn \l__nmsl_prev_tl { d }
          }
          { \__nmsl_atom: \tl_set:Nn \l__nmsl_prev_tl { o } }
      }
  }
\cs_new_protected:Npn \__nmsl_cs:N #1
  {
    \str_set:Nx \l__nmsl_name_str { \cs_to_str:N #1 }
    \str_case:VnF \l__nmsl_name_str
      {
        { pm } { \__nmsl_sign: }
        { mp } { \__nmsl_sign: }
        { lbrace } { \__nmsl_open: }
        { langle } { \__nmsl_open: }
        { lbrack } { \__nmsl_open: }
        { rbrace } { \__nmsl_close: }
        { rangle } { \__nmsl_close: }
        { rbrack } { \__nmsl_close: }
        { frac } { \__nmsl_fraction: }
        { tfrac } { \__nmsl_fraction: }
        { dfrac } { \__nmsl_fraction: }
        { cfrac } { \__nmsl_fraction: }
        { nmfrac } { \__nmsl_fraction: }
        { nmslash } { \__nmsl_fraction: }
        { , } { } { ; } { } { ! } { } { : } { } { > } { }
        { quad } { } { qquad } { } { thinspace } { } { medspace } { } { thickspace } { }
        { negthinspace } { } { relax } { } { protect } { } { nobreak } { }
        { displaystyle } { } { textstyle } { } { scriptstyle } { }
        { left } { } { right } { } { big } { } { Big } { } { bigg } { } { Bigg } { }
        { bigl } { } { bigr } { } { Bigl } { } { Bigr } { } { biggl } { } { biggr } { }
        { Biggl } { } { Biggr } { }
      }
      { \__nmsl_other_cs:N #1 }
  }
\cs_new_protected:Npn \__nmsl_fraction:
  {
    \int_compare:nNnT \l__nmsl_depth_int = 0 { \bool_set_true:N \l__nmsl_frac_bool }
    \__nmsl_atom:
    \tl_set:Nn \l__nmsl_prev_tl { c }
  }
\cs_new_protected:Npn \__nmsl_other_cs:N #1
  {
    \str_set:Nn \l__nmsl_spec_str { x }
    \bool_lazy_and:nnT
      { \token_if_macro_p:N #1 }
      { ! \token_if_protected_macro_p:N #1 }
      {
        \str_set:Nx \l__nmsl_spec_str { \cs_parameter_spec:N #1 }
        \str_if_in:NnT \l__nmsl_name_str { @ } { \str_set:Nn \l__nmsl_spec_str { x } }
        \tl_set:No \l__nmsl_exp_tl {#1}
        \tl_if_head_eq_meaning:VNT \l__nmsl_exp_tl \protect
          { \str_set:Nn \l__nmsl_spec_str { x } }
      }
    \bool_lazy_all:nTF
      {
        { \str_if_empty_p:N \l__nmsl_spec_str }
        { \int_compare_p:nNn \l__nmsl_level_int < { 16 } }
      }
      {
        \int_incr:N \l__nmsl_level_int
        \exp_args:No \__nmsl_scan:n {#1}
        \int_decr:N \l__nmsl_level_int
      }
      { \__nmsl_atom: \tl_set:Nn \l__nmsl_prev_tl { c } }
  }
\cs_new_protected:Npn \nmslash #1#2
  {
    \__nmsl_measure:n {#1}
    \bool_lazy_or:nnTF { \l__nmsl_sum_bool } { \l__nmsl_frac_bool } { (#1) } { #1 }
    /
    \__nmsl_measure:n {#2}
    \bool_lazy_any:nTF
      {
        { \l__nmsl_sign_bool }
        { \l__nmsl_frac_bool }
        { \int_compare_p:nNn \l__nmsl_atoms_int > { 1 } }
      }
      { (#2) } { #2 }
  }
\ExplSyntaxOff
\renewcommand{\nmfrac}[2]{\nmslash{#1}{#2}}
\AtBeginEnvironment{equation}{\renewcommand{\nmfrac}[2]{\frac{#1}{#2}}}
\AtBeginEnvironment{align}{\renewcommand{\nmfrac}[2]{\frac{#1}{#2}}}

\newcommand{\inlf}[1]{{\renewcommand{\frac}[2]{\nmslash{##1}{##2}}#1}}
\newcommand{\inlp}[1]{{\renewcommand{\nmfrac}[2]{(\nmslash{##1}{##2})}#1}}
\newcommand{\inlfp}[1]{{\renewcommand{\frac}[2]{(\nmslash{##1}{##2})}#1}}
\newcommand{\longsep}{$\par\hskip-5.5em$\displaystyle}
\newcommand{\spacedmatrix}[1]{{\renewcommand{\arraystretch}{1.5}#1}}
\let\LEFTsaved\left
\let\RIGHTsaved\right
\newcommand{\fixopen}[1]{\mathopen{\hbox{$\displaystyle\LEFTsaved#1\vphantom{\frac{d^a}{N_R}}\RIGHTsaved.$}}}
\newcommand{\fixclose}[1]{\mathclose{\hbox{$\displaystyle\LEFTsaved.\vphantom{\frac{d^a}{N_R}}\RIGHTsaved#1$}}}
\newenvironment{longeq}{\begin{equation}\begin{minipage}[c]{0.86\textwidth}\raggedright
  \leftskip=5.5em\lineskip=3pt plus 1pt\lineskiplimit=3pt\binoppenalty=300\let\left\fixopen\let\right\fixclose\longeqbig\hskip-5.5em$\displaystyle}%
  {$\end{minipage}\end{equation}\ignorespacesafterend}
\makeatletter
\newcommand{\longeqbig}{\def\bBigg@##1##2{{\@mathmeasure\z@{\nulldelimiterspace\z@}%
  {\LEFTsaved##2\vcenter to##1\big@size{}\RIGHTsaved.}\box\z@}}}
\makeatother
\newcommand{\keepparens}{\def\left##1##2\right##3{\mathinner{\hbox{$\displaystyle\LEFTsaved##1##2\RIGHTsaved##3$}}}}
\newcommand{\Trc}{\mathbf{T}^2_{r-c}}
\newcommand{\Tt}{\mathbf{T}^2_{t}}
\newcommand{\MSbar}{\overline{\mathrm{MS}}}

\newcommand{\nmfracsmall}{\renewcommand{\nmfrac}[2]{\tfrac{##1}{##2}}}
\newcommand{\narrowcols}{\setlength{\tabcolsep}{3pt}}
\newcommand{\widecols}{\setlength{\tabcolsep}{5pt}\setlength{\extrarowheight}{2pt}}
\newcommand{\widecolstight}{\setlength{\tabcolsep}{5pt}\setlength{\extrarowheight}{1.5pt}}
\newcommand{\fracrows}[1]{\renewcommand{\arraystretch}{#1}}
\newcommand{\fracrowsS}{\fracrows{1.5}}

\newcommand{\fracrowsXL}{\fracrows{1.9}}
\newcommand{\blockgap}{\noalign{\vskip\smallskipamount}}
\newcommand{\rowsapartmain}{\setlength{\extrarowheight}{2pt}}
\newcommand{\rowsapartwide}{\setlength{\extrarowheight}{4.5pt}}
\let\WtwoTopFraction\topfraction
\let\WtwoTextFraction\textfraction
\let\WtwoFloatPageFraction\floatpagefraction
\renewcommand{\topfraction}{0.95}
\renewcommand{\textfraction}{0.05}
\renewcommand{\floatpagefraction}{0.8}

\title{The signature-changing soft-gluon term\\
as a possible source of non-planar $\zeta_2$ terms at small $x$\\
in the non-singlet quark-plus-antiquark splitting function}
\author{
  Olasantan Ebenezer Adelaja \qquad
  Alex Prygarin\thanks{Corresponding author.}\\[2pt]
  Claudelle Capasia Madjuogang Sandeu\\[6pt]
  \small Department of Physics, Ariel University, Ariel 40700, Israel\\[2pt]
  \small \texttt{olasanta.adelaja@msmail.ariel.ac.il}\\[1pt]
  \small \texttt{alexanderp@ariel.ac.il}\\[1pt]
  \small \texttt{claudell.madjuoga@msmail.ariel.ac.il}}
\date{}
\newcommand{\breakbeforemathsigns}{%
  \binoppenalty=10000
  \relpenalty=10000
  \mathcode`\+="8000
  \mathcode`\-="8000
  \begingroup\lccode`\~=`\+\lowercase{\endgroup\def~}{\discretionary{}{}{}\mathchar"202B\relax}
  \begingroup\lccode`\~=`\-\lowercase{\endgroup\def~}{\discretionary{}{}{}\mathchar"2200\relax}
}
\newcommand{\countword}[1]{\ifcase#1\relax zero\or one\or two\or three\or four\or five\or six\or seven\or eight\or nine\else#1\fi}
\MakeRobust{\inlf}\MakeRobust{\inlp}\MakeRobust{\inlfpfirst}\MakeRobust{\inlfp}
\usepackage{longtable}

\newcommand{\fracrowsSupp}{\renewcommand{\arraystretch}{1.7}}
\newcolumntype{P}{p{3.0cm}}\newcolumntype{Q}{p{5.2cm}}
\newcommand{\whyPtitle}{Why \texorpdfstring{\protect\boldmath$P^+$}{P+} and not \texorpdfstring{\protect\boldmath$P^-$}{P-}}
\makeatletter
\renewcommand\section{\@startsection{section}{1}{\z@}{-2.5ex plus -1ex minus -.2ex}{1.2ex plus .2ex}{\normalfont\Large\bfseries}}
\renewcommand\subsection{\@startsection{subsection}{2}{\z@}{-2ex plus -.8ex minus -.2ex}{.8ex plus .2ex}{\normalfont\large\bfseries}}
\makeatother
\AtBeginDocument{%
  \setlength{\abovedisplayskip}{8pt plus 2pt minus 2pt}%
  \setlength{\belowdisplayskip}{8pt plus 2pt minus 2pt}%
  \setlength{\abovedisplayshortskip}{0pt plus 2pt}%
  \setlength{\belowdisplayshortskip}{5pt plus 2pt minus 2pt}%
}
\begin{document}
\maketitle

\begin{abstract}
A relation underlying resummations at small momentum fraction $x$ predicts the highest logarithms of quark-plus-antiquark flavor-difference splitting functions from lower orders. Complete three- and four-loop results violate it at third-highest logarithms by pi-squared terms suppressed for many colors, of unexplained origin and color structure. At four loops Moch and Vogt suggested a squared imaginary phase. We retain the signature-changing soft-gluon term also in positive signature in Kirschner and Lipatov's double-logarithmic equations for forward quark-antiquark scattering, without derivation, with their printed negative-signature sign and adjoint Born term. We show that the proposed equations reproduce both violations in full color for quark-plus-antiquark flavor differences at zero power of $x$, fixed coupling and first order in pi squared, for defining-representation quarks of special unitary groups with at least three colors. The proposed equations at third-highest logarithms use one wave per exchanged color and signature, the retained term at lowest-order strength as the positive-signature color-neutral channel's only new source of these violations, no such terms from gluons across the softest exchanged quark-antiquark pair at comparable or larger transverse momentum, and unchanged quark-pair normalization in the splitting function. This agreement gives \nmWordCountConsXzero{} independent tests of one system with no continuous coefficient adjusted to the violations, although corresponding non-quartic quark-minus-antiquark terms are missed. In the proposed equations, the retained term weighted by this phase is a possible origin of both violations. The equations predict these full-color five-loop quark-plus-antiquark terms if the relation otherwise holds and no new source first acts at five loops. Confirmation would support the retained term as the origin of the violations through five loops.
\end{abstract}

\section{Introduction}
\label{sec:intro}
Splitting functions govern how quark and gluon distributions change with resolution scale. Dokshitzer~\cite{Dokshitzer:1977sg} calculated the structure functions of deep inelastic scattering and electron-positron annihilation in quantum chromodynamics (QCD) at leading-logarithmic accuracy. Lipatov~\cite{Lipatov:1996sx} reviewed perturbative QCD at small momentum fractions. The four-loop splitting functions are needed for precision evolution at the third order beyond leading. The approximations of Moch et~al.~\cite{Moch:2017uml}, based on moments and endpoint limits, have their largest uncertainty at small momentum fraction $x$, as discussed by Gehrmann, von~Manteuffel, Sotnikov and Yang~\cite{Gehrmann:2026dai}. The complete four-loop calculation of Gehrmann, von~Manteuffel, Sotnikov and Yang~\cite{Gehrmann:2026qbl,Gehrmann:2026dai} removes this uncertainty for the quark-to-quark functions and tests predictions of the highest small-$x$ logarithms from lower orders.
The small-$x$ logarithms predicted from lower orders constrain these four-loop approximations and check complete calculations~\cite{Gehrmann:2026dai}. The question here is why a relation between successive orders fails in terms suppressed for many colors, and whether their origin fixes a contribution at the next order.

The relation concerns the splitting function for quark-plus-antiquark flavor differences. Positive and negative signature mean symmetry and antisymmetry under crossing, the exchange of incoming and outgoing particles, also called crossing-even and crossing-odd. The relation generalizes the positive-signature double-logarithmic equation of Kirschner and Lipatov~\cite{Kirschner:1983di} and first appeared in Ref.~\cite{Velizhanin:2014dia}. Davies, Kom, Moch and Vogt~\cite{Davies:2022ofz} stated the relation, eq.~(\ref{eq:relation}), except for color-suppressed terms containing even zeta values and possibly quartic group invariants. The complete three-loop results of Moch, Vermaseren and Vogt~\cite{Moch:2004pa} and four-loop results~\cite{Gehrmann:2026qbl} obey it at the highest and second highest logarithmic powers at $x^0$, meaning terms with no power of $x$. They also obey it at the third highest power for rational and flavor-dependent terms and for planar terms, those leading for many colors.

Gehrmann, von~Manteuffel, Sotnikov and Yang~\cite{Gehrmann:2026qbl} found deviations from the prediction of Davies, Kom, Moch and Vogt~\cite{Davies:2022ofz} for the four-loop $x^0$ coefficients at the third highest logarithmic power. These deviations are proportional to $\pi^2$ in all five color structures without the number of quark flavors $n_f$ and cancel for many colors. Moch and Vogt~\cite{Moch:2026qsw} gave the differences from the double-logarithmic predictions at even powers of $x$ for the quark-plus-antiquark function and odd powers for the quark-minus-antiquark function, including the quartic invariant formed from traces of four color generators. Moch and Vogt suggested a squared imaginary phase as a possible origin of these differences. Gehrmann, von~Manteuffel, Sotnikov and Yang~\cite{Gehrmann:2026cwg} also asked whether a phase causes these violations. Their origin and color structure remain unexplained.
Manashov, Moch and Shumilov~\cite{Manashov:2025kgf} describe the origin of the non-planar poles in QCD and in $\mathcal{N}=4$ super-Yang-Mills theory as an open question.
The non-planar poles are also discussed in Ref.~\cite{Velizhanin:2022seo}, without an account of their origin. Manashov, Moch and Shumilov~\cite{Manashov:2026qtv} relate the regularity at zero spin of the quadratic combination of anomalous dimensions on the left of the relation, eq.~(\ref{eq:relation}), to trajectory-shadow mixing in scalar and Gross-Neveu models. Chakraborty and Shumilov~\cite{Chakraborty:2026efj} obtain such regularity at the critical point through three loops for even spin, but not odd spin, in the Gross-Neveu-Yukawa model, without gauge fields or a $\zeta_2$ term. The three-loop QCD violation was given in eq.~(11) of Ref.~\cite{Velizhanin:2014dia} and reported by Shumilov~\cite{Shumilov:2026kke}, without a physical origin.

\begin{figure}[!htbp]
  \centering
  \includegraphics[width=0.95\textwidth]{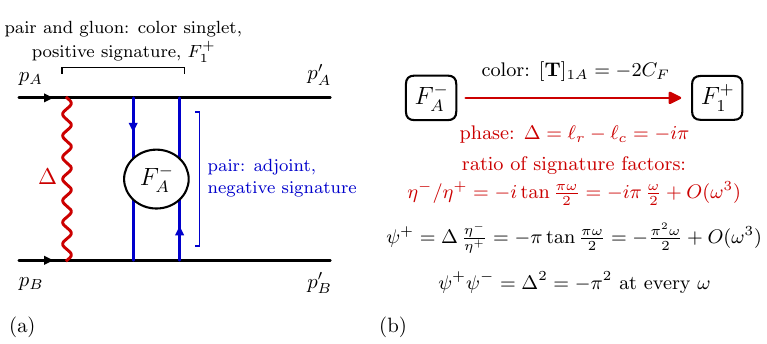}
  \caption{Signature change in forward scattering. In panel (a), the circle is the negative-signature adjoint wave $F_A^-$ of the exchanged quark-antiquark pair and the rest of the ladder graph. The softest gluon (wavy line) joins the external quark lines outside the pair and forms the positive-signature color singlet $F_1^+$ with it. Panel (b), where $\mathbf{T}$ denotes $\Trc$, gives the factors on $F_A^-$ in the equation for $F_1^+$, eq.~(\ref{eq:singlet-row}). They are the singlet-adjoint color entry $[\Trc]_{1A}=\nmSingletAdjoint$ of the phase operator, which mixes exchanged color representations, the phase $\Delta$ and the signature ratio $\eta^-/\eta^+$. Here $\Delta$ is the direct minus the crossed energy logarithm, and $\eta^\pm$ are the signature factors combined with the energy phase, eq.~(\ref{eq:sign-eta-main}), whose ratio $\eta^-/\eta^+$ equals $\xi^-/\xi^+$. The weight $\psi^+$ is $\Delta$ times the ratio without the color entry, and $\psi^-$ is the corresponding weight into negative signature.
The phase $\Delta=-i\pi$ and the ratio $\eta^-/\eta^+=-i\pi\omega/2+O(\omega^3)$ give $\psi^+=\nmPsiPlusLeading+O(\omega^3)$ in the normalization of Sect.~\ref{sec:picture}, in which the lowest-order amplitudes of both signatures coincide before the signature factors are applied.
}
  \label{fig:mechanism-w2}
\end{figure}

The physical picture is forward flavor exchange, such as $u\bar u\to d\bar d$, at large energy and small momentum transfer. The exchanged quark-antiquark pair can be a color singlet or an adjoint in SU($N_c$). A soft gluon attached to the external quarks can change the exchanged color and the signature together. Its energy logarithm supplies an imaginary phase, as shown in Fig.~\ref{fig:mechanism-w2}.
In this paper we propose extending the double-logarithmic equations of Kirschner and Lipatov beyond double-logarithmic accuracy by keeping the signature-changing soft-gluon term (the retained term) also in every positive-signature equation.
 The singlet equation couples it to the negative-signature adjoint wave. Projection onto positive signature supplies a second phase in the normalization of Sect.~\ref{sec:picture}, so the term is proportional to $\pi^2$ and enters two logarithms below the leading logarithm, at next-to-next-to-leading logarithmic accuracy (NNLL). Section~\ref{sec:equations} defines the proposal, including its assumptions, in Table~\ref{tab:conditions-main}.

Kirschner and Lipatov~\cite{Kirschner:1982xw,Kirschner:1982qf,Kirschner:1983di} derived the double-logarithmic equations by separating graphs according to their softest virtual particle.\footnote{Double logarithms for fermion exchange in quantum electrodynamics were obtained by Frolov, Gorshkov, Gribov and Lipatov~\cite{Gorshkov:1966qed,Gorshkov:1967qed,Gorshkov:1967ee}. We cite Ref.~\cite{Gorshkov:1967qed} from the credits in Ref.~\cite{Kirschner:1983di} and the paper of Bartels, Ermolaev and Ryskin~\cite{Bartels:1995iu}, and Refs.~\cite{Gorshkov:1967ee,Kirschner:1980tx} from the credits in Ref.~\cite{Kirschner:1983di}. Kirschner's non-abelian extension~\cite{Kirschner:1980tx} is credited on p.~124 of Ref.~\cite{Kirschner:1983di}.}
They retained signature change in negative signature and set it aside in positive signature
as negligible at double-logarithmic accuracy,
on p.~270 of Ref.~\cite{Kirschner:1982xw} and p.~134 of Ref.~\cite{Kirschner:1983di}. Kirschner~\cite{Kirschner:1983cut} set the signature-changing term aside in positive signature for quark exchange because it has an $i\pi$ constant in place of the energy logarithm, ``one logarithm less'', in eq.~(19) of Ref.~\cite{Kirschner:1983cut} on p.~525. For color-singlet exchange Kirschner set the signature-changing term aside as ``inessential in the cases considered here'' on p.~137 of Ref.~\cite{Kirschner:1985cb}. Kirschner obtained, for forward quark scattering with gluon exchange, the positive-signature singlet at first order in $i\pi$ from the full reggeized negative-signature adjoint amplitude, not from a double-logarithmic partial wave, eq.~(13) of Ref.~\cite{Kirschner:1983cut}. The adjoint terms retained for transversity by Kirschner, Mankiewicz, Sch\"afer and Szymanowski~\cite{Kirschner:1996jj} and for the photon structure function by Ermolaev, Kirschner and Szymanowski~\cite{Ermolaev:1998jv} feed a negative-signature adjoint wave into a positive-signature equation at double-logarithmic accuracy, with the signature roles interchanged near angular momentum~$-1$.

Ermolaev, Greco and Troyan~\cite{Ermolaev:2000sg} include signature-changing terms in both signatures with running coupling in eq.~(28) of Ref.~\cite{Ermolaev:2000sg}.
 Their positive-signature contribution from the signature-changing graphs in eq.~(32) of Ref.~\cite{Ermolaev:2000sg} vanishes at fixed coupling in the double-logarithmic approximation with the octet at Born level.
They justify the Born approximation of the octet by the decrease with energy of the positive-signature octet feeding the negative-signature singlet, on p.~13 of Ref.~\cite{Ermolaev:2000sg}. They do not discuss the negative-signature octet feeding the positive-signature singlet.
The Born approximation removes the loop poles responsible for the NNLL violations from three loops on. Fixed coupling does not remove them.
 Section~\ref{sec:equations} gives the approximation and shows which of their equations agree with the common sign used here.

To the best of our knowledge, no earlier paper evaluates quark-antiquark signature change beyond double-logarithmic accuracy using the all-order double-logarithmic negative-signature adjoint wave instead of its Born term.

Borden and Kovchegov~\cite{Borden:2023ugd} find that the double-logarithmic polarized flavor-singlet gluon anomalous dimension at large~$N_c$ differs from the flavor-singlet infrared evolution of Bartels, Ermolaev and Ryskin~\cite{Bartels:1996wc} from four loops. No fixed-order result distinguishes these two flavor-singlet four-loop predictions. Borden and Kovchegov~\cite{Borden:2023ugd} suggest hard gluons outside the ladder graphs as a possible explanation, without reaching a firm conclusion. This comparison concerns another channel, but motivates checking gluons that are not the softest particle.

For gluon exchange, Korchemskaya and Korchemsky~\cite{Korchemskaya:1994qp} obtain a small Wilson-loop eigenvalue proportional to $\pi^2/\ln(s/m^2)$ that controls high-energy singlet exchange for massive quarks of mass $m$. Del~Duca et~al.~\cite{DelDuca:2011ae} find NNLL breaking of single-Reggeon factorization.
Caron-Huot, Gardi and Vernazza~\cite{Caron-Huot:2017fxr} identify the two-loop three-Reggeon term of the signature-odd amplitude, proportional to a squared imaginary phase, with a Regge cut, a branch singularity produced by exchange of several reggeized particles.
 Falcioni et~al.~\cite{Falcioni:2021dgr} develop the pole and cut separation.
Falcioni, Gardi, Milloy and Vernazza~\cite{Falcioni:2020lvv} compute the four-loop NNLL signature-odd gluon-exchange amplitude in full color, with transitions between one and three Reggeons. The result is entirely non-planar and contains quartic invariants. Caron-Huot, Gardi, Reichel and Vernazza~\cite{Caron-Huot:2017zfo,Caron-Huot:2020grv} compute the positive-signature two-Reggeon gluon-exchange amplitude at next-to-leading logarithmic accuracy (NLL). These results concern gluon-exchange amplitudes and powers of $\ln(s/|t|)$, not quark-antiquark exchange or powers of $\ln x$ in a splitting function.

The proposed equations express the violations through one wave, eq.~(\ref{eq:S}) in Sect.~\ref{sec:equations}.
 The combination $S=P^+(P^+-\omega)+2aC_F$ has NNLL $\zeta_2$ poles whose coefficients are the violations $S_n$ for $n\ge3$. Here $P^+$ is the quark-plus-antiquark flavor-difference splitting function, $a=\alpha_s/(4\pi)$ is the coupling, $N$ is the Mellin moment, $\omega=N$ at $x^0$, $\zeta_2=\pi^2/6$, and $\kappa$ is the common sign of signature change.
\[
 S(\omega)\big|_{\rm NNLL\ poles}
 =\nmReductionCoef\,\kappa\,a\zeta_2 C_F\big[\omega F_A^-(\omega)\big]_{\rm poles} .
\]
The pole part on the right excludes the Born term. At each order, $S_{n+1}=\nmReductionCoef\kappa\zeta_2C_Fd_n$ for $n\ge2$, where $d_n$ is the coefficient of $a^n\omega^{1-2n}$ in $F_A^-$. Thus the $n$-loop adjoint wave fixes the violation at $n+1$ loops in the proposed equations.
The Born and regular one-loop terms give the undetermined boundary coefficients at two loops and one loop, respectively. The small-$N$ assumption means that the relation holds at NNLL apart from the retained term. It is needed to turn the higher-order violations into splitting-function coefficients. The all-order violations from the proposed equations would replace the continued three-loop non-planar terms of the NNLL resummation of Davies, Kom, Moch and Vogt~\cite{Davies:2022ofz}, at fixed coupling and first order in $\zeta_2$, using the complete four-loop result and assuming no new source beyond four loops.
Related resummations of the $1/x$ double logarithms in semi-inclusive electron-positron annihilation were obtained by Vogt~\cite{Vogt:2011jv} and Kom, Vogt and Yeats~\cite{Kom:2012hd}.

Sections~\ref{sec:relation-w4} and \ref{sec:equations} define the relation and the proposed equations. Sections~\ref{sec:threefour} to \ref{sec:amplitude} compare the complete results, show why adjoint corrections leave the poles $S_n$ of $S$ unchanged, and check the amplitude. Section~\ref{app:towercore} gives the five-loop coefficient and the sequence $S_{n\ge3}$ of NNLL $\zeta_2$ violations at all orders, called the tower. The appendices give the derivations. The ancillary notes, \texttt{notes.pdf}, are among the arXiv ancillary files. Their numbering uses the prefix N, beginning with Sect.~N1, eq.~(N1) and Table~N1. Their Sect.~N1 gives the fuller literature comparison.

We show that the proposed equations reproduce every color structure of the three- and four-loop violations in the quark-plus-antiquark flavor-difference function at $x^0$, including the quartic invariant, at fixed coupling and first order in $\zeta_2$, with the sign $\kappa=+1$ that Kirschner and Lipatov~\cite{Kirschner:1983di} printed for the negative-signature term, used for both signatures, and the adjoint Born term $b_A^-=b_A$ they printed in SU($N_c$).
This agreement gives \nmWordCountConsXzero{} independent tests of one system, with no continuous coefficient adjusted to the violations (Sect.~\ref{sec:threefour}), although the equations do not reproduce two groups of non-quartic NNLL $\zeta_2$ terms whose source remains unidentified, those of the quark-minus-antiquark function $P^-$ at $x^0$ and positive even powers of $x$ and those of $P^+$ at odd powers (Sect.~\ref{sec:protection-w4}).
Within the proposed equations, the retained term, weighted by a squared imaginary phase of the form Moch and Vogt suggested~\cite{Moch:2026qsw}, is a possible origin of the three- and four-loop violations of the quark-plus-antiquark function at $x^0$.
 The proposed equations also predict the $\zeta_2$ coefficient at the third highest logarithmic power of the five-loop quark-plus-antiquark function, with no power of $x$, in general color, eq.~(\ref{eq:fivegen-w2}). This prediction uses the small-$N$ assumption and assumes no new source first acting at five loops. Confirmation would constrain five-loop approximations beyond the large-$N_c$ reconstruction of Herzog et~al.~\cite{Herzog:2018kwj}.

\section{The relation and its violations}
\label{sec:relation-w4}
We write $P^\pm=P_{\rm ns}^\pm$ for the even- and odd-moment splitting functions of the flavor differences $q_i+\bar q_i-(q_k+\bar q_k)$ and $q_i-\bar q_i-(q_k-\bar q_k)$.
Their signatures are positive and negative, respectively.
The coupling is $a=\alpha_s/(4\pi)$, and $P^{(n)}$ is the coefficient of $a^{n+1}$ in $P$.
The leading logarithm (LL) has the highest power of $\ln x$ at each order. Next-to-leading logarithms (NLL) have one power less, and NNLL have two powers less.
The Mellin moment $N$ converts $a^n\ln^{m-1}(1/x)/(m-1)!$ at $x^0$ into $a^nN^{-m}$.
The LL, NLL and NNLL poles at order $a^n$ are therefore proportional to $N^{1-2n}$, $N^{2-2n}$ and $N^{3-2n}$.
The relation for $P^+$ is
\begin{equation}
  P^+(N)\,\big(P^+(N)-N+\beta(a)/a\big)=O(N^0),
  \label{eq:relation}
\end{equation}
where $\beta(a)=-\beta_0a^2-\dots$ is the beta function.
The prediction at each order follows by inserting the complete lower orders on the left.

The quartic invariant is $d_F^{abcd}d_A^{abcd}/N_R$ in the conventions of van Ritbergen, Schellekens and Vermaseren~\cite{vanRitbergen:1998pn}, where $N_R$ is the dimension of the quark representation.
The variable $\omega$ is the angular momentum conjugate to the energy logarithm and equals $N$ at $x^0$.
We define \mbox{$S=P^+(P^+-\omega)+2aC_F$} and denote by $S_n$ the $\zeta_2$ part of its $a^n\omega^{4-2n}$ coefficient, including $\zeta_2$.
Thus $S_n$ contains the NNLL coefficient of the $n$-loop function $P^{(n-1)}$ and products of lower orders.
The harmonic sum $S_1(N)$ used in Sect.~\ref{app:towercore} is distinct from these coefficients $S_n$.
The equation $S=0$ is the positive-signature double-logarithmic equation.
The coefficients $S_n$ with $n\ge3$ measure the NNLL violations of eq.~(\ref{eq:relation}) because its beta-function terms contain no $\zeta_2$ at NNLL.

The complete results~\cite{Moch:2004pa,Gehrmann:2026qbl} give
\begin{align}
  S_3&=\nmSthreeGen , \label{eq:S3}\\
  S_4&=\nmSfourGen . \label{eq:S4}
\end{align}

The extraction and complete coefficients are in App.~\ref{app:core-data}, Table~\ref{tab:datatable}.
The violations contain no rational or $n_f$-dependent part.
In SU($N_c$), $S_3=O(N_c)$ is suppressed by the inverse square of the number of colors relative to planar three-loop terms, while $S_4=O(N_c^0)$ is suppressed by the inverse fourth power relative to planar four-loop terms.

\section{The proposed equations}\label{sec:equations}
\label{sec:picture}
The proposed equations are the double-logarithmic system of Kirschner and Lipatov~\cite{Kirschner:1983di} with signature change retained in every positive-signature row. Table~\ref{tab:conditions-main} defines the proposal. The retained term at lowest-order strength is the only new source of NNLL poles of $S$ in the positive-signature singlet equation. Gluons across the softest pair at comparable or larger transverse momentum add no crossing-even non-planar $\zeta_2$ NNLL term. The map from the cutoff wave to the $\MSbar$ splitting function has the quark-pair normalization and no further NNLL term. One wave per exchanged representation and signature suffices for evolution at NNLL. The representation conditions require each channel in the quark representation times its conjugate to occur once, with the complex-representation signature assignment also used for real and pseudo-real representations. Fundamental SU($N_c$) quarks satisfy them for $N_c\ge3$.

The softest gluon attaches to the external quarks through the one-loop dipole operator, which Kirschner and Lipatov used at double-logarithmic accuracy~\cite{Kirschner:1983di}.
Its form beyond that accuracy is given by Kidonakis, Oderda and Sterman~\cite{Kidonakis:1998nf}, Catani~\cite{Catani:1998bh}, Becher and Neubert~\cite{Becher:2009cu,Becher:2009qa} and Gardi and Magnea~\cite{Gardi:2009qi,Gardi:2009zv}.
\begin{table}[tbp]
  \centering
  \caption{Definition and status of the proposed equations. The representation conditions are stated in the text and App.~\ref{app:core-color}. An input confirmed by the complete results is not counted as a test. Further assumptions and the calculations that would decide them are given in App.~\ref{app:assumptions}.}\label{tab:conditions-main}\label{tab:inputs-compact}
  \footnotesize
  \widecols\rowsapartwide
  \begin{tabular}{>{\raggedright\arraybackslash}p{2.4cm}>{\raggedright\arraybackslash}p{4.4cm}>{\raggedright\arraybackslash}p{3.6cm}>{\raggedright\arraybackslash}p{4.7cm}}
  \hline
  Name & Statement & Status & What would decide it\tabularnewline
  \hline
  Dipole kernel & Softest gluon acts through the one-loop dipole operator with its phases &
Derived at double-logarithmic accuracy by Kirschner and Lipatov~\cite{Kirschner:1983di}, asserted beyond it, consistent with the four-quark amplitudes at one to three loops (Sect.~\ref{sec:amplitude}) & Derivation beyond double-logarithmic accuracy\tabularnewline
  Direct logarithm & Graphs without crossed-channel discontinuity depend on the direct logarithm alone at leading power & Asserted, consistent with the four-quark amplitudes at one to three loops (Sect.~\ref{sec:amplitude}) & Derivation beyond double-logarithmic accuracy\tabularnewline
  Sign $\kappa=\nmKappa$ & Common sign of both signature-changing weights & Discontinuity derivation with the asserted dipole kernel and direct-logarithm dependence. Confirmed by the complete results & $P^-$ and four-quark leading logarithms confirm the printed negative-signature sign\tabularnewline
  Adjoint Born term & $b_A^-=b_A$, nonzero for the presence and sign comparisons & Printed by Kirschner and Lipatov~\cite{Kirschner:1983di} in SU($N_c$), confirmed by the complete results & $S_3$ confirms presence, $S_4$ tests sign\tabularnewline
  Non-singlet Born and pair constants & One-loop pole $\inlf{\nmOneLoopPole}$ of $F_1^+$, giving $\nmOneLoopSterm$ in $S$, fixed by $P^{(0)}$.
Unit quark-pair weights in both channels of the double-logarithmic equations of Kirschner and Lipatov~\cite{Kirschner:1983di} & Born weight fixed, pair terms derived at double-logarithmic accuracy & One-loop residue and double-logarithmic derivation\tabularnewline
  Single-source assumption & $\psi^+$ keeps its leading coefficient. No other new source of NNLL poles of $S$ in the positive-signature singlet equation. Its extension in full beyond four loops is used only for the five-loop and tower results & Assumed & NNLL derivation of the positive-signature singlet equation\tabularnewline
  No-hard-gluon assumption & No crossing-even non-planar $\zeta_2$ NNLL term from gluons across the softest pair at comparable or larger transverse momentum & Assumed, partly computed &
Calculation of the weight of a gluon with transverse momentum comparable to that of the softest pair, attached to the external legs at one end only, and of the three-loop configurations with two gluons that are not the softest parton (App.~\ref{app:status})
\tabularnewline
  Normalization condition & Map from the cutoff wave to the $\MSbar$ splitting function has $c_+=\ell_+$ and no further NNLL term & Imposed. The absence of a planar violation requires $c_+=\ell_+$ if the difference is color-independent and no other map or quark-pair term is present & NNLL Compton and on-shell quark-pair integrals\textsuperscript{a}\tabularnewline
  One-wave assumption & One partial wave per $t$-channel representation and signature suffices for evolution at NNLL & Assumed at NNLL &
NNLL derivation with composites containing reggeized gluons, not examined here\tabularnewline
  \hline
  \end{tabular}
\par\smallskip
{\raggedright\textsuperscript{a}The positive-signature double-logarithmic forms of the NNLL Compton and on-shell quark-pair integrals are eqs.~(22) and (29) of the arXiv version of Ref.~\cite{Ermolaev:1995fx} by Ermolaev, Manayenkov and Ryskin, with the signature label from eq.~(5.12) of Ref.~\cite{Kirschner:1983di} (App.~\ref{sm:sign-projection}).\par}
\end{table}
Each pair of legs contributes its color-charge product times the logarithm of its energy invariant.
The direct and crossed energy invariants are $s_r$ and $s_c$.
The forward limit at leading power gives $s_c\simeq-s_r$.
Crossing exchanges them, so crossing-even and crossing-odd mean symmetric and antisymmetric under $s_r\leftrightarrow s_c$.
We denote the direct and crossed energy logarithms by $\ell_r=\ln(-s_r-i0)$ and $\ell_c=\ln(-s_c-i0)$, with invariants in units of $\mu^2$, where $\mu$ is the transverse cutoff.
Their difference is $\Delta=\ell_r-\ell_c$ and their mean is $\tilde L=(\ell_r+\ell_c)/2$.
The physical region $s_r>0$, $s_c<0$ gives the direct logarithm its $-i\pi$ and makes the crossed logarithm real, so $\Delta=-i\pi$.
The diagonal part of the dipole operator gives Sudakov suppression and vanishes on the color singlet.
Its phase part mixes singlet and adjoint.
Because the phase changes sign when the energy channels are exchanged, it also reverses signature.
We use the normalization where the lowest-order amplitudes of both signatures coincide before the signature factors are applied.
Figure~\ref{fig:mechanism-w2} shows how the softest gluon converts a negative-signature adjoint quark-antiquark pair into a positive-signature color singlet.

Each partial wave in this normalization enters the amplitude multiplied by a signature factor, defined by
\begin{equation}
  \xi^{\pm}(\omega)=\nmSigFactor .
  \label{eq:xi-main}
\end{equation}
\makeatletter%
\protected@edef\@currentlabel{\theequation}\label{eq:xi}%
\makeatother%
The mean logarithm in the physical region is $\tilde L=L-i\pi/2$, with $L=\ln(s_r/\mu^2)$.
The signature factors combine with the energy dependence at $s_r>0$ to give
\begin{equation}
  e^{\omega L}\xi^\pm=\tfrac12\big(e^{\omega\ell_r}\pm e^{\omega\ell_c}\big)=e^{\omega\tilde L}\eta^\pm ,\qquad
  \eta^+=\cosh\frac{\omega\Delta}{2},\qquad \eta^-=\sinh\frac{\omega\Delta}{2}.
  \label{eq:sign-eta-main}
\end{equation}
\makeatletter%
\protected@edef\@currentlabel{\theequation}\label{eq:sign-eta}%
\makeatother%
The energy phase is $e^{\omega(L-\tilde L)}=e^{i\pi\omega/2}$, removed from $e^{\omega L}$ when the latter is written as $e^{\omega\tilde L}$, so $\eta^\pm=e^{i\pi\omega/2}\xi^\pm$.

The signature factors combined with the energy phase are therefore \mbox{$\eta^+=\cos(\pi\omega/2)$} and \mbox{$\eta^-=-i\sin(\pi\omega/2)$}, with the same ratio $\eta^-/\eta^+=\xi^-/\xi^+$.
Projection of a signature-changing contribution divides by the factor of the receiving wave.
The resulting weight therefore contains a ratio of signature factors as well as the soft-gluon phase.

For negative signature the phase product is
\begin{equation}
  \psi^-=\Delta\,\frac{\eta^+}{\eta^-}=(-i\pi)\Big(i\cot\frac{\pi\omega}{2}\Big)
  =\pi\cot\frac{\pi\omega}{2}=\frac{2}{\omega}+O(\omega),
  \label{eq:psiminus}
\end{equation}
which is double-logarithmic and of the same order as the Born term.
The positive-signature phase product is
\begin{equation}
  \psi^+=\Delta\,\frac{\eta^-}{\eta^+}=(-i\pi)\Bigl(-i\tan\frac{\pi\omega}{2}\Bigr)
  =-\pi\tan\frac{\pi\omega}{2}=-\frac{\pi^2\omega}{2}+O(\omega^3).
  \label{eq:psiplus}
\end{equation}

The two weights in this normalization can also be written
\begin{equation}
 \psi^+=\Delta\tanh\frac{\omega\Delta}{2},\qquad
 \psi^-=\Delta\coth\frac{\omega\Delta}{2},\qquad
 \psi^+\psi^-=\Delta^2.
 \label{eq:phase-hyperbolic-w3}
\end{equation}
Both weights are even in $\Delta$, so the two physical regions give the same equations.

The positive-signature weight is smaller in magnitude than the negative-signature weight by the factor $\pi^2\omega^2/4$ near zero angular momentum, so it first contributes at NNLL with $\pi^2=6\zeta_2$.
The phase in this normalization is $-i\pi$ and the ratio is $-i\pi\omega/2$ at leading order. Their product supplies the squared phase suggested by Moch and Vogt~\cite{Moch:2026qsw}. The Born waves coincide in both signatures in this normalization because they come from a single pole in $s_r$. Planar ladder graphs have no crossed-channel discontinuity. If graphs without a crossed-channel discontinuity depend on the direct energy logarithm alone, the waves of planar ladder graphs also coincide in both signatures at leading power. The dependence on the direct energy logarithm alone is the asserted input of Table~\ref{tab:conditions-main} and App.~\ref{sec:sign}. A signature normalization $n_\pm$ constant or dependent on $\omega$, and nonzero at the origin, multiplies $\psi^+$ by $n_-/n_+$ and $\psi^-$ by its inverse. Thus $\psi^+$ itself and the split into the factors $-i\pi$ and $-i\pi\omega/2$ depend on the relative signature normalization. The identity $\psi^+\psi^-=\Delta^2=-\pi^2$ holds at every $\omega$ and is unchanged by a change of normalization of the signature factors.
The factors $n_-/n_+$ and $n_+/n_-$ cancel in the product.
The induced changes in the waves, Born terms, quark-pair terms, Sudakov terms and splitting-function map leave $S$ and every $S_n$ unchanged (App.~\ref{sec:reduction}). The $P^-$ leading logarithms fix the leading $\psi^-$ in this normalization.

The normalization of the positive-signature singlet wave in the splitting function is
\begin{equation}
  P^+=(1+c_+\pi^2\omega^2)F_1^+ .
  \label{eq:equations-cplus}
\end{equation}

The double-logarithmic equations do not determine $c_+$ or the NNLL quark-pair weight $\ell_+$. The normalization condition in Table~\ref{tab:conditions-main} sets them equal.

The Mellin integral of the partial wave times its signature factor gives the forward amplitude. At double-logarithmic accuracy the singlet waves give $P^\pm$~\cite{Kirschner:1983di}. The NNLL map to $P^+$ requires the normalization condition.

The effect in $P^+$ is non-planar because both sources of the negative-signature quark adjoint wave decrease as $1/N_c$ in SU($N_c$).
These sources are its Born weight and its coupling to the positive-signature singlet.
The change at order $a^n$, with coupling $a=\alpha_s/(4\pi)$, grows at most as $N_c^{n-2}$, compared with $N_c^n$ for planar double logarithms.

The infrared evolution equations separate the softest gluon from the softest quark-antiquark pair~\cite{Kirschner:1983di}.
Their partial wave $F_R^\pm(\omega)$ describes a pair in exchanged color representation $R$ and signature~$\pm$.
The singlet is $R=1$, and the adjoint is $R=A$.
The constants $\ell_\pm$ parametrize the NNLL normalization of the quark-pair term.
The fixed-coupling system with the signature-changing term retained is
\begin{equation}
  \omega F_R^\pm=2a\,b_R+2aC_R\Big[\frac{dF_R^\pm}{d\omega}+\phi^\pm F_R^\pm\Big]
  +\big(1+\ell_\pm\pi^2\omega^2\big)\big(F_R^\pm\big)^2
  +2a\,\kappa\,\psi^\pm\sum_{R'}\big[\Trc\big]_{RR'}F_{R'}^\mp ,
  \label{eq:system}
\end{equation}
where the Born weight $b_R=\inlf{\nmBornWeight}$ describes one-gluon exchange, and $C_R$ is the quadratic Casimir of the exchanged representation.
The quark Casimir is $C_F$ and the adjoint Casimir is $C_A$.
Incoming legs $1$ and $2$ are the quark and antiquark and carry the direct invariant $s_r$.
Legs $1$ and $3$ carry the momentum transfer $t$, and legs $1$ and $4$ the crossed invariant $s_c$.
The phase operator $\Trc=\mathbf T_1\cdot\mathbf T_2-\mathbf T_1\cdot\mathbf T_4=(\mathbf T_s^2-\mathbf T_u^2)/2$ mixes the representations, where $\mathbf T_i$ acts on the color index of leg $i$, and $\mathbf T_s^2$ and $\mathbf T_u^2$ are the Casimirs of the pairs $(1,2)$ and $(1,4)$.
The phase operator $\Trc$ equals $\mathbf T_{s-u}^2$ of eq.~(4.10) of Ref.~\cite{Caron-Huot:2017fxr} after interchange of legs $3$ and $4$.

The terms on the right describe one-gluon exchange, Sudakov suppression, the softest quark-antiquark pair and signature change, in that order.
The logarithmic derivatives of the signature factors combined with the energy phase are
\begin{equation}
  \phi^+=-\frac{\pi}{2}\tan\frac{\pi\omega}{2},\qquad \phi^-=\frac{\pi}{2}\cot\frac{\pi\omega}{2},
  \label{eq:phases}
\end{equation}
and $\kappa=\nmKappa$ is the sign derived in App.~\ref{sec:sign}.
Only the leading terms of the tangent and cotangent about $\omega=0$ enter the results (App.~\ref{sec:reduction}).
We write $F_R^{\pm(n)}$ for the order-$a^n$ term of $F_R^\pm$.
At leading logarithm, the positive-signature waves obey $F_R^{+(2)}/F_R^{+(1)}=2a(b_R-C_R)/\omega^2$, so the positive-signature amplitude $2a^2b_R(b_R-C_R)\tilde L^2$ contains the Sudakov factor $1-aC_R\tilde L^2$ relative to the Born term, with $\tilde L=L-i\pi/2$.

The leading part of the Sudakov term of eq.~(\ref{eq:system}) vanishes on the negative-signature Born term $2ab_R/\omega$, while its NNLL remainder is $-2aC_R\zeta_2ab_R$.
The first negative-signature loop terms at double-logarithmic accuracy therefore come from the quark-pair and signature-changing terms.
At double-logarithmic accuracy, $\phi^+$ and $\psi^+$ vanish, while $\phi^-$ and $\psi^-$ retain their poles at zero.
The system then reduces to the equations of Kirschner and Lipatov~\cite{Kirschner:1983di}, including the color-singlet equations of Kirschner and Lipatov and of Bartels, Ermolaev and Ryskin~\cite{Bartels:1995iu}. Its NNLL extension is assumed.
The representation conditions give a complete decomposition of the forward amplitude without helicity flip. Extending these waves to evolution with gluons that are not the softest parton is assumed. Composites containing reggeized gluons could require further waves at NNLL. A new wave in the positive-signature singlet row would be another source excluded by the proposal (App.~\ref{app:status}).

The singlet has no Sudakov term because $C_1=0$.
Its phase-operator row has only the adjoint entry $[\Trc]_{1A}=\nmSingletAdjoint$ with the representation conditions stated above and in App.~\ref{app:core-color}.
The singlet row is therefore
\begin{equation}
  \omega F_1^+=2aC_F+\big(1+\ell_+\pi^2\omega^2\big)\big(F_1^+\big)^2+4a\kappa C_F\,\pi\tan\frac{\pi\omega}{2}\,F_A^- .
  \label{eq:singlet-row}
\end{equation}

Substitution of eq.~(\ref{eq:equations-cplus}) and elimination of $(F_1^+)^2$ give, at $\kappa=\nmKappa$ and with the normalization condition,
\begin{equation}
  \begin{aligned}
  S(\omega)\equiv P^+\big(P^+-\omega\big)+2aC_F
  &=-\big(1+c_+\pi^2\omega^2\big)\,4aC_F\,\pi\tan\frac{\pi\omega}{2}\,F_A^-(\omega)
   -2a\,c_+\pi^2\omega^2\,C_F\\
  &=\nmReductionCoef\,a\,\zeta_2\,C_F\,\omega\,F_A^-(\omega)\nmSoneCplusShift\,a\,\omega^2
   +O(\zeta_2^2).
  \end{aligned}
  \label{eq:S}
\end{equation}

Equation~(\ref{eq:S}) has no LL or NLL poles because the double-logarithmic $S$ vanishes and the weight of $F_A^-$ contains $\zeta_2\omega$.
The NNLL coefficients of the complete results contain $\zeta_2$ only linearly (Table~\ref{tab:datatable}).
In the proposed equations, $\zeta_2^2$ first enters two powers of $\omega$ beyond NNLL, through the $\omega^3$ term of $\psi^+$, the factor $(1+c_+\pi^2\omega^2)$ of the map in eq.~(\ref{eq:equations-cplus}), and the $\zeta_2$ terms of the negative-signature adjoint equation, whose coefficients are not derived.
A running-coupling term brings $\beta_0$ and hence $n_f$, which the violations lack.
The NNLL pole comparisons therefore concern the terms linear in $\zeta_2$ without running-coupling contributions. The sum of the non-planar NNLL poles to all orders (Sect.~\ref{app:towercore}) and $S$ at finite $\omega$ are studied at fixed coupling and first order in $\zeta_2$.

The normalization condition cancels the $(F_1^+)^2$ terms and removes $c_+$ from every $S_{n\ge2}$ at first order in $\zeta_2$. The term $\nmSoneCplusShift\,a\,\omega^2$ changes only the regular one-loop coefficient $S_1$.
The expansion is about $\omega=O(\sqrt a)$, outside the physical even moments.

Equation~(28) of Ref.~\cite{Ermolaev:2000sg} is written with running coupling, as stated in Sect.~\ref{sec:intro}. It gives the last term of eq.~(\ref{eq:singlet-row}) and its negative-signature partner with a common sign if evaluated with the full signature factors at fixed coupling.
 The choice $\kappa=+1$ for an octet with Born weight $b_A$ agrees with eqs.~(31), (36), (37) and (38) of Ref.~\cite{Ermolaev:2000sg}, whereas eqs.~(30) and (32) print the opposite overall sign. Equation~(28) of Ref.~\cite{Ermolaev:2000sg} alone does not fix it. The Born approximation gives
\[
 4a\kappa C_F\,\pi\tan\frac{\pi\omega}{2}\,\frac{2ab_A}{\omega}
 =4\pi^2\kappa C_Fb_Aa^2+O(\omega^2).
\]
This is regular in $\omega$ and changes the undetermined two-loop boundary coefficient $S_2$, not the poles, and Sect.~\ref{sec:protection-w4} compares this $S_2$ with the complete-result value. The loop terms of $F_A^-$ give the poles $S_{n\ge3}$.

The positive-signature contribution of the signature-changing graphs vanishes in eq.~(32) of Ref.~\cite{Ermolaev:2000sg}, as stated in Sect.~\ref{sec:intro}. This zero is the fixed-coupling limit of eq.~(31) of Ref.~\cite{Ermolaev:2000sg} with the Born octet of eq.~(29) of Ref.~\cite{Ermolaev:2000sg} and $s\sim\mu^2$ omitted. The negative-signature part of that Born octet vanishes at fixed coupling. The first-order signature factors in eq.~(18) of Ref.~\cite{Ermolaev:2000sg} also give zero, even for a non-Born octet.
The phase-space $\pi^2$ terms mentioned on p.~17 of Ref.~\cite{Ermolaev:2000sg} are excluded here by the single-source assumption.

The small-$N$ assumption is needed for the five-loop splitting function and the tower. The five-loop prediction assumes no new source first acting at five loops, and the tower assumes none beyond four loops. The shifted-pole assumption applies the proposed equations at $N=-k$, multiplying the Born weights by the ratio $\lambda_k$ of the one-loop residue there to its value at $N=0$. The ratio $\lambda_k$ equals $2$ for $k\ge2$. The status and deciding calculations of these assumptions are stated in App.~\ref{app:assumptions}.

\section{Three and four loops}
\label{sec:threefour}\label{app:count}
The NNLL poles follow from the loop coefficients of the double-logarithmic negative-signature adjoint wave,
\begin{equation}
  F_A^-=\sum_{n\ge1}d_n\,a^n\omega^{1-2n},\qquad S_{n+1}=\nmReductionCoef\,\kappa\,\zeta_2C_Fd_n\quad(n\ge2).
  \label{eq:Sn-dn}
\end{equation}
The Born term and the next two coefficients are
\[
 d_1=2b_A^-,\qquad d_2=4(b_A^-)^2+8\kappa M_1,\qquad
 d_3=4(b_A^--C_A)\,d_2+16\kappa\,{(3M_2-2C_FM_1)} .
\]
Here $M_p\equiv\sum_{R'}[\Trc]_{AR'}\,b_{R'}^p$ are adjoint moments of the positive-signature Born weights.
The quark-pair term acts on the Born term of $F_A^-$ with weight $b_A^-$, the Sudakov term has coefficient $C_A$, and the feed from positive signature has sign $\kappa$.
The positive-signature waves enter $d_3$ at second order, $F_R^+=2ab_R/\omega+4a^2b_R(b_R-C_R)/\omega^3$, and $C_R=2(C_F-b_R)$ turns their adjoint sum, weighted by the adjoint row of the phase operator as in $M_p$, into three times $M_2$ minus twice $C_FM_1$.
The color reduction gives
\begin{equation}
  \begin{aligned}
  M_1&=\nmMoneGen ,\\
  M_2&=\nmMtwoGen .
  \end{aligned}
  \label{eq:M12}
\end{equation}
In SU($N_c$), $[\Trc]_{11}=0$, $[\Trc]_{1A}=\nmSingletAdjoint$, $[\Trc]_{A1}=-1/N_c$ and $[\Trc]_{AA}=2/N_c-N_c/2$, with $b_1=C_F$ and $b_A=-1/(2N_c)$, give $M_1=-(N_c^2+2)/(4N_c^2)$ and $M_2=-(N_c^2-2)(2N_c^2+1)/(8N_c^3)$.

The expressions for $d_2$ and $d_3$ give $S_3$ and $S_4$ in eqs.~(\ref{eq:S3}) and (\ref{eq:S4}) at $\kappa=+1$ and $b_A^-=b_A$.
The general-color reduction is in App.~\ref{app:core-color}.
The leading $O(N_c)$ terms in $d_3$ from the Sudakov and quark-pair action cancel the feed from positive signature for both signs of $\kappa$.
The result is odd in $N_c$ and equals \inlf{$\nmDthreeSUN$} at $\kappa=\nmKappa$.
If the single-source assumption holds and the overall coefficient in eq.~(\ref{eq:S}) is fixed, the quartic term $-2d_F^{abcd}d_A^{abcd}/(N_RC_F)$ in $M_2$ gives $-96\kappa d_F^{abcd}d_A^{abcd}/(N_RC_F)$ in $d_3$ and hence $\nmQuarticOfSFour\,\zeta_2d_F^{abcd}d_A^{abcd}/N_R$ in $S_4$, independent of $\kappa$ and $b_A^-$.
The squared Born weights in $M_2$ have the color of the planar two-gluon ladder graph.
Neither $S_3$ nor $S_4$ contains $n_f$ or $d_F^{abcd}d_F^{abcd}/N_R$. The proposed equations give no $n_f$ in $S_n$ at fixed coupling, where $\beta_0$ is absent and $c_+=\ell_+$ cancels in $S_{n\ge2}$ at first order in $\zeta_2$. The remaining coefficient $S_1$ contains no $n_f$ because $c_+$ is a pure number by the color-degree rule at relative order zero, defined below in this section and argued from double-logarithmic two-quark cuts but assumed at NNLL.
No $S_n$ contains an invariant formed by contracting two quark $d_F$ tensors with each other. Products of separately contracted invariants occur from $S_7$ onward (App.~\ref{app:core-color}). The quartic invariant reduces to a combination of $C_FC_A^3$ and $C_F^2C_A^2$ in SU($N_c$). It is independent in SO(5) and SO(7), where direct solutions give the coefficients of eqs.~(\ref{eq:S3}) and (\ref{eq:S4}).

Kirschner and Lipatov~\cite{Kirschner:1983di} print $\kappa=+1$ for the negative-signature term in eq.~(4.4) of Ref.~\cite{Kirschner:1983di}, together with the adjoint Born term used here.
Section~\ref{sec:equations} compares this common sign with the equations of Ermolaev, Greco and Troyan~\cite{Ermolaev:2000sg}. Only the derivation that the positive-signature term has this same sign, from discontinuities in App.~\ref{sec:sign}, is new.
The sign was first fixed by the four-loop $P^-$ leading logarithm, eq.~(3.16) of Ref.~\cite{Moch:2026qsw}. The derivation from discontinuities came later.
The two-loop leading logarithm of $P^-$ already selects $\kappa=\nmKappa$, as the coefficients for both signs in App.~\ref{sec:sign} show.
The $C_FC_A^2$ component of $S_3$ confirms the Born term's presence.
The equation without the adjoint Born term gave an $S_3$ that disagreed with the complete result, so the Born term was added. The comparison with $S_4$ followed the fixing of both inputs, so it tests an order not used for that change.
Table~N4 of the ancillary notes shows the complete results and the equations with each input varied separately.

The comparisons comprise \countword{\nmCountSthreeRaw} components of $S_3$ and \countword{\nmCountSfourRaw} of $S_4$, as counted in Table~N5 of the ancillary notes.
The four-loop $x^0$ difference is the complete $\zeta_2x^0\ln^4x$ coefficient of $P^{(3)+}$ minus the NNLL prediction of eq.~(3.3) of Ref.~\cite{Davies:2022ofz}, also eq.~(4.37) of Ref.~\cite{Davies:2016bwb}, the thesis of Davies.
This prediction differs from that of eq.~(\ref{eq:relation}). Equations~(13) and (14) of Ref.~\cite{Gehrmann:2026cwg} give the same difference from the complete results.
The \countword{\nmCountXzeroRaw} coefficients of $A,B,C,D$ in the four-loop $x^0$ difference, eq.~(\ref{eq:xzerodiff}), are combinations of the components of $S_3$ and $S_4$.
The \countword{\nmCountCovoneRaw} components at $x^1$ equal those at $x^0$, and the \countword{\nmCountCovtwoRaw} components from $x^2$ are fixed linear combinations for every variant.
These \nmWordCountCovtwoRaw{} components at $\lambda_k=2$ are fixed-order data independent of the $x^0$ data, covering $P^+$ at even $k\ge2$ and $P^-$ at odd $k\ge3$. Their linear relations to $x^0$ hold for every variant, so they test the shifted-pole assumption but do not distinguish the sign and Born variants. They are not counted among the three tests at $x^0$.
The quartic comparison in $P^{(3)-}$ comprises \countword{\nmCountPminusquarticRaw} number at $x^0$ and \countword{\nmCountUncquarticRaw} number from $x^2$. The assumption on $d$-invariants in $P^-$ is that the NNLL terms missing from the negative-signature singlet and positive-signature adjoint equations give no $d$-invariant in $P^-$ through order $a^5$.
Both numbers check this assumption at four loops, order $a^4$. Five-loop results for $P^-$ would test its order-$a^5$ part (the further assumptions in App.~\ref{app:assumptions}).
The coefficients $V_n^-$ replace $P^+$ by $P^-$ in the definition of $S_n$. At a shifted pole, $V_n$ denotes the $\zeta_2$ part of the coefficient of $a^n\omega^{4-2n}$ in eq.~(\ref{eq:xk-S}), with $\omega=N+k$.
The compared coefficients are listed in Table~\ref{tab:comparison-w2}.

A reproduced number counts as an independent test only if it fixes no input and no linear relation valid for every variant fixes it from other compared numbers.
The variants reverse $\kappa$ and rescale the negative-signature adjoint Born term, including reversal and removal, with the overall coefficient in eq.~(\ref{eq:S}) fixed and the normalization condition imposed.
The planar combinations giving $N_c^3$ in $S_3$ and $N_c^4$ in $S_4$ in SU($N_c$), the $N_c^2$ combination of $S_4$, and the relation
\begin{equation}
 \left[S_4/\zeta_2\right]_{d_F^{abcd}d_A^{abcd}/N_R}
 =\nmQuarticRelCFCAsq\left[S_3/\zeta_2\right]_{C_FC_A^2}
 +\nmQuarticRelCFsqCA\left[S_3/\zeta_2\right]_{C_F^2C_A}
 \label{eq:quartic-relation-w4}
\end{equation}
hold for every variant, also when the overall coefficient in eq.~(\ref{eq:S}) is varied.
Every row of Table~N4 of the ancillary notes satisfies this last relation.
The fixed overall coefficient in eq.~(\ref{eq:S}) predicts the quartic component $\nmQuarticOfSFour$ of $S_4/\zeta_2$ for every variant. Its agreement is a test of that coefficient. Equation~(\ref{eq:quartic-relation-w4}) then fixes the $C_F^2C_A$ component of $S_3$, which is not counted again. The other two tests are the $C_FC_A^3$ and $C_F^2C_A^2$ components of $S_4$.

The complete results satisfy these four relations, leaving four independent combinations.
The $C_FC_A^2$ component of $S_3$ confirms the adjoint Born term, leaving \nmWordCountConsXzero{} tests of one system, not separate calculations.
No continuous coefficient is adjusted to the violations once the normalization condition is imposed and no further NNLL map or quark-pair term is admitted.
The fitted family is described in Sect.~\ref{sec:protection-w4}.
The color-degree rule assigns relative-order-$a^k$ constants pure-number color factors at $k=0$ and homogeneous degree-$k$ polynomials in $C_A,C_F$ and, where present, $n_fT_F$ otherwise. It is argued from two-quark cuts at double-logarithmic accuracy and assumed at NNLL. NNLL pair cuts, Born and quark-pair corrections would decide it (Sect.~N6.11 of the ancillary notes).

The count stays \nmWordCountConsDegOne{} if normalization constants and channel-diagonal quark-pair constants of color degree one are admitted without the channel-mixing, signature-mixed or singlet-adjoint pair terms defined in App.~\ref{app:assumptions}. It falls to \nmWordCountConsCoefFree{} if the overall coefficient in eq.~(\ref{eq:S}) is free (Sect.~N6.4 of the ancillary notes), and to zero for arbitrary functions of group invariants (Sect.~N6.11 of the ancillary notes).

The alternative $\kappa=-1$ with a channel-mixing term $\zeta_2ah(F_A^+)^2$, $h=\nmKappaDegH$, reproduces $S_3$ and $S_4$ in general color.
The $P^-$ leading logarithms exclude this sign independently of $h$, and the no-hard-gluon assumption excludes the added term.
The alternatives differ at five loops, giving $S_5=\nmSfiveKmHSUthree$ for $\kappa=-1$ against $S_5=\nmSfiveSUthree$ in SU(3), if no new source first acts at five loops.

The abelian limit relates the violations to the leading logarithms of $P^-$. Setting $C_A$ and the structure-constant invariants to zero makes both Born weights $C_F$, removes the Sudakov terms and makes the two phase-operator rows act equally on the coincident positive-signature waves. The rows themselves remain different, but both give $\nmSingletAdjoint F^+$ because their row sums coincide. The double-logarithmic equations give $F_A^-=F_1^-$ for either sign of $\kappa$. This is the equation value of $P^-_{\rm LL}$, equal to the known complete-result coefficients at two through four loops only for $\kappa=+1$. Equation~(\ref{eq:Sn-dn}) gives $S_{n+1}^{\rm ab}=\nmReductionCoef\kappa\zeta_2C_F[P^{(n-1)-}]_{\rm LL}^{\rm ab}$ for $n\ge2$. Here the right side is the coefficient of $a^n\omega^{1-2n}$, not the whole wave, and $S_1,S_2$ are excluded.
The complete-result values $S_3^{\rm ab}=\nmSthreeAbelian=\nmReductionCoef\zeta_2 C_F(\nmLLminusTwoAbelian)$ and $S_4^{\rm ab}=\nmSfourAbelian=\nmReductionCoef\zeta_2 C_F(\nmLLminusThreeAbelian)$ satisfy this relation for $\kappa=+1$, with $\nmLLminusTwoAbelian$ and $\nmLLminusThreeAbelian$ the complete-result two- and three-loop abelian $P^-$ LL coefficients.
 The abelian relation adds no independent test because its three- and four-loop components are among the compared components of $S_3$ and $S_4$. The proposed equations give $S_5^{\rm ab}=\nmSfiveAbelian=\nmReductionCoef\zeta_2 C_F(\nmLLminusFourAbelian)$ if no new source first acts at five loops.

\begin{table}[t]
  \centering
  \caption{Complete-result and equation values with vanishing $\ell_\pm$ and $c_-$ and $P^-=F_1^-$, with $c_+=\ell_+$ for $P^+$. The $S_n$ and $V_n$ rows give coefficients of $\zeta_2$ in the three-loop basis $(C_FC_A^2,\,C_F^2C_A,\,C_F^3)$ and four-loop basis $(C_FC_A^3,\,C_F^2C_A^2,\,C_F^3C_A,\,C_F^4,\,d_F^{abcd}d_A^{abcd}/N_R)$.
The rows at $x^2$ use the proposed equations and the shifted-pole assumption at $N=-k$.
The last two rows give the full displayed terms, with \mbox{$K=\nmKdef$}. Their two-loop equation values use \mbox{$\eta^+=\cos(\pi\omega/2)$} and omit the $\pi^2\omega^2$ part $\pi^2\omega^2\beta_e$ of the singlet Born term, the correction $a\zeta_2v$ relative to the singlet Born term, the map constant $d$ and map coefficient $e$ of Table~\ref{tab:constants-w4}. Both quartic $P^-$ agreements require the assumption on $d$-invariants in $P^-$. At $x^0$ the agreement also requires the leading $\psi^+$ in the positive-signature adjoint equation.}\label{tab:comparison-w2}
  \scriptsize
  \widecolstight
  \begin{tabular}{>{\raggedright\arraybackslash}p{2.6cm}>{\raggedright\arraybackslash}p{4.4cm}>{\raggedright\arraybackslash}p{4.4cm}>{\raggedright\arraybackslash}p{3.3cm}}
  \hline
  Quantity & Complete results & Equations & Status\\
  \hline
  $S_3/\zeta_2$, $x^0$ & $\nmSthreeVec$ & $\nmSthreeVec$ & Reproduced\\
  $S_4/\zeta_2$, $x^0$ & $\nmSfourVec$ & $\nmSfourVec$ & Reproduced\\
  $V_3/\zeta_2$, $P^+$ at $x^2$ & $\nmXkDataVthreePlusXtwo$ & $\nmXkVthreeCovLamTwo$ & Reproduced\\
  $V_4/\zeta_2$, $P^+$ at $x^2$ & $\nmXkDataVfourPlusXtwo$ & $\nmXkVfourCovLamTwo$ & Reproduced\\
  $V_3^-/\zeta_2$, $x^0$ & $\nmXkDataVthreeMinusXzero$ & $\nmXkVthreeUncLamOne$ & Not reproduced\\
  $V_4^-/\zeta_2$, $x^0$ & $\nmXkDataVfourMinusXzero$ & $\nmXkVfourUncLamOne$ & Only the quartic reproduced\\
  $V_3/\zeta_2$, $P^-$ at $x^2$ & $\nmXkDataVthreeMinusXtwo$ & $\nmXkVthreeUncLamTwo$ & Not reproduced\\
  $V_4/\zeta_2$, $P^-$ at $x^2$ & $\nmXkDataVfourMinusXtwo$ & $\nmXkVfourUncLamTwo$ & Only the quartic reproduced\\
  \hline
  $\zeta_2/N$ term of $P^{(1)+}$ & $\nmPoneZetaData/N$ & $\nmPonePlusZetaOwn/N$ & Boundary constant, not determined\\
  $P^{(1)-}-P^{(1)+}$ & $\nmPoneSigDiffData$ & $\nmPoneSigDiffOwn$ &
$N^{-3}$ agrees. The $\zeta_2/N$ boundary constant is undetermined, with the opposite sign here. The $N^{-2}$ and rational $N^{-1}$ terms are missed.\\
  \hline
  \end{tabular}
\end{table}

\section{\whyPtitle}
\label{sec:protection-w4}
The positive-signature result requires less of the adjoint equations than the negative-signature result does.
The coefficient of $F_A^-$ in eq.~(\ref{eq:S}) already supplies $\zeta_2\omega$, so only the double-logarithmic part of this wave reaches $S_{n\ge3}$ at NNLL.
Corrections of relative order $\omega$ or $a$ to the negative-signature equations or positive-signature adjoint equation enter $S$ beyond NNLL or at order $\zeta_2^2$.
In contrast, the negative-signature singlet couples to $F_A^+$ through $\psi^-$ of order $1/\omega$.
NNLL corrections in the positive-signature adjoint equation can therefore reach $P^-$ at NNLL while leaving the poles of $S$ unchanged.
At NNLL the negative-signature comparison has a different outcome.
The NNLL map constant $c_-$ enters $P^-=(1+c_-\pi^2\omega^2)F_1^-$.
Table~\ref{tab:comparison-w2} gives the general-color coefficients at three and four loops with vanishing $\ell_\pm$ and $c_-$ and $P^-=F_1^-$.
All four non-quartic components at four loops are missed.
In general color, no choice of the numbers $\ell_+$, $\ell_-$ and $c_-$ reproduces both orders (App.~\ref{sec:lim-pminus}).
The non-quartic NNLL $\zeta_2$ terms of $P^-$ at positive even $k$ and $P^+$ at odd $k$ are missed.
The non-quartic differences, complete results minus equations, for $P^-$ at even $k$ and $P^+$ at odd $k$, including $P^-$ at $x^0$, are proportional to $C_F(C_A-2C_F)$ in the convention $\ell_\pm=0$, $P^-=F_1^-$.
The same factor multiplies $S_3$.
It is equivalent to large-$N_c$ suppression for these $n_f$-free color structures without a quartic invariant, so the shared factor gives no evidence that the $P^-$ differences and $S_3$ share an origin.
These differences for $P^-$ at $x^0$ lose this factor for $\ell_-\ne c_-$ (App.~\ref{sec:lim-pminus}).
The helicity-dependent non-singlet function of the quark-plus-antiquark combinations, $\Delta P_{\rm ns}^+$ entering $g_1$, equals $P_{\rm ns}^-$ in $\MSbar$ at two and three loops, as stated below eq.~(2.9) and above eq.~(4.8) of Ref.~\cite{Moch:2014sna} by Moch, Vermaseren and Vogt.
This helicity function is included in the same $P^-$ comparison and supplies no separate test.
The polarized (helicity) non-singlet intercept obtained in the dipole formalism by Kovchegov, Pitonyak and Sievert~\cite{Kovchegov:2016zex} is computed at large~$N_c$.
There it agrees with the large-$N_c$ limit of the result of Ref.~\cite{Bartels:1995iu}, so it neither includes nor tests the $1/N_c^2$-suppressed terms considered here.

The fitted NLL term $a(\nmNLLtermCoef)F_A^+$ in the positive-signature adjoint equation reproduces the three- and four-loop NLL coefficients of $P^-$. It leaves the rational and $n_f$-dependent NNLL poles unreproduced (App.~\ref{sec:lim-pminus}).
Three corrections to the positive-signature equations proportional to $\Tt$ leave the NNLL $\zeta_2$ differences unreproduced. The operator $\Tt$ in eq.~(\ref{eq:dipole-split}) is the quadratic Casimir of the exchanged representation, with eigenvalue $C_R$, zero on the singlet and $C_A$ on the adjoint. The forms are a relative Sudakov correction $a\zeta_2(p_AC_A+p_FC_F)$, an additive $q a\zeta_2\omega C_RF_R^+$ term, and an additive $a^2\zeta_2C_R(p_AC_A+p_FC_F)\omega^{-1}F_R^+$ term. Each form has at most two pure-number coefficients. Here $p_A,p_F,q$ are free numbers for each tested form. They are distinct from the coefficients $p_A,p_F$ of the quark-pair constant $k_P$ in Table~\ref{tab:amp-nnll}, defined in Table~\ref{tab:constants-w4}, and from the five-loop source coefficient $q$ in App.~\ref{app:five}. The convention is $\kappa=+1$, $b_A^-=b_A$, $\ell_+=\ell_-=c_-=0$, $c_+=\ell_+$ and $P^-=F_1^-$.
None of the three forms reproduces even $V_3^-$ or $V_4^-$ separately in general color at first order in $\zeta_2$ and NNLL, with or without the fitted NLL term.
 Their contributions to $V_3^-$ and $V_4^-$ through $F_A^+$ and $\psi^-$ vanish in the abelian limit for polynomial Casimir coefficients, where the differences remain $\nmVthreeMinusAbelianDiff$ and $\nmVfourMinusAbelianDiff$. This result does not cover coefficients singular as $C_A\to0$, such as $1/C_A$, or arbitrary adjoint corrections. Power counting leaves every NNLL $S_{n\ge3}$, the $P^-$ LL and NLL terms, and the quartic $V_4^-$ unchanged. This preservation adds no test. The forms also leave unreproduced the non-quartic $\lambda_k=2$ differences for $P^-$ at even $k\ge2$ and $P^+$ at odd $k\ge3$ under the shifted-pole assumption. More general adjoint corrections fitted at $x^0$ miss them too if the added terms are kept unchanged at the shifted poles (App.~\ref{sec:lim-pminus}).

The two-loop $\zeta_2/N$ terms in Table~\ref{tab:comparison-w2} are boundary contributions the equations leave undetermined. Eliminating $(F_1^+)^2$ with eq.~(\ref{eq:singlet-row}) makes a singlet Born correction or an order-$a^2$ constant source enter $S$ only as a constant, whereas loop terms of $F_A^-$ give poles.
Here $a\zeta_2v$ is the relative singlet Born correction, and $\pi^2\omega^2\beta_e$ is the $\pi^2\omega^2$ part of the singlet Born term (Table~\ref{tab:constants-w4}).
The $\omega^0$ part of the signature-changing term acting on the Born term of $F_A^-$, with $\ell_\pm=c_+=0$ and without $\beta_e$, $v$, the map constant $d$ and the map coefficient $e$, gives $S_2=\nmStwoOwn$ and the $C_A$ part of the one-loop singlet residue, $\nmResDiffSingletCA\,\kappa\zeta_2C_A$ per Born term, which is absent from the one-loop four-quark finite remainder (App.~\ref{sec:sign}, after eq.~(\ref{eq:sign-mismatch})). The complete-result value is $S_2=\nmStwoData$.
The $C_A$ part of $v$, $v_A=\nmBornCorrMatchVA$, removes both $C_A$ contributions without changing any $S_n$ with $n\ge3$ in the fitted family with $k_P=d=e=0$ of Sect.~N7.1 of the ancillary notes.

Table~\ref{tab:constants-w4} defines the NNLL normalization, Born, map and quark-pair constants $c_+=\ell_+$, $\alpha$, $\beta_e$, $v$, $r$, $k_P$, the map constant $d$ and map coefficient $e$.
The fitted family has $k_P=d=e=0$. The coefficients $S_1$, $S_2$ and the one- to three-loop $\zeta_2$ amplitude coefficients in SU($N_c$) fix $v=\nmBornCorrMatchV$ and $r=\nmConstSetR$ and give $c_+=\ell_+=\nmConstFamilyCplus$ and $\alpha=\nmConstFamilyAlpha$, with $\beta_e$ free. These constants are fitted, not derived. Here $\alpha$ is the $\pi^2\omega^2$ part of a common signature-factor normalization and the amplitude map has scalar factor $1-a\zeta_2r$ (Sect.~N7.1 of the ancillary notes).
This family leaves every $S_n$ with $n\ge3$ unchanged and gives no further evidence for the retained term within the proposed NNLL form.

The positive-signature singlet could contain a further source like those required by the failure in $P^-$.
The evidence against such a source in the configurations computed is the leading-power calculation of gluons across the softest pair in App.~\ref{app:status}.
The configuration with one end on an external leg and the two-gluon configurations still require the no-hard-gluon assumption. Forms outside the examined list rest on the single-source assumption (App.~\ref{app:status}).

Non-planar $\zeta_2$ terms both with and without $n_f$ violate eq.~(\ref{eq:relation}) at N$^3$LL, one logarithm below NNLL~\cite{Moch:2004pa,Gehrmann:2026qbl,Kniehl:2025ttz}.
The coefficient of $a^4N^{-4}$ in the complete four-loop $P^{(3)+}$ minus the prediction of eq.~(\ref{eq:relation}) with complete lower orders is
\[
 \begin{aligned}
 &\nmNthreeLLnfFreeGen\\
 &{}+n_f\zeta_2(\nmNfExcessGen) .
 \end{aligned}
\]
Its flavor-independent and flavor-dependent parts in SU(3) are $\nmNthreeLLnfFreeSUthree$ and $n_f\zeta_2(\nmNfExcessSUthree)$, respectively.
The four-loop $n_f^2$ terms satisfy eq.~(\ref{eq:relation}) at every power of $\ln x$ except for $\zeta_2$ terms with color factor $C_F(C_A-2C_F)$, as stated in eq.~(4.20) and the text below it of Ref.~\cite{Davies:2016jie} by Davies et~al.

The equations give no prediction at N$^3$LL, for example for $\zeta_2x^0\ln^3x$ at four loops, because an undetermined relative $\omega$ correction to the adjoint Born term contributes, as may running coupling and gluons across the softest pair at comparable or larger transverse momentum.
An extension retaining these contributions would be tested by the complete four-loop $\zeta_2x^0\ln^3x$ coefficients.

The shifted-pole assumption gives the same $\zeta_2$ agreement for $P^+$ at even $k$ and $P^-$ at odd $k$ in $x^k$ (App.~\ref{sec:xk}). The rational NNLL terms also violate the relation at $x^k$ for $k\ge1$ in $P^+$ at even $k$ and $P^-$ at odd $k$. These rational terms and the lower logarithmic orders are outside the agreement.
The small-$N$ assumption at $N=-k$, $k\ge1$, covers only the $\zeta_2$ terms outside the retained term because the rational NNLL terms violate the relation there.
The equations omit the valence contribution proportional to $d^{abc}d_{abc}\,n_f$, the $s$ part $P_{\rm v}-P^-$ of the valence splitting function. These terms do not reach the NNLL poles $S_n$.

\section{A consistency check from the forward four-quark amplitude}
\label{sec:amplitude}
The forward amplitude and $P^+$ depend on the same partial wave $F_1^+$, through Mellin inversion and the splitting-function map.
The finite remainders of Caola et~al.~\cite{Caola:2021rqz} through three loops provide a comparison in SU($N_c$).
We identify the cutoff amplitude with the finite remainder of minimal infrared subtraction at $\mu^2=|t|$, up to the finite map described in App.~\ref{sm:amp-map}.
The identification is assumed beyond one loop.
The finite-map conditions require the maximal-weight map to be the exponential of a sum of dipole operators, each multiplied by a polynomial in $\ln(-s_{ij})$, with the three-loop quadrupole correction of Almelid, Duhr and Gardi~\cite{Almelid:2015jia} in the high-energy form of Ref.~\cite{Caron-Huot:2017fxr} and Falcioni et~al.~\cite{Falcioni:2021buo} (the dipole condition), to be even under exchange of the energy channels, and to contain no logarithm-dependent color-identity part (proportional to the identity in color space).
The one-loop leading logarithms agree in every channel, which motivates this choice of finite remainder.
The $\epsilon^0$ coefficient of the renormalized amplitude equals this remainder at one loop. The $\epsilon^0$ coefficient of the renormalized amplitude, in the conventions of Ref.~\cite{Caola:2021rqz}, differs from eq.~(\ref{eq:system}) in all \countword{\nmAmpEpsZeroLLDiffCount} leading-logarithmic coefficients at two and three loops, because infrared subtraction combines poles with positive powers of $\epsilon$ in lower-loop amplitudes.
The comparison therefore depends on the choice of infrared subtraction. This dependence on infrared subtraction is distinct from the factorization scheme of the splitting function.
The undetermined constants do not reach the three $\zeta_2$ coefficients compared below, the two-loop $C_A^2$ and three-loop $C_A^3$ and $C_A^2C_F$ coefficients, if the constants have the color degrees stated after eq.~(\ref{eq:amp-rel-full-w3}). The subtraction operator's Coulomb phase $i\pi\Trc$ gives all of the difference between the renormalized $\epsilon^0$ coefficient and the finite remainder in the two-loop $C_A^2$ and three-loop $C_A^3$ coefficients and more than the whole difference in the three-loop $C_A^2C_F$ coefficient (Sect.~N9.1 of the ancillary notes).

The quantity $A_R^\pm$ is the forward amplitude of signature $\pm$ in exchanged representation $R$, and $A_R^{(1)}$ is its Born term.
All $\nmAmpLLCount$ leading-logarithmic coefficients of the finite remainders at one through three loops agree with the equations at $\kappa=+1$ in both color channels and both signatures (Table~\ref{tab:amp-ll}). At $\kappa=-1$, all \countword{\nmAmpLLOddCount} negative-signature entries disagree with the finite remainders, as the last column of Table~\ref{tab:amp-ll} shows.

\begin{table}[htb]
  \centering
  \caption{Coefficients of the leading logarithms of the forward flavor-exchange amplitude at $L$ loops in SU($N_c$), in the
  $t$-channel singlet and adjoint of eq.~(\ref{eq:amp-proj}), divided by the Born term $A_R^{(1)}$ of the channel, at
  $\mu^2=|t|$. The coefficients of $a^L\tilde L^{2L}$ (positive signature) and of $a^L\Delta\tilde
  L^{2L-1}$ (negative signature) of the finite remainders of Ref.~\cite{Caola:2021rqz} are equal to those of
  eq.~(\ref{eq:system}) at $\kappa=\nmKappa$. The last column gives the negative-signature coefficients of
  eq.~(\ref{eq:system}) at $\kappa=-1$.}\label{tab:amp-ll}
  \footnotesize
  \widecols\fracrowsXL\rowsapartmain\everymath{\displaystyle}
  \begin{tabular}{lccc}
  \hline
  Channel & $a^L\tilde L^{2L}$ in $A_R^+/A_R^{(1)}$ & $a^L\Delta\tilde L^{2L-1}$ in $A_R^-/A_R^{(1)}$
    & The same, $\kappa$ reversed \\
  \hline
  \multicolumn{4}{l}{one loop}\\
  \quad singlet & $\nmAmpLLOneSingEven$ & $\nmAmpLLOneSingOdd$ & $\nmAmpLLKmOneSing$ \\
  \quad adjoint & $\nmAmpLLOneAdjEven$ & $\nmAmpLLOneAdjOdd$ & $\nmAmpLLKmOneAdj$ \\
  \multicolumn{4}{l}{two loops}\\
  \quad singlet & $\nmAmpLLTwoSingEven$ & $\nmAmpLLTwoSingOdd$ & $\nmAmpLLKmTwoSing$ \\
  \quad adjoint & $\nmAmpLLTwoAdjEven$ & $\nmAmpLLTwoAdjOdd$ & $\nmAmpLLKmTwoAdj$ \\
  \multicolumn{4}{l}{three loops}\\
  \quad singlet & $\nmAmpLLThreeSingEven$ & $\nmAmpLLThreeSingOdd$ & $\nmAmpLLKmThreeSing$ \\
  \quad adjoint & $\nmAmpLLThreeAdjEven$ & $\nmAmpLLThreeAdjOdd$ & $\nmAmpLLKmThreeAdj$ \\ \blockgap
  \hline
  \end{tabular}
\end{table}

We denote by $A_L[C_A^{L-j}C_F^j]$ the coefficient of that color factor in the $\zeta_2a^L\tilde L^{2L-2}$ term of $A_1^+/A_1^{(1)}$.
The entry $A_1[C_A]$ is the one-loop $\zeta_2C_A$ constant.
The coefficients $s_n[\dots]$ are those of $S_n/(\zeta_2C_F)$ after reducing the quartic invariant to $C_F$ and $C_A$ in SU($N_c$).
The common partial wave gives
\begin{equation}
  A_2[C_A^2]=-\nmRelCoefQuarter\,s_3[C_A^2] ,
  \label{eq:amp-rel}
\end{equation}
and at three loops
{\allowdisplaybreaks
\begin{align}
  A_3[C_A^3]&=-\nmRelCoefFortyEighth\,s_4[C_A^3] ,\nonumber\\
  A_3[C_A^2C_F]&=-\nmRelCoefFortyEighth\,s_4[C_A^2C_F]-\nmRelCoefTwelfth\,s_3[C_A^2] ,\nonumber\\
  A_3[C_AC_F^2]&=\nmRelFourCoefTwoThirds\,A_2[C_AC_F]-\nmRelFourCoefThird\,A_1[C_A]+\nmRelCoefTwelfth\,s_3[C_AC_F]-\nmRelCoefFortyEighth\,s_4[C_AC_F^2] .
  \label{eq:amp-rel-full-w3}
\end{align}}%

These relations reproduce the amplitude coefficients $\nmAmpTwoLoopCA$, $\nmAmpThreeLoopCA$, $\nmAmpThreeLoopCACF$ and $\nmAmpThreeLoopCACFsq$.
The first \nmWordCountAmpCfreeRaw{} follow from $S_3$ and $S_4$ if the finite-map conditions hold and the undetermined constants of the finite map and the equations have color factors of degree one at relative order $a$ and none at order $a^0$. Appendix~\ref{sm:amp-map} and the fitted family of Sect.~\ref{sec:protection-w4} identify these constants. Their contributions to the positive-signature singlet amplitude are only $C_F^L$ or $rC_F^{L-1}$ when they have the stated color degrees, as shown by eq.~(N6) in Sect.~N9.1 of the ancillary notes.
The last relation also uses the one-loop $C_A$ and two-loop $C_AC_F$ amplitude entries and requires color-independent $\ell_+$ and $\alpha$.
The relations hold for every $\kappa$, every scale of $\psi^+$ and both signs of the negative-signature adjoint Born term and with it absent, if the finite-map conditions hold.
They test the map and the common partial wave and add no independent determination of the retained term after $S_3$ and $S_4$ are fixed.

The two-loop $C_A^2$ coefficient splits as $\nmAmpTwoLoopCA=\nmAmpTwoLoopIRpart+\nmAmpTwoLoopPairPart$.
The first part comes from the phase acting twice on the Born term and follows from exponentiating the one-loop phase independently of eq.~(\ref{eq:system}), for either sign of $\kappa$.
The remaining part comes from the phase acting on the adjoint quark-pair term and is proportional to $b_A^2$.
The amplitude therefore tests that the source acts inside the positive-signature singlet equation. It does not identify the source.

The negative-signature adjoint leading logarithms in Table~\ref{tab:amp-ll} determine the same $d_n$ as the poles of $S$. They contain no $\psi^+$ and test $\kappa$ independently of the retained term.
The coefficient $\mathcal{A}_L$ of $a^L\Delta\tilde L^{2L-1}$ in $A_A^-/A_A^{(1)}$ in Table~\ref{tab:amp-ll} is the negative-signature adjoint leading logarithm at $L$ loops.
The subscript $L$ in $A_L[\cdot]$ and $\mathcal A_L$ is a loop order, whereas $R$ in $A_R^+$ labels the exchanged representation.
The residue rule of App.~\ref{sm:sign-evolution}, with $\eta^-=\omega\Delta/2+O(\omega^3)$ applied to $F_A^-=\sum_n d_na^n\omega^{1-2n}$, gives the coefficient $d_{L+1}\Delta\tilde L^{2L-1}/[2(2L-1)!]$ of $a^{L+1}$ in $A_A^-$.
Division by the adjoint Born term $A_A^{(1)}=2ab_A$ gives $\mathcal{A}_L=d_{L+1}/[4b_A(2L-1)!]$, so eq.~(\ref{eq:Sn-dn}) gives the amplitude form in SU($N_c$),
\begin{equation}
 S_{L+2}=-48\,\kappa\,\zeta_2\,C_F\,b_A\,(2L-1)!\,\mathcal{A}_L ,
 \label{eq:amp-form-w4}
\end{equation}
whose one-, two- and three-loop entries give $S_3$, $S_4$ and $S_5$, respectively, with no new source first acting at five loops, in agreement with eqs.~(\ref{eq:S34SUN}) and (\ref{eq:S5SUN}).
Equation~(\ref{eq:amp-form-w4}) restates the pole relation and adds no test.

\section{Five loops and the tower}
\label{app:towercore}
We write $[f]_{\zeta_2x^k\ln^mx}$ for the coefficient of $\zeta_2x^k\ln^mx$ in $f$.
The invariant $d_A^{abcd}d_A^{abcd}/N_A$ uses the adjoint representation, of dimension $N_A$.
The proposed equations predict the general-color coefficient of $\zeta_2x^0\ln^6x$ in the five-loop quark-plus-antiquark splitting function from the coefficients through $S_5$, with $S_1$ and $S_2$ taken from the complete results,
\begin{longeq}\label{eq:fivegen-w2}
  \Big[P^{(4)+}\Big]_{\zeta_2x^0\ln^6x}=\nmFiveLoopXzeroGen ,
\end{longeq}
or $\nmFiveLoopXzeroSUthree$ in SU(3).
The prediction uses the small-$N$ assumption and assumes no new source first acting at five loops. In SU(3), its part fixed by $S_1$ through $S_4$ is $\nmFiveLoopXzeroRelPart$ and its part from $S_5$ is $\nmFiveLoopXzeroTowerPart$. A complete SU(3) coefficient determines $S_5$ in SU(3). The value predicted here differs by $\nmFiveLoopMinusNNLLpred$ from the earlier NNLL prediction of Davies, Kom, Moch and Vogt, eq.~(3.4) of Ref.~\cite{Davies:2022ofz}.
The difference is the sum of the shift $\nmFiveLoopXzeroUpdateShift$ and the shift $\nmFiveLoopXzeroTowerPart$ from $S_5$. The first shift equals the part fixed by $S_1$ through $S_4$ (with the complete four-loop input) minus eq.~(3.4) of Ref.~\cite{Davies:2022ofz}. It is the sum of the continued four-loop $x^0$ difference and the different continuation of the three-loop term (Sect.~N11.4 of the ancillary notes).
Appendix~\ref{app:five} gives $S_5$ and $S_6$ and their color reductions. Table~N8 of the ancillary notes gives the decomposition of the five-loop coefficient into the part fixed by $S_1$ through $S_4$ and the part from $S_5$.

\begin{figure}[tbp]
  \centering
  \includegraphics[width=0.57\textwidth]{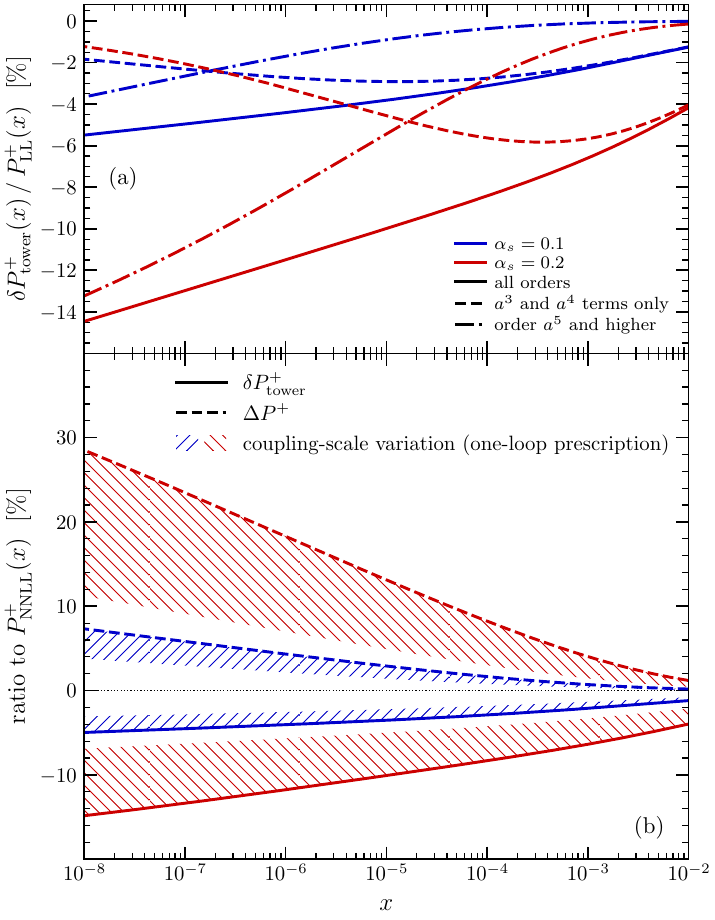}
  \caption{Panel (a) shows the SU(3) non-planar $\zeta_2$ tower's $x^0$ change $\delta P^+_{\rm tower}$ in percent of the leading-logarithmic $P^+_{\rm LL}$. The lines in both panels are at fixed coupling and first order in $\zeta_2$. Solid curves include all orders, dashed $a^3$ and $a^4$, dash-dotted the part from $a^5$ onward, which the three- and four-loop results do not test. This part needs the single-source assumption in full and, as a change of $P^+$, the small-$N$ assumption. Panel (b) divides $\delta P^+_{\rm tower}$ (solid) and $\Delta P^+$ (dashed) by $P^+_{\rm NNLL}$ of eq.~(3.2) of Ref.~\cite{Davies:2022ofz}. Here $\Delta P^+$ is the change when the tower replaces the continuation of that expression's three-loop term. Blue and red denote $\alpha_s(Q^2)=\nmAlphaSOne$ and $\nmAlphaSTwo$. The hatched bands in panel (b) show how the ratios depend on the coupling scale for one prescription, using the one-loop coupling $\nmRunCouplingDef$ at $Q^2e^\ell$ in numerator and denominator. Only this coupling uses nonzero $\beta_0$, at $n_f=\nmNfRun$, while eq.~(3.2) of Ref.~\cite{Davies:2022ofz} keeps $\beta_0=0$. The limit $\ell_0$ in $0\le\ell\le\ell_0$ depends on $x$ and $\alpha_s$ (the coupling-scale prescription in Sect.~N10.2 of the ancillary notes). The curves are not phenomenological predictions, and
the bands are not uncertainty estimates
 and do not estimate missing kernel or matching terms.}
  \label{fig:tower_size}
\end{figure}

The coefficients of $C_FC_A^4$, $C_A\,d_F^{abcd}d_A^{abcd}/N_R$ and $C_F\,d_A^{abcd}d_A^{abcd}/N_A$ are $\nmFiveLoopTestCFCAfour$, $\nmFiveLoopTestCAdFA$ and $\nmFiveLoopTestCFdAA$.
Only $S_5$ contributes to them, and all vanish in the NNLL prediction of Davies, Kom, Moch and Vogt~\cite{Davies:2022ofz}.
The defining representations of SU($N_c$), SO($N$) and Sp($N$) obey a relation among the five-loop invariants, eq.~(\ref{eq:fiveloop-relation}), that prevents these representations from separating the three coefficients.
A general-color five-loop calculation could test them separately.
An additional non-planar source first acting at this order would preserve the lower-order comparisons and change the prediction as in eq.~(\ref{eq:five-shift}).

The complete coefficients of $\zeta_3x^0\ln x$ in $P^{(2)+}$ and $\zeta_3x^0\ln^3x$ in $P^{(3)+}$ are $\nmZetaThreeThreeVec$ and $\nmZetaThreeFourVec$ in the bases of Table~\ref{tab:comparison-w2}. Both satisfy eq.~(\ref{eq:relation}), including the vanishing flavor-dependent parts. The harmonic sum $S_1(N)=\sum_{i=1}^Ni^{-1}$ has the expansion $\nmHarmonicSoneSmallN$ about $N=0$.
If the four-loop NNLL violation entered as a separate term proportional to the harmonic sum $S_1(N)$, it would enter $P^{(3)+}$ as $-(S_4/\zeta_2)S_1(N)N^{-6}$, whose $\zeta_2$ part gives $-S_4/24$ at $\zeta_2x^0\ln^4x$.
 Its $\zeta_3$ part, from $-\zeta_3N^2$, would add $-S_4/(6\zeta_2)$, or $\nmZetaThreeContrastFour$ in the four-loop basis of Table~\ref{tab:comparison-w2}, to the $\zeta_3x^0\ln^3x$ coefficient (Sect.~N5.2 of the ancillary notes gives the alternative based on the $x^0$ difference). The absence of this addition one logarithm below NNLL is consistent with a phase origin, while cancellations among other harmonic sums remain possible. Harmonic sums do not distinguish a phase $\pi^2$ from $\psi'(1)=\zeta_2$, where $\psi$ is the digamma function, distinct from the weights $\psi^\pm$, so the $\zeta_2$ terms alone neither support nor exclude the signature-changing term (Sect.~N5.1 of the ancillary notes).

The next term of $\pi\tan(\pi\omega/2)$ in the singlet equation, eq.~(\ref{eq:singlet-row}), expanded in eq.~(\ref{eq:tancot}), is of order $\zeta_2^2\omega^3$. Its first pole in $S$ is $a^4\omega^{-2}$, corresponding to $\zeta_2^2x^0\ln^2x$ in $P^{(3)+}$, two logarithms below NNLL. Terms not determined by the proposed equations contribute at the same accuracy, so the four-loop results do not test this next tangent term. A calculation including them could compare the full $\omega$ dependence of $\pi\tan(\pi\omega/2)F_A^-$ with the continued harmonic sums, testing the phase interpretation against an origin in the harmonic sum $S_1(N)$ alone (Sect.~N5.3 of the ancillary notes).

The tower consists of the non-planar $\zeta_2$ NNLL poles of $S$ to all orders, generated by the double-logarithmic negative-signature adjoint wave at fixed coupling and first order in $\zeta_2$.
The change of the splitting function is
\begin{equation}
  \delta P^+_{\rm tower}=-\frac{S_{\ge3}}{\sqrt{\omega^2-\omega_0^2}} ,
  \label{eq:tower-shift-w2}
\end{equation}
where $S_{\ge3}$ denotes the terms $S_na^n\omega^{4-2n}$ with $n\ge3$, and $\omega_0=\nmOmegaZeroDef$ is the LL branch point.
The tower uses the proposed equations with no new source first acting beyond four loops. Its interpretation as a change of $P^+$ also requires the small-$N$ assumption.
Appendix~\ref{app:tower} states its convergence and branch-point behavior, derived in Sect.~N10.1 of the ancillary notes. Figure~\ref{fig:tower_size} shows its size in SU(3) at fixed coupling and first order in $\zeta_2$.

The change $\delta P^+_{\rm tower}$ ranges from $\nmSizeOneXtwo$ to $\nmSizeOneXeight$ percent as $x=10^{\nmXmaxExp}$ decreases to $10^{\nmXminExp}$ at $\alpha_s=\nmAlphaSOne$ relative to the leading-logarithmic $P^+_{\rm LL}$.
The range is $\nmSizeTwoXtwo$ to $\nmSizeTwoXeight$ percent at $\alpha_s=\nmAlphaSTwo$ (Fig.~\ref{fig:tower_size}).
The tower reduces $P^+$ without dominating it over the plotted range.
The first-order ratio $\delta P^+_{\rm tower}/P^+_{\rm LL}$ grows like $\ln(1/x)$. It stays small while $|\delta\omega_0|\ln(1/x)$ is small. Here $\delta\omega_0$ is the negative first-order $\zeta_2$ shift of the square-root branch point $\omega_0$ of $P^+$. Section~N10.2 of the ancillary notes defines the shift and gives the sizes.
The tower's $a^3$ and $a^4$ terms already belong to the three- and four-loop splitting functions. Replacing the continued three-loop term in the NNLL expression of Davies, Kom, Moch and Vogt~\cite{Davies:2022ofz} by this tower raises that expression at small $x$ in the fixed-coupling SU(3) example of Fig.~\ref{fig:tower_size}.

An earlier all-order NNLL expression containing $\zeta_2$ was obtained from the three-loop results by Vogt et~al.~\cite{Vogt:2012gb}. It is a closed form for $P^+$ itself in $\sqrt{N^2-\omega_0^2}$. The tower here consists of the violations $S_n$, determined by $F_A^-$.

The small-$x$ logarithms do not dominate the complete three-loop non-singlet functions at measured momentum fractions~\cite{Moch:2004pa,Davies:2022ofz}.

\FloatBarrier
\section{Conclusions}
\label{sec:conclusions}
Our main result is that the proposed equations reproduce every color structure of the three- and four-loop violations at $x^0$ for quark-plus-antiquark flavor differences, at first order in $\zeta_2$ and fixed coupling, with the sign $\kappa=+1$ that Kirschner and Lipatov~\cite{Kirschner:1983di} printed for the negative-signature term, used for both signatures, and the adjoint Born term $b_A^-=b_A$ they printed in SU($N_c$). The quartic invariant is included. No continuous coefficient is adjusted to the violations. The retained soft-gluon term couples the positive-signature singlet to the double-logarithmic negative-signature adjoint wave. Its phase and signature projection, in the normalization of Sect.~\ref{sec:picture}, supply the squared imaginary phase suggested by Moch and Vogt~\cite{Moch:2026qsw}, giving, within the proposed equations, a possible origin of the violations in the complete results of Moch, Vermaseren and Vogt~\cite{Moch:2004pa} and Gehrmann, von~Manteuffel, Sotnikov and Yang~\cite{Gehrmann:2026qbl}.

Table~\ref{tab:conditions-main} distinguishes the derived, asserted, assumed and imposed parts of the NNLL extension. The proposed equations do not reproduce the non-quartic NNLL $\zeta_2$ terms of the quark-minus-antiquark function $P^-$ at $x^0$ and positive even powers of $x$ or $P^+$ at odd powers, as described in Sect.~\ref{sec:protection-w4}. The two-loop boundary and one-loop singlet constants are undetermined and belong to the fitted family described there. The amplitude relations check the common wave and map for the finite remainder with the assumed identification, after the violations are fixed. They do not test the retained term.

A complete SU(3) five-loop $\zeta_2x^0\ln^6x$ coefficient of $P^+$ determines $S_5$ in SU(3), while a general-color calculation can separate the $C_FC_A^4$, $C_A\,d_F^{abcd}d_A^{abcd}/N_R$ and $C_F\,d_A^{abcd}d_A^{abcd}/N_A$ coefficients supplied only by $S_5$ and absent from the earlier NNLL prediction of Davies, Kom, Moch and Vogt in eq.~(3.4) of Ref.~\cite{Davies:2022ofz}.
The proposed equations also supply a replacement for the continuation of the three-loop non-planar terms in the NNLL resummation of Ref.~\cite{Davies:2022ofz}, at fixed coupling and first order in $\zeta_2$, with the small-$N$ assumption and no new source beyond four loops. This replacement begins with the four-loop $x^0$ difference between the complete $\zeta_2x^0\ln^4x$ coefficient of Gehrmann, von~Manteuffel, Sotnikov and Yang~\cite{Gehrmann:2026qbl} and the NNLL prediction of eq.~(3.3) of Ref.~\cite{Davies:2022ofz}, defined in Sect.~\ref{sec:threefour}. The replacement raises that resummation at small $x$ in the fixed-coupling SU(3) example of Fig.~\ref{fig:tower_size}. These fixed-coupling SU(3) values are not phenomenological predictions.

We propose testing whether the retained term is the quark-antiquark-exchange counterpart of the three-Reggeon cut proportional to a squared imaginary phase in the gluon-exchange calculation of Caron-Huot, Gardi and Vernazza~\cite{Caron-Huot:2017fxr}. Fadin and Lipatov~\cite{Fadin:2017nka} describe the factorization-breaking terms at two and three loops in the negative-signature color-octet gluon-exchange amplitude by a three-Reggeon cut together with the pole. Identifying the retained term as the quark-antiquark-exchange counterpart would require the pair-exchange coefficient in a Regge-cut, Wilson-line or Glauber calculation. The gluon-exchange calculations do not derive it.

The four-loop non-planar breaking of the $\mathcal{N}=4$ super-Yang-Mills counterpart of eq.~(\ref{eq:relation}) was found in Ref.~\cite{Kniehl:2021ysp}. It is a related instance of non-planar breaking, but adjoint SU($N_c$) fields with $N_c\ge3$ fail the representation condition that each exchanged channel occur once. We claim no result for that theory.

The open problem is an NNLL derivation of the positive-signature singlet equation. It must decide the contributions of gluons across the softest pair, including a gluon of comparable transverse momentum with only one end on an external leg. A nearer calculation is the one-loop cutoff amplitude, which would fix the singlet Born correction and the scalar finite-map correction of Table~\ref{tab:constants-w4}, discussed in Sect.~\ref{sec:protection-w4} and Sect.~N7.1 of the ancillary notes.

We show that the violation at $n+1$ loops is $S_{n+1}=\nmReductionCoef\kappa\zeta_2C_Fd_n$ for $n\ge2$ in the proposed equations and that these products reproduce every color structure of the complete three- and four-loop violations of the quark-plus-antiquark function at $x^0$, including the quartic invariant, at fixed coupling and first order in $\zeta_2$. Here $d_n$ is the coefficient of $a^n\omega^{1-2n}$ in the double-logarithmic negative-signature adjoint wave.
We propose a general-color calculation of the five-loop $\zeta_2x^0\ln^6x$ coefficient to test eq.~(\ref{eq:fivegen-w2}), which uses the small-$N$ assumption and assumes no new source first acting at five loops.
Agreement with this five-loop coefficient would support the retained term as the origin of the violations through five loops. Disagreement would show that a new source of NNLL poles of $S$ first acts at five loops or that the small-$N$ assumption fails at five loops, that is, eq.~(\ref{eq:relation}) is violated at NNLL beyond the retained term's $S_5$.
Agreement would also provide a non-planar constraint for five-loop approximations beyond the large-$N_c$ reconstruction~\cite{Herzog:2018kwj}.

\section*{Data and program availability}

The arXiv ancillary files include the ancillary notes as \texttt{notes.pdf} and the programs and output for the paper and its appendices, including Figs.~\ref{fig:mechanism-w2} and \ref{fig:tower_size}.
\texttt{README.txt} gives their contents, software versions and instructions.
\texttt{numbers/provenance.csv} has one row per \texttt{numbers/numbers.tex} number macro, with computation and source.
Values typed in the text, such as matrix~(\ref{eq:hard-color-matrix-w3}), have no row.
The ancillary programs recompute the violations, comparisons, $x^0$ difference and five- and six-loop coefficients.

The published inputs include the splitting functions of Moch, Vermaseren and Vogt~\cite{Moch:2004pa,Moch:2014sna}, their expansions in Appendix~A of Ref.~\cite{Davies:2022ofz}, the four-loop files \begin{NoHyper}\url{Pqq3NSp.m}\end{NoHyper}, \begin{NoHyper}\url{Pqq3NSm.m}\end{NoHyper} and their expansions about $x=0$ from \href{https://arxiv.org/abs/2604.09534v2}{\begin{NoHyper}\url{arXiv:2604.09534v2}\end{NoHyper}}~\cite{Gehrmann:2026qbl}, and the amplitude files of Caola et~al.~\cite{Caola:2021rqz}.
\nolinkurl{make_numbers.py} computes values or reads them from stored outputs, checks them, writes the macro and provenance files, and reads its published inputs listed in \nolinkurl{numbers/published/MANIFEST.tsv}.
It reads eqs.~(3.15), (3.16) and the $\ln^4x$, $\ln x$ terms of eq.~(3.14) of Ref.~\cite{Moch:2026qsw} from the LaTeX source \href{https://arxiv.org/abs/2605.03889v1}{\begin{NoHyper}\url{arXiv:2605.03889v1}\end{NoHyper}}.
The \href{https://arxiv.org/abs/2609.17267v1}{\begin{NoHyper}\url{arXiv:2609.17267v1}\end{NoHyper}} source supplies the $\zeta_2x^0\ln^4x$ coefficients of eq.~(12) of Ref.~\cite{Gehrmann:2026cwg}, checked against \begin{NoHyper}\url{Pqq3NSpXto0.m}\end{NoHyper}, and the $x^0$ difference of eqs.~(13), (14) for comparison in Sect.~\ref{sec:threefour}.
The section ``Published inputs (not redistributed)'' of \texttt{README.txt} describes every published input and its origin.

The manifest lists \nolinkurl{make_numbers.py}'s published inputs separately from published files, their conversions and derived files that belong to individual calculations in \begin{NoHyper}\url{numbers/deposit/}\end{NoHyper}.
It gives names and origins, and checksums except for one journal PDF checked against its quoted sentences by \nolinkurl{make_numbers.py}.
Only the manifest is included, not its listed files.
The published files themselves are not included in the ancillary files.
Some outputs and programs contain coefficients computed or transcribed from them, for example \begin{NoHyper}\url{numbers/deposit/xk/x01_coeffs.txt}\end{NoHyper}.
\nolinkurl{make_numbers.py} and some program, data and output files in \begin{NoHyper}\url{numbers/deposit/}\end{NoHyper} quote short published passages.

The file \nolinkurl{three_loop_attachment_colors.txt} gives the attachment colors per Born term for three-loop placements of two added gluons on the box, in the notation of Sect.~N6.2 of the ancillary notes.
In each listed placement, each added gluon joins leg one or three or a soft quark line to leg two or four or a soft quark line.
The file omits rung attachments, gluons with both ends on legs one and three or both on legs two and four, and configurations where the two gluons meet at a three-gluon vertex.
The file \nolinkurl{compared_numbers.txt} lists, for each compared quantity, the number and class of compared coefficients, their relations and the discrete inputs behind the count in Sect.~\ref{sec:threefour}.

\section*{Acknowledgments}

We thank Sergey Bondarenko for useful discussions.
On the authors' instructions, Claude Opus~5.5 (Anthropic) and GPT-6 Astra (OpenAI) carried out the derivations and computations and wrote the ancillary programs that compute the numbers and draw Figs.~\ref{fig:mechanism-w2} and \ref{fig:tower_size}.
Most results were recomputed by a separate instance of a model or by the other model, not given the expected values, and checked by the authors.
The text of the article and of its ancillary notes was drafted and revised with Claude Opus~5.5 (Anthropic), GPT-6 Astra and GPT-6.1 Sol (OpenAI) on the authors' instructions and checked by the authors.
The authors checked and verified all derivations, computations and results, reviewed the text, and take full responsibility for the content of the paper.

\appendix
\let\topfraction\WtwoTopFraction
\let\textfraction\WtwoTextFraction
\let\floatpagefraction\WtwoFloatPageFraction

\section{Signature factors, the map to \texorpdfstring{\protect\boldmath$P^+$}{P+} and the reduction}
\label{app:core}\label{app:sign}

\subsection{Signature projection of the soft-gluon phase}
\label{sm:sign-projection}

The relation of the partial waves to the splitting functions follows from eq.~(5.12) of Ref.~\cite{Kirschner:1983di} for two color-singlet currents and eqs.~(3.14) and (3.19) of Ref.~\cite{Bartels:1995iu} for $g_1$.
The fixed-coupling double-logarithmic photon-quark equation for the unpolarized non-singlet structure function is eq.~(22) of the arXiv version of Ref.~\cite{Ermolaev:1995fx}.
Its solution, eq.~(36) of the arXiv version of Ref.~\cite{Ermolaev:1995fx}, has the $Q^2$ exponent $f_0(\omega)/(8\pi^2)$, where $f_0$ is the on-shell forward quark amplitude of Ref.~\cite{Kirschner:1983di} satisfying eq.~(29) of the arXiv version of Ref.~\cite{Ermolaev:1995fx}.
The photon-quark equation and its solution have no signature label and do not treat $P^-$, and the positive-signature identification follows from eq.~(5.12) of Ref.~\cite{Kirschner:1983di}.

The cutoff evolution of Kirschner and Lipatov~\cite{Kirschner:1983di} for the softest gluon uses the one-loop dipole operator, with the leg labels of Sect.~\ref{sec:equations},
\begin{equation}
  \Gamma=\nmDipoleOperator .
  \label{eq:dipole}
\end{equation}
The sum runs over pairs of external legs~\cite{Kidonakis:1998nf,Catani:1998bh,Becher:2009cu,Gardi:2009qi,Becher:2009qa,Gardi:2009zv}.
Here $\mathbf{T}_i$ is the color charge of leg $i$, and $s_{ij}$ is the invariant of legs $i$ and $j$.

The $t$-channel logarithm vanishes at $\mu^2=|t|$, the scale of the amplitude comparison.
Color conservation, with $\ell_r=\ln(-s_r-i0)$ and $\ell_c=\ln(-s_c-i0)$, gives
\begin{equation}
 \Gamma=2a\left[-\Tt\,\tilde L+\Trc\,\Delta\right],\qquad
 \tilde L=\tfrac12(\ell_r+\ell_c),\qquad\Delta=\ell_r-\ell_c .
 \label{eq:dipole-split}
\end{equation}
The singlet row of the phase operator has only the adjoint entry $\nmSingletAdjoint$ if the adjoint occurs once in the quark representation times its conjugate (App.~\ref{app:core-color}).

\subsection{The equations and reduction}
\label{sm:sign-reduction}

The Born weights are $b_R=\inlf{\nmBornWeight}$, with $b_1=C_F$ and \inlf{$b_A=\nmBornAdjSUN$} in SU($N_c$).
At double-logarithmic accuracy, eq.~(\ref{eq:system}) reduces to eqs.~(4.1) and (4.4) of Ref.~\cite{Kirschner:1983di} and includes eqs.~(3.6) and (3.7) of Ref.~\cite{Bartels:1995iu}.
The NNLL form of eq.~(\ref{eq:system}) is proposed, not derived.

The signature normalization gives the Born terms, and at double-logarithmic accuracy the quark-pair terms, the same form in both signatures up to a factor common to both~\cite{Kirschner:1982xw,Kirschner:1983di}.
The normalization condition removes $c_+$ and $\ell_+$ from $S_{n\ge2}$ at first order in $\zeta_2$.

\label{sec:reduction}
The positive-signature singlet row is eq.~(\ref{eq:singlet-row}) of Sect.~\ref{sec:equations}, which at $\kappa=+1$ becomes
\begin{equation}
  \omega F_1^+=2aC_F+\big(1+\ell_+\pi^2\omega^2\big)\big(F_1^+\big)^2
  +4aC_F\,\pi\tan\frac{\pi\omega}{2}\,F_A^- .
  \label{eq:singlet}
\end{equation}
Equation~(\ref{eq:S}) follows at fixed coupling from the map of eq.~(\ref{eq:equations-cplus}) in Sect.~\ref{sec:equations}, the normalization condition $c_+=\ell_+$ and \mbox{$\pi\tan(\pi\omega/2)=3\zeta_2\,\omega+O(\zeta_2^2\,\omega^3)$}.

The double-logarithmic $F_A^-$ is $f_V^-$ of eq.~(4.4) of Ref.~\cite{Kirschner:1983di}.
Appendix~\ref{app:status} treats possible new crossing-odd terms in this equation.

On the physical branch, $\partial S/\partial P^+=2P^+-\omega=-\sqrt{\omega^2-\omega_0^2}$.
A change $\delta S$ therefore shifts $P^+$ by $-\delta S/\sqrt{\omega^2-\omega_0^2}$ at first order in $\zeta_2$, with the LL branch point $\omega_0$ defined in Sect.~\ref{app:towercore}.
Multiplication of $\xi^\pm$ by signature normalizations $n_\pm(\omega)$ nonzero at the origin divides the waves of each signature by their own $n_\pm$, multiplies $\psi^+$ by $n_-/n_+$ and $\psi^-$ by its inverse.
The matching changes of the Born, quark-pair and Sudakov terms and of the map leave $S$ and every $S_n$ unchanged, for constant or \mbox{$\omega$-dependent}~$n_\pm$.

\subsection{Double-logarithmic limit}
\label{sm:sign-evolution}
At double-logarithmic accuracy, $\phi^+\to0$, $\phi^-\to1/\omega$, $\psi^+\to0$ and $\psi^-\to2/\omega$, and the quark-pair weights reduce to one.
Equation~(\ref{eq:system}) then becomes Kirschner and Lipatov's system for both signatures, eqs.~(4.1) and (4.4) of Ref.~\cite{Kirschner:1983di}, where the adjoint is labeled $V$. This system contains eqs.~(3.6) and (3.7) of Ref.~\cite{Bartels:1995iu}.
The positive-signature singlet equation is $\omega F_1^+=2aC_F+(F_1^+)^2$, whose solution\hfill\mbox{}
\begin{equation}
  F_1^+(\omega)=\frac12\Big(\omega-\sqrt{\omega^2-8aC_F}\Big)
  \label{eq:LL}
\end{equation}
is the LL $P^+$~\cite{Kirschner:1983di,Blumlein:1995jp}.

{\lineskiplimit=\smallskipamount\lineskip=\smallskipamount\relax For $F_A^-$ this limit is the second line of eq.~(4.4) of Ref.~\cite{Kirschner:1983di}, written in general color in eq.~(\ref{eq:adjoint-general-w3}).
The representation conditions of Sect.~\ref{sec:equations} apply. The amplitude at each order is the residue at $\omega=0$ of $e^{\omega\tilde L}\eta^\tau F_R^\tau$, as in eq.~(\ref{eq:sign-eta-main}).
The contour of its Mellin integral is to the right of the singularities of $F_R^\tau$.
The expansions of $\phi^\pm$ and $\psi^\pm$ are eqs.~(\ref{eq:sign-phases}) and (\ref{eq:tancot}) in App.~\ref{sm:sign-discontinuities}.\par}

\section{The sign of the signature-changing term}
\label{sec:sign}
The discontinuities of the Born and ladder graphs fix $\kappa$ without input from the complete results or the amplitudes. The derivation uses the two asserted inputs of Table~\ref{tab:conditions-main}, the dipole kernel beyond double-logarithmic accuracy and dependence on the direct energy logarithm alone at leading power for graphs without a crossed-channel discontinuity.

\subsection{Discontinuities and the sign}\label{sm:sign-discontinuities}
Color algebra and the dipole operator fix the signature-changing term, the last term of eq.~(\ref{eq:system}), up to $\kappa$. Both weights arise from $\Trc\Delta$, so the sign of the $\psi^-$ term fixes that of the $\psi^+$ term. Their product $\psi^+\psi^-=\Delta^2=-\pi^2$ is invariant under crossing and signature normalization. Expanding the same amplitude in $s_c$ instead of $s_r$ reverses $F^-$ but leaves $F^+$ unchanged.

Partial waves follow from the discontinuity in the expansion variable. The Born gluon has a pole only in $s_r$ because its momentum is the difference of the momenta at the ends of one quark line. It therefore gives equal waves in both signatures in the $s_r$ expansion, in the normalization of eq.~(\ref{eq:xi-main}) up to a common factor. Kirschner and Lipatov state on p.~137 of Ref.~\cite{Kirschner:1983di} that the Born contributions coincide for both signatures at lowest order because of this single right-hand singularity. They state on pp.~268--269 of Ref.~\cite{Kirschner:1982xw} that the one-loop ladder graph has the same degeneracy. Planar ladder graphs, formed by iterating the quark-pair term on the Born term, have discontinuities in $s_r$ and $t$ but none in $s_c$. Their leading-power dependence on $\ln(-s_r)$ alone is the second asserted input. The Born and double-logarithmic pair terms are consequently signature independent in the $s_r$ expansion.

The soft operator must use the $s_r$-expansion labeling, in which both signature-changing terms come from $\Trc\Delta$, and the same physical region for the logarithms and signature factors. In the region $s_r>0$, the phase term $\Trc\Delta$ of the dipole operator, projected with the signature ratio $\eta^+/\eta^-$, gives the product of $2a$, the phase $-i\pi$, the ratio $\eta^+/\eta^-=i\cot(\pi\omega/2)$ and the color entry $\nmSingletAdjoint$ for the coefficient of $F_A^+$. It equals $-4aC_F\pi\cot(\pi\omega/2)$, tending to $-8aC_F/\omega$. The coefficient multiplying $F_A^+$ in the signature-changing term of the negative-signature singlet equation of eq.~(\ref{eq:system}) is $2a\kappa\psi^-[\Trc]_{1A}=-4a\kappa C_F\,\pi\cot(\pi\omega/2)$. Comparing it with the coefficient derived from the dipole phase gives $\kappa=+1$, the sign printed for the signature-changing term in the negative-signature equations in the double-logarithmic limit in eq.~(4.4) of Ref.~\cite{Kirschner:1983di} and eq.~(3.6) of Ref.~\cite{Bartels:1995iu}. Using one region's phase with the other's signature factors gives $\kappa=-1$.

The one-loop amplitude formed from the dipole operator and planar box is compared below with the residues of eq.~(\ref{eq:system}) in the $\tilde L$-dependent terms, testing the signature projection of the dipole operator and planar box. The positive-signature logarithms agree for both signs, but the negative-signature logarithm agrees only for $\kappa=+1$. The check does not test the two asserted inputs or the NNLL coefficient of $\psi^+$, which contributes only a constant here.

At one loop, the soft gluon in eq.~(\ref{eq:dipole}) acts on the Born amplitude $M_0=2a\sum_Rb_RP_R$, with the projectors $P_R$ defined in App.~\ref{app:core-color}.
We write $\ell'=\ln(s/\mu'^2)$, where $\mu'$ is the running cutoff and $s=|s_r|$.
A pair $(i,j)$ contributes $\mathbf T_i\cdot\mathbf T_j\,\big(\ell'+\ell_r-\ln(s/\mu^2)\big)$ to the integrand for invariant $s_r$ and $\mathbf T_i\cdot\mathbf T_j\,\big(\ell'+\ell_c-\ln(s/\mu^2)\big)$ for invariant $s_c$.
Here $\ell_r$ and $\ell_c$ are the energy logarithms of Sect.~\ref{sec:picture}.
Color conservation gives $\mathbf T_1\cdot\mathbf T_2+\mathbf T_3\cdot\mathbf T_4=-\tfrac12\Tt+\Trc$ and $\mathbf T_1\cdot\mathbf T_4+\mathbf T_2\cdot\mathbf T_3=-\tfrac12\Tt-\Trc$.
In the region $s_r>0$,
\begin{equation}
  \begin{aligned}
  M^{(1)}&=2a\int_{c}^{L}d\ell'\,\Bigl[\bigl(-\tfrac12\Tt+\Trc\bigr)(\ell'-i\pi)
   +\bigl(-\tfrac12\Tt-\Trc\bigr)\ell'\Bigr]M_0\\
  &=2a\Bigl[-\tfrac12\Tt\,\tilde L^2+\Trc\,\Delta\,\tilde L\Bigr]M_0+\mathcal{C},
  \end{aligned}
  \label{eq:sign-M1}
\end{equation}
where $\mathcal C$ is independent of $\tilde L$ and depends on the lower limit $c$.
The region $s_c>0$ gives the same result with $\Delta=+i\pi$.

In channel $R$, the operator $\Trc$ takes $M_0$ to $2aX_R$, where
\begin{equation}
  X_R=\sum_{R'}\big[\Trc\big]_{RR'}\,b_{R'} ,\qquad X_1=\big[\Trc\big]_{1A}\,b_A=\nmKdef ,
  \label{eq:sign-XR}
\end{equation}
and $X_A$ is $M_1$ of eq.~(\ref{eq:color-M1}), or $\inlf{\nmXadjSUN}$ in SU($N_c$).
The planar box $(2ab_R)^2\ell_r^2/2$, with $\ell_r=\tilde L+\Delta/2$, completes the order-$a^2$ amplitude in channel $R$,
\begin{equation}
  A_R^{(2)}=2a^2\Big[b_R(b_R-C_R)\,\tilde L^2+\big(b_R^2+2X_R\big)\Delta\tilde L\Big]+\text{const}.
  \label{eq:sign-A2}
\end{equation}
The even and odd parts of eq.~(\ref{eq:sign-A2}) in $\Delta$ give the two signatures at fixed color operators and $\tilde L$,
\begin{equation}
  A_R^{+(2)}=2a^2 b_R(b_R-C_R)\,\tilde L^2+\text{const},\qquad
  A_R^{-(2)}=2a^2\big(b_R^2+2X_R\big)\Delta\tilde L+\text{const}.
  \label{eq:sign-A2pm}
\end{equation}

The factors $\eta^\pm$ combining the signature factors with the energy phase of eq.~(\ref{eq:sign-eta-main}) in Sect.~\ref{sec:picture} have the cosine and sine forms given there for $s_r>0$.
The phases of eq.~(\ref{eq:system}) are $\phi^\pm=\eta^{\pm\prime}/\eta^\pm$ and $\psi^\pm=\Delta\,\eta^\mp/\eta^\pm$.
At $\Delta=-i\pi$ they are eqs.~(\ref{eq:psiminus}), (\ref{eq:psiplus}) and (\ref{eq:phases}), the last in Sect.~\ref{sec:equations}.
Their expansions near $\omega=0$ are\hfill\mbox{}
\begin{equation}
  \begin{aligned}
  \phi^+&=\frac{\Delta^2\omega}{4}+O(\omega^3),&
  \phi^-&=\frac{1}{\omega}+\frac{\Delta^2\omega}{12}+O(\omega^3),\\
  \psi^+&=\frac{\Delta^2\omega}{2}+O(\omega^3),&
  \psi^-&=\frac{2}{\omega}+\frac{\Delta^2\omega}{6}+O(\omega^3).
  \end{aligned}
  \label{eq:sign-phases}
\end{equation}
All four are even in $\Delta$, so the two physical regions give the same equations.
The expansions of the weights $\psi^-$ and $\psi^+$ of eqs.~(\ref{eq:psiminus}) and (\ref{eq:psiplus}), the latter with its minus sign, follow for $\Delta^2=-6\zeta_2$ from
\begin{equation}
  \pi\tan\frac{\pi\omega}{2}=3\zeta_2\,\omega+\tfrac32\zeta_2^2\,\omega^3+\dots,\qquad
  \pi\cot\frac{\pi\omega}{2}=\frac{2}{\omega}-\zeta_2\,\omega-\tfrac{1}{10}\zeta_2^2\,\omega^3-\dots,
  \label{eq:tancot}
\end{equation}
which give the orders in $\zeta_2$.

Equation~(\ref{eq:system}) gives the following result with these expansions, $F_R^{\pm(1)}=2ab_R/\omega$ and quark-pair weight $1+\ell_\pm\pi^2\omega^2$,
\begin{equation}
  \begin{aligned}
  F_R^{+(2)}&=\frac{4a^2b_R(b_R-C_R)}{\omega^3}+\frac{a^2\Delta^2(C_Rb_R+2\kappa X_R)+4\ell_+\pi^2a^2b_R^2}{\omega}+O(\omega),\\
  F_R^{-(2)}&=\frac{4a^2(b_R^2+2\kappa X_R)}{\omega^3}+\frac{a^2\Delta^2(C_Rb_R+2\kappa X_R)}{3\,\omega}
   +\frac{4\ell_-\pi^2a^2b_R^2}{\omega}+O(\omega).
  \end{aligned}
  \label{eq:sign-F2}
\end{equation}
The Sudakov term does not contribute to the leading pole of $F_R^{-(2)}$ because $\phi^-=1/\omega+O(\omega)$.
The terms $4\ell_\pm\pi^2a^2b_R^2/\omega$ come from the quark-pair weight acting on $(F_R^{\pm(1)})^2$.
The residues of $e^{\omega\tilde L}\eta^\pm F_R^{\pm(2)}$ at $\omega=0$ are
\begin{equation}
  \begin{aligned}
  \mathrm{Res}^+_R&=2a^2b_R(b_R-C_R)\,\tilde L^2+\tfrac12a^2\Delta^2\big(b_R^2+C_Rb_R+4\kappa X_R\big)+4\ell_+\pi^2a^2b_R^2,\\
  \mathrm{Res}^-_R&=2a^2\big(b_R^2+2\kappa X_R\big)\Delta\tilde L .
  \end{aligned}
  \label{eq:sign-Res2}
\end{equation}
The $1/\omega$ terms of $F_R^{-(2)}$ have zero residue because they multiply $\eta^-$, which vanishes at $\omega=0$.

The positive-signature terms depending on $\tilde L$ agree with eq.~(\ref{eq:sign-A2pm}) for both signs of $\kappa$, whereas the negative-signature difference is
\begin{equation}
  A_R^{-(2)}-\mathrm{Res}^-_R=4a^2(1-\kappa)X_R\,\Delta\tilde L=4i\pi a^2(\kappa-1)X_R\,\tilde L
  \qquad(s_r>0),
  \label{eq:sign-mismatch}
\end{equation}
which vanishes for $\kappa=+1$ and excludes $\kappa=-1$ when $X_R\ne0$.
This comparison uses the dipole operator and planar box already present in eq.~(\ref{eq:system}) and tests their signature projection, with the positive-signature term checking the LL Sudakov and quark-pair normalizations and the negative-signature term checking $\phi^-=1/\omega+O(\omega)$, equal quark-pair weights and $\kappa$.
An $\omega^0$ term in $\psi^+$ would give an absent term linear in $\tilde L$, but the coefficient $\Delta^2\omega/2$ contributes only a constant at this order and is not tested.

The constants depend on the lower integration limits, $\ell_+$ and a one-loop Born correction, so they do not determine $\kappa$.
The residue $\mathrm{Res}^+_R$ minus the dipole and box form $2a^2b_R(b_R-C_R)(\tilde L^2+\Delta^2/4)$ is $a^2\Delta^2(C_Rb_R+2\kappa X_R)+4\ell_+\pi^2a^2b_R^2$.
In the singlet, its $C_A$ part relative to the Born term, $\nmResDiffSingletCA\,\kappa\zeta_2C_A$, is absent from the finite remainder, whose singlet $\zeta_2$ constant is \inlp{$\nmAmpOneLoopZetaConst$} (App.~\ref{sm:amp-map}, eq.~(\ref{eq:amp-H1}) and Table~\ref{tab:amp-nnll}).
The $\zeta_2/N$ term of $P^{(1)+}$ follows from the $1/\omega$ coefficient of $F_1^{+(2)}/a^2$ plus $4c_+\pi^2C_F^2$ from the map of eq.~(\ref{eq:equations-cplus}).
Its numerator is \mbox{$\nmPonePlusZetaOwn+4(\ell_++c_+)\pi^2C_F^2$} at $\kappa=+1$, against $\nmPoneZetaData$ in eq.~(4.6) of Ref.~\cite{Moch:2004pa} (Sect.~\ref{sec:protection-w4}, Table~\ref{tab:comparison-w2} and Sect.~N7.1 of the ancillary notes).

The two-loop leading pole of $P^-$ independently selects the printed sign. Its coefficient is \mbox{$\nmLLminusTwo$} for $\kappa=+1$, compared with $\nmPminusLLtwoKm$ for $\kappa=-1$, while the complete result is \mbox{$\nmLLminusTwo$}~\cite{Moch:2004pa}. The complete coefficient follows from the $P^{(1)+}$ coefficient $\nmPplusLLtwo$ and the signature difference in eqs.~(4.6) and (4.7) of Ref.~\cite{Moch:2004pa}. Appendix~\ref{sec:Pminus} gives the three- and four-loop leading logarithms. Sections~N3.5 and N3.4 of the ancillary notes give the one-loop amplitude residues and the sign-even and sign-odd parts of the violations.

\section{Color algebra and the coefficients \texorpdfstring{\protect\boldmath$d_n$}{d n}}
\label{app:core-color}

Quark line A runs from leg $1$ to $2$, with invariant $s_r$, and line B from $4$ to $3$. The invariant of legs $1$ and $3$ is $t$.
The $t$-channel projectors $P_R$ obey $P_RP_{R'}=\delta_{RR'}P_R$, with multiplication along the two quark lines.
Their sum is the unit color flow $U=\sum_RP_R$.
We write $[\mathbf{T}]_{RR'}$ for the coefficient of $P_R$ in $\mathbf{T}P_{R'}$.
The one-gluon exchange tensor is $B=-\mathbf{T}_1\cdot\mathbf{T}_3\,U=\sum_Rb_RP_R$, where $b_R=\inlf{\nmBornWeight}$.

The operators $\mathbf{T}_s^2$ and $\mathbf{T}_u^2$ are the Casimir operators for the pairs $(1,2)$ and $(1,4)$.
Color conservation gives $\mathbf{T}_s^2+\Tt+\mathbf{T}_u^2=4C_F$ and $\Trc=(\mathbf{T}_s^2-\mathbf{T}_u^2)/2$, with $\Tt P_R=C_RP_R$.
The matrix of $\Trc$ satisfies $\dim(R)\,[\Trc]_{RR'}=\dim(R')\,[\Trc]_{R'R}$.

We require each irreducible representation in the quark representation times its conjugate to occur once.
Equation~(\ref{eq:system}) assumes this by assigning one partial wave per $t$-channel representation.
Every quark representation used here, the fundamental of SU($N_c$) and the representations of the color check in this appendix, meets the condition.
It excludes adjoint quarks in SU(3), for example, whose product contains the adjoint twice.
The two quark lines have different flavors.
We therefore also assume the signature assignment of complex representations in eq.~(\ref{eq:system}) for real and pseudo-real quark representations.
The general-color expressions of this appendix rest on both conditions. The lowest Regge exponents at zero momentum transfer are $j=\nmReggeJpair$ for a quark-antiquark pair with any number of gluons and $j=\nmReggeJfour$ for four fermions, so four-fermion exchange first contributes at $x^1$. Evolution with one wave per channel and signature is therefore assumed for a quark-antiquark pair with gluons that are not the softest partons.

Color conservation, $\mathbf{T}_1\cdot(\mathbf{T}_1+\mathbf{T}_2+\mathbf{T}_3+\mathbf{T}_4)=0$, turns $\Trc$
into $C_F+2\mathbf{T}_1\cdot\mathbf{T}_2+\mathbf{T}_1\cdot\mathbf{T}_3$. The unit color flow $U$ is a singlet and one-gluon exchange $B$ an adjoint for legs $1$ and $2$. Hence
$\mathbf{T}_1\cdot\mathbf{T}_2\,U=-C_FU$, and $t^ct^at^c=(C_F-C_A/2)t^a$ gives
$\mathbf{T}_1\cdot\mathbf{T}_2\,B=-b_AB$. The rows of $\Trc$ obey the following relations by \mbox{$\mathbf{T}_1\cdot\mathbf{T}_3\,P_R=-b_RP_R$},
\begin{equation}
  \sum_{R'}\big[\Trc\big]_{RR'}=\nmRowSum ,\qquad
  \sum_{R'}\big[\Trc\big]_{RR'}\,b_{R'}=\nmRowSumBorn .
  \label{eq:color-rows}
\end{equation}
The singlet row of $\Trc$ has one nonzero entry, in the adjoint column.
The first relation of eq.~(\ref{eq:color-rows}) at $R=1$ gives $[\Trc]_{1A}=\nmSingletAdjoint$. The symmetry of
$\dim(R)[\Trc]_{RR'}$ then gives $[\Trc]_{A1}=-2C_F/N_A$.

If legs $1$ and $4$ are the energy channel, every entry of $\Trc$ changes sign, and the Born and quark-pair terms of negative signature change sign with it (App.~\ref{sm:sign-discontinuities}).
For the adjoint row, the second relation of eq.~(\ref{eq:color-rows}) gives $M_1$ of eq.~(\ref{eq:M12}) in Sect.~\ref{sec:threefour},
\begin{equation}
  M_1\equiv\sum_{R'}\big[\Trc\big]_{AR'}\,b_{R'}=\nmMoneGen .
  \label{eq:color-M1}
\end{equation}

The positive-signature singlet equation gives $S=\nmReductionCoef\,\kappa\,a\,\zeta_2\,C_F\,\omega F_A^-$ at first order in $\zeta_2$ for a general sign $\kappa$.
We obtain the following expression from $d_1=2b_A^-$ and $d_2=4(b_A^-)^2+8\kappa M_1$ of eq.~(\ref{eq:Sn-dn}) and $M_1$ from eq.~(\ref{eq:color-M1})
\begin{equation}
  S_3=\nmReductionCoef\,\kappa\,\zeta_2C_Fd_2=\nmSthreeBornForm .
  \label{eq:color-S3}
\end{equation}
The result holds for either sign of $\kappa$, where $b_A^-=b_A$ is the negative-signature adjoint Born weight (Sect.~\ref{sec:threefour}).
At $\kappa=+1$ this is eq.~(\ref{eq:S3}), whereas at $\kappa=-1$ it is $\nmSthreeKmGen$.

The coefficient of $F_A^-$ that gives the four-loop $S_4$ is $d_3=4(b_A^--C_A)d_2+16\kappa(3M_2-2C_FM_1)$, with $M_2$ of eq.~(\ref{eq:color-M2}).
The first term combines the quark-pair term acting on the Born term of $F_A^-$, of weight $b_A^-$, with the Sudakov term.
The Sudakov coefficient $C_A$ equals $2(C_F-b_A)$ with the positive-signature weight $b_A$.
The moment $M_2$ is the adjoint component of $\Trc$ acting on the planar two-gluon ladder graph $(\mathbf{T}_1\cdot\mathbf{T}_3)^2U$.
On this graph, $\mathbf{T}_1\cdot\mathbf{T}_2$ represents a gluon on line A enclosing both rungs.
Its adjoint component is one quark loop with eight generators,
\mbox{$M_2=C_Ab_A^2/2-(4/N_A)\,{\rm Tr}(t^dt^at^ct^dt^et^ct^at^e)$}.
The same quark-loop argument gives, with $N_A=2N_RC_F$,
\begin{equation}
  M_p=\frac{C_A}{2}\,b_A^p-\frac{2}{N_RC_F}\,\Lambda_p ,\qquad
  \Lambda_p={\rm Tr}\big(t^dt^{a_1}\cdots t^{a_p}t^dt^et^{a_p}\cdots t^{a_1}t^e\big) .
  \label{eq:color-Mp}
\end{equation}

The generator $T_i^a$ acts on the $i$th open adjoint index of a quark-line segment $Z$.
A gluon enclosing that segment gives\hfill\mbox{}
\begin{equation}
 t^aZt^a=\left[C_F-\frac12\left(\sum_iT_i\right)^2\right]Z .
 \label{eq:trace-enclosure-w3}
\end{equation}
We define $O=F^a\otimes F^a$ for two open indices, with $(F^a)_{bc}=-if^{abc}$, and $\widehat d_R{}_{ab,cd}=d_R^{abcd}$.
The definition $G_j=\operatorname{Tr}(t^{c'}t^{e'}t^et^c)[O^j]_{ce,c'e'}$ gives
\begin{equation}
 \Lambda_p=\sum_{j=0}^{p}{p\choose j}(C_F-C_A)^{p-j}(-1)^jG_j .
 \label{eq:trace-binomial-w3}
\end{equation}
The four-generator trace reduces to its symmetric part $d_F^{abcd}$ and
$-T_F(f^{abx}f^{cdx}-f^{adx}f^{bcx})/6$, with the terms containing one symmetric three-index tensor and one structure constant vanishing in $G_j$.
Here $T_FN_A=N_RC_F$.
The remaining contraction on the product of two adjoint representations is
\begin{equation}
 G_j=\operatorname{Tr}(\widehat d_F O^j)
 -\frac{T_F}{6}\left[\operatorname{Tr}O^{j+1}-\left(-\frac{C_A}{2}\right)^jC_AN_A\right],\qquad
 O^2=\widehat d_A-\frac{C_A}{6}O+\frac{C_A^2}{6}P_A ,
 \label{eq:trace-box-w3}
\end{equation}
where $P_A$ projects onto the antisymmetric adjoint in this product.
Thus
\[
 G_0=N_RC_F^2,\qquad G_1=-T_FC_A^2N_A/4,\qquad
 G_2=d_F^{abcd}d_A^{abcd}+T_FC_A^3N_A/12 .
\]
Equation~(\ref{eq:trace-binomial-w3}) then gives the result at $p=2$,
\begin{equation}
  \Lambda_2=N_R\big(C_F^4-2C_AC_F^3+\tfrac{3}{2}C_A^2C_F^2-\tfrac{5}{12}C_A^3C_F\big)+d_F^{abcd}d_A^{abcd} .
  \label{eq:Lambda-two}
\end{equation}
These identities hold in every simple Lie algebra and express $\Lambda_2$ in $C_F$, $C_A$, the dimensions $N_R$ and $N_A$ and $d_F^{abcd}d_A^{abcd}$.
The same contractions determine the higher moments below.
Substitution into eq.~(\ref{eq:color-Mp}) gives $M_2$ of eq.~(\ref{eq:M12}) in Sect.~\ref{sec:threefour},
\begin{equation}
  M_2=\nmMtwoGen ,
  \label{eq:color-M2}
\end{equation}
and, with $S_4=\nmReductionCoef\,\kappa\,\zeta_2C_Fd_3$, gives eq.~(\ref{eq:S4}) at $\kappa=+1$.
The quartic invariant enters $S_4$ only through $M_2$, with the same coefficient for either sign of $\kappa$.

Equations~(\ref{eq:S3}) and (\ref{eq:S4}) become the following expressions for the gauge group SU($N_c$), where $d_F^{abcd}d_A^{abcd}/N_R=\inlf{\nmDFASUN}$,
\begin{equation}
  S_3=\nmSthreeSUN ,\qquad S_4=\nmSfourSUN ,
  \label{eq:S34SUN}
\end{equation}
that is $S_3=\nmSthreeSUthree$ and $S_4=\nmSfourSUthree$ in SU(3), with
$d_F^{abcd}d_A^{abcd}/N_R=\nmDFASUthree$.

Equation~(\ref{eq:system}) assigns the positive-signature weight, $b_A^-=b_A$, in every group.
The weight $b_A^-$ enters $S_3$ only quadratically, eq.~(\ref{eq:color-S3}), while $\nmSfourRevDiffGen$ in general color.
Section~\ref{sec:threefour} gives the presence and sign comparisons and cites the SU($N_c$) adjoint Born term. The presence and sign comparisons do not apply in G$_2$, where the adjoint Born weight vanishes. Direct group solutions check the color algebra and do not supply independent perturbative calculations.
Both comparisons require nonzero $b_A$.
The equations give $S_3=\nmCtrlKappaSthreeSUthree$ in SU(3) with $\kappa=-1$, compared with $\nmSthreeSUthree$ in the complete results.
This test also requires the negative-signature adjoint Born term because $\kappa$ multiplies only its squared weight in $S_3$.

In SU($N_c$), the Born weights are $b_1=C_F$ and \inlf{$b_A=\nmBornAdjSUN$}, and in the basis $(1,A)$,
\begin{equation}
  \Trc=\spacedmatrix{\nmTrcSUN}
  \label{eq:TrcSUN}
\end{equation}
(Table~\ref{tab:color-SOSp} and the group data below give other groups).

In general color, eq.~(\ref{eq:adjoint-general-w3}) below gives
\begin{equation}
  \begin{aligned}
  &\nmDrecFour ,\\
  &\nmDrecFive .
  \end{aligned}
  \label{eq:color-d45}
\end{equation}
Since $C_R=2(C_F-b_R)$, each channel sum in the recursion is an adjoint moment $M_p$ of Sect.~\ref{sec:threefour}.
Thus $d_4$ and $d_5$ depend on the group only through $C_F$, $C_A$ and $M_1$ to $M_4$.

The trace identities of eqs.~(\ref{eq:trace-enclosure-w3}) to (\ref{eq:Lambda-two}) also reduce $\Lambda_p$ through $p=4$ to $C_F$, $C_A$, the representation dimensions and the quartic and higher contractions in eq.~(\ref{eq:color-M34}).
For $p=3$ and $p=4$,
\begin{longeq}\label{eq:color-M34}
  \keepparens M_3=\nmMthreeGen ,\longsep\keepparens M_4=\nmMfourGen ,
\end{longeq}
which in SU($N_c$) are \inlf{$M_3=\nmMthreeSUN$} and \inlf{$M_4=\nmMfourSUN$}.
They give eqs.~(\ref{eq:S5}) and (\ref{eq:S6}) at $\kappa=+1$ by eq.~(\ref{eq:color-d45}) and $S_{n+1}=\nmReductionCoef\,\kappa\,\zeta_2C_Fd_n$.

We use the quartic invariants of Ref.~\cite{vanRitbergen:1998pn} with ${\rm Tr}(t^at^b)=\delta^{ab}/2$ in the quark representation.
As in Ref.~\cite{Moch:2026qsw}, $d_F^{abcd}d_A^{abcd}$ is divided by the quark dimension $N_R=N_c$.
In SU($N_c$) this gives $\inlf{\nmDFASUN}$.
The value decreases by the factor $N_A/N_R=2C_F$ if $N_R$ is replaced by $N_A$.
Equation~(\ref{eq:S4}) then fails to match the complete results.
In SU($N_c$), $d_A^{abcd}d_A^{abcd}/N_A=\inlf{\nmDAASUN}$ and $d_F^{abcd}d_A^{abef}d_A^{cdef}/N_R=\inlf{\nmDFAASUN}$ as well.

For general color, the double-logarithmic negative-signature adjoint row of eq.~(\ref{eq:system}) is
\begin{equation}
 \omega F_A^-=2ab_A+2aC_A\left(\frac{d}{d\omega}+\frac1\omega\right)F_A^-
 +(F_A^-)^2+\frac{4a\kappa}{\omega}\sum_{R'}[\Trc]_{AR'}F_{R'}^+ .
 \label{eq:adjoint-general-w3}
\end{equation}

The quark representation enters $\Lambda_p$ only through its four-generator trace.
Each moment $M_p$, at every order, is linear in quark invariants formed by contracting one $d_F^{abcd}/N_R$ with fully symmetric adjoint tensors, such as $d_F^{abcd}d_A^{abcd}/N_R$ and $d_F^{abcd}d_A^{abef}d_A^{cdef}/N_R$, with products of Kronecker deltas excluded because their contractions reduce to polynomials in $C_F$ and $C_A$ that enter the coefficients.
No invariant with $d_F^{a_1\cdots a_6}$ or with two $d_F$ tensors contracted with each other, such as $d_F^{abcd}d_F^{abcd}/N_R$, appears in any $S_n$.
The recursion for the $d_n$ is quadratic in the $d_k$ through $(F_A^-)^2$.
The coefficients $S_3$ to $S_6$ stay linear in these invariants because $M_1$, $d_1$ and $d_2$ contain none of them, so the products of the $d_k$ in $d_3$ to $d_5$ stay linear in them.
Products of these invariants occur from $S_7$ on, with the first term in $(d_F^{abcd}d_A^{abcd}/N_R)^2/C_F$ in $S_7$ coming from the square of $M_2$ through $d_3^2$ in $d_6$.
The ancillary files list the products in $S_7$ and $S_8$.

The product with the conjugate contains three representations, $1$, $A$ and $X$, for the vector representation of SO($N$) and the defining representation of Sp($N$). The two relations of
eq.~(\ref{eq:color-rows}), the entry $[\Trc]_{A1}=-2C_F/N_A$ and the symmetry of $\dim(R)[\Trc]_{RR'}$ then
determine the matrix, with \inlf{$[\Trc]_{AX}=\nmTAXthree$}. Table~\ref{tab:color-SOSp} gives it together with $d_F^{abcd}d_A^{abcd}/N_R$, $S_3$ and $S_4$, for $N=3$ and $N\ge5$ in SO($N$) and for even $N$ in
Sp($N$). At $N=2$ the Sp($N$) expressions give the SU(2) values.

\begin{table}[htb]
  \centering
  \caption{Channels of the product of the quark representation and its conjugate for the vector representation of SO($N$), with $N=3$ or $N\ge5$, and the defining representation of Sp($N$), with $N$ even, in list order, with ${\rm Tr}(t^at^b)=\delta^{ab}/2$ in the quark representation. For each channel $R$, the entries are its dimension, the Casimir $C_R$, the Born weight $b_R=\nmBornWeight$, the same in both signatures, and the row $R$ of the matrix $[\Trc]_{RR'}$. The last three lines give the quartic invariant and the coefficients $S_3$ and $S_4$ of eqs.~(\ref{eq:S3}) and (\ref{eq:S4}) in these groups.}\label{tab:color-SOSp}
  \small
  \fracrowsSupp
  \begin{tabular}{lll}
    \hline
    & SO($N$), vector & Sp($N$), defining \\
    \hline
    channels & singlet, adjoint, symmetric traceless & singlet, antisymmetric traceless, adjoint \\
    dimensions & $\nmSONdims$ & $\nmSpNdims$ \\
    $C_R$ & $\nmSONcasimirs$ & $\nmSpNcasimirs$ \\
    $b_R$ & $\nmSONborn$ & $\nmSpNborn$ \\
    \noalign{\smallskip}
    $[\Trc]_{RR'}$ & $\nmTrcSON$ & $\nmTrcSpN$ \\[\smallskipamount]
    $d_F^{abcd}d_A^{abcd}/N_R$ & $\nmDFASON$ & $\nmDFASpN$ \\
    $S_3$ & $\nmSthreeSON$ & $\nmSthreeSpN$ \\
    $S_4$ & $\nmSfourSON$ & $\nmSfourSpN$ \\
    \hline
  \end{tabular}
\end{table}

The group data for Sect.~\ref{sec:threefour} and for App.~\ref{sec:fiveloop} include fundamental and higher representations of low-rank classical groups and $\mathrm{G}_2$, $\mathrm{F}_4$, $\mathrm{E}_6$, $\mathrm{E}_7$ and $\mathrm{E}_8$, $\nmRepsExtCount$ in all.
One program computes the full $\Trc$ matrix in the quark representation times its conjugate using explicit representation matrices.
It then solves the equations in rational arithmetic without rounding.
A separately written program in another computer-algebra system obtains the same group data for $\nmRepsIndependentDistinct$ of these representations from characters alone.
Every solution agrees with eqs.~(\ref{eq:S3}), (\ref{eq:S4}), (\ref{eq:S5}) and (\ref{eq:S6}).

\FloatBarrier
\section{The complete results and the \texorpdfstring{\protect\boldmath$x^0$}{x0} difference}
\label{app:core-data}
\begin{table}[!htbp]
  \centering
  \caption{Coefficients of $x^0\ln^mx$ in the non-singlet splitting functions $P^{(n)\pm}$, with $P^\pm=\sum_n a^{n+1}P^{(n)\pm}$ and $a=\alpha_s/(4\pi)$, for the leading logarithms (LL, $\nmDTmLL$), the next-to-leading logarithms (NLL, $\nmDTmNLL$) and the next-to-next-to-leading logarithms (NNLL, $\nmDTmNNLL$). The columns headed ``coefficient of $\zeta_2$'' and ``part without $\zeta_2$'' give the two parts separately. No other transcendental number occurs at these orders. Each block's first line gives its source. At $N=0$ a term $x^0\ln^mx$ gives the pole $\nmDTMellinRule$.}\label{tab:datatable}
  \footnotesize
  \narrowcols\rowsapartwide
  \begingroup
  \breakbeforemathsigns
  \renewcommand{\nmInvFA}{d_F^{abcd}d_A^{abcd}/N_R}
  \begin{tabular}{llp{0.27\textwidth}p{0.5\textwidth}}
    \hline
    & & coefficient of $\zeta_2$ & part without $\zeta_2$ \tabularnewline
    \hline
    \multicolumn{4}{l}{$P^{(1)+}$, two loops, eq.~(4.6) of Ref.~\cite{Moch:2004pa}} \tabularnewline
     & LL & \raggedright $\nmDTPOnePlusLtwoZeta$ & \raggedright $\nmDTPOnePlusLtwoRat$ \tabularnewline
     & NLL & \raggedright $\nmDTPOnePlusLoneZeta$ & \raggedright $\nmDTPOnePlusLoneRat$ \tabularnewline
     & NNLL & \raggedright $\nmDTPOnePlusLzeroZeta$ & \raggedright $\nmDTPOnePlusLzeroRat$ \tabularnewline\blockgap
    \multicolumn{4}{l}{$P^{(1)-}$, two loops, eq.~(4.7) of Ref.~\cite{Moch:2004pa}} \tabularnewline
     & LL & \raggedright $\nmDTPOneMinusLtwoZeta$ & \raggedright $\nmDTPOneMinusLtwoRat$ \tabularnewline
     & NLL & \raggedright $\nmDTPOneMinusLoneZeta$ & \raggedright $\nmDTPOneMinusLoneRat$ \tabularnewline
     & NNLL & \raggedright $\nmDTPOneMinusLzeroZeta$ & \raggedright $\nmDTPOneMinusLzeroRat$ \tabularnewline\blockgap
    \multicolumn{4}{l}{$P^{(2)+}$, three loops, eq.~(4.9) of Ref.~\cite{Moch:2004pa}} \tabularnewline
     & LL & \raggedright $\nmDTPTwoPlusLfourZeta$ & \raggedright $\nmDTPTwoPlusLfourRat$ \tabularnewline
     & NLL & \raggedright $\nmDTPTwoPlusLthreeZeta$ & \raggedright $\nmDTPTwoPlusLthreeRat$ \tabularnewline
     & NNLL & \raggedright $\nmDTPTwoPlusLtwoZeta$ & \raggedright $\nmDTPTwoPlusLtwoRat$ \tabularnewline\blockgap
    \multicolumn{4}{l}{$P^{(2)-}$, three loops, eq.~(4.10) of Ref.~\cite{Moch:2004pa}} \tabularnewline
     & LL & \raggedright $\nmDTPTwoMinusLfourZeta$ & \raggedright $\nmDTPTwoMinusLfourRat$ \tabularnewline
     & NLL & \raggedright $\nmDTPTwoMinusLthreeZeta$ & \raggedright $\nmDTPTwoMinusLthreeRat$ \tabularnewline
     & NNLL & \raggedright $\nmDTPTwoMinusLtwoZeta$ & \raggedright $\nmDTPTwoMinusLtwoRat$ \tabularnewline\blockgap
    \multicolumn{4}{l}{$P^{(3)+}$, four loops, file \texttt{Pqq3NSpXto0.m} of Ref.~\cite{Gehrmann:2026qbl}} \tabularnewline
     & LL & \raggedright $\nmDTPThreePlusLsixZeta$ & \raggedright $\nmDTPThreePlusLsixRat$ \tabularnewline
     & NLL & \raggedright $\nmDTPThreePlusLfiveZeta$ & \raggedright $\nmDTPThreePlusLfiveRat$ \tabularnewline
     & NNLL & \raggedright $\nmDTPThreePlusLfourZeta$ & \raggedright $\nmDTPThreePlusLfourRat$ \tabularnewline\blockgap
    \multicolumn{4}{l}{$P^{(3)-}$, four loops, file \texttt{Pqq3NSmXto0.m} of Ref.~\cite{Gehrmann:2026qbl}} \tabularnewline
     & LL & \raggedright $\nmDTPThreeMinusLsixZeta$ & \raggedright $\nmDTPThreeMinusLsixRat$ \tabularnewline
     & NLL & \raggedright $\nmDTPThreeMinusLfiveZeta$ & \raggedright $\nmDTPThreeMinusLfiveRat$ \tabularnewline
     & NNLL & \raggedright $\nmDTPThreeMinusLfourZeta$ & \raggedright $\nmDTPThreeMinusLfourRat$ \tabularnewline
    \noalign{\vskip\smallskipamount}\hline
  \end{tabular}

\endgroup
\end{table}

The extraction at $x^0$ uses the one-loop expansion at $N=0$,
\begin{equation}
  P^{(0)}(N)=\nmDTPzeroLaurent ,
  \label{eq:datatable-P0}
\end{equation}
which follows from $P^{(0)}(N)=\nmDTPzeroDigamma$, where $\psi$ is the digamma function, distinct from the weights $\psi^\pm$, and $\gamma_E$ is Euler's constant.

Each $x^0\ln^mx$ term gives a Laurent coefficient through the pole $\nmDTMellinRule$ at $N=0$.
We write the $x^0\ln^mx$ coefficient in $P^{(n)}$ as $c^{(n)}_m=r^{(n)}_m+\zeta_2\,z^{(n)}_m$, with $r^{(n)}_m$ and $z^{(n)}_m$ independent of $\zeta_2$.
The LL coefficients, $\nmDTmLL$, and NLL coefficients, $\nmDTmNLL$, contain no $\zeta_2$.
NNLL corresponds to $\nmDTmNNLL$.

Equation~(\ref{eq:datatable-P0}) and these poles, inserted into $S(\omega)$ of eq.~(\ref{eq:S}) at $\omega=N$, give
\begin{align}
  S_3&=\nmDTformSthree , \label{eq:datatable-S3}\\
  S_4&=\nmDTformSfour , \label{eq:datatable-S4}
\end{align}
using the $P^+$ entries of Table~\ref{tab:datatable}.
All entries refer to the splitting functions $P$, not to the anomalous dimensions $\gamma=-P$.
The quantity $S_3$ uses the two- and three-loop entries, and $S_4$ those through four loops.
The superscript of $c^{(n)}_m$ labels $P^{(n)}$, the coefficient of $a^{n+1}$, while the subscript of $S_n$ counts powers of $a$.
The $C_F\,c^{(1)}_2$ and $C_F\,c^{(2)}_4$ terms come from multiplying the $\zeta_2$ part linear in $N$ in eq.~(\ref{eq:datatable-P0}) by the leading logarithms.

The two- and three-loop entries are extracted from eqs.~(4.6), (4.7), (4.9) and (4.10) of Ref.~\cite{Moch:2004pa}, with $P^-$ formed from $P^+$ and the signature differences, and the four-loop entries from \begin{NoHyper}\url{Pqq3NSpXto0.m}\end{NoHyper} and its negative-signature counterpart of \mbox{\href{https://arxiv.org/abs/2604.09534v2}{\begin{NoHyper}\url{arXiv:2604.09534v2}\end{NoHyper}}}~\cite{Gehrmann:2026qbl}.
The entries of Table~\ref{tab:datatable} for $P^{(2)\pm}$ are also the coefficients $D^\pm_0$,
$D^\pm_1$ and $D^\pm_2$ of eqs.~(4.15) and (4.17) of Ref.~\cite{Moch:2004pa}.
In SU(3) the four-loop
entries agree with the expansion printed in Appendix~B of Ref.~\cite{Gehrmann:2026qbl}, whose $\zeta_2$
terms require $d_F^{abcd}d_A^{abcd}/N_R=\nmDFASUthree$.

\label{sec:xzerodiff}
In $x$ space, the violations enter the $\zeta_2\,x^0\ln^4x$ coefficient of $P^{(3)+}$ (Sect.~\ref{sec:threefour}).
We use the coefficient notation $[f]$ of Sect.~\ref{app:towercore}, here for general $x^k\ln^mx$.
The complete coefficient multiplied by $\zeta_2$ equals the coefficient predicted by eq.~(\ref{eq:relation}), also multiplied by $\zeta_2$, plus the violation $-S_4/24$. The prediction of eq.~(\ref{eq:relation}), evaluated with complete lower orders, and the NNLL prediction of eq.~(3.3) of Ref.~\cite{Davies:2022ofz} continued from three loops have the common part $\zeta_2(\nmNNLLpredCFfourPart)$, but their $S_3$ parts are \inlf{$\nmRelSthreePart$} and \inlf{$\nmNNLLpredSthreePart$}, respectively. Their difference gives the second term of eq.~(\ref{eq:xzerodiff}).
The $x^0$ difference defined in Sect.~\ref{sec:threefour} is
\begin{longeq}\label{eq:xzerodiff}
  \zeta_2\Big[P^{(3)+}_{\rm complete}-P^{(3)+}_{\rm NNLL}\Big]_{\zeta_2x^0\ln^4x}
  =\nmXzeroDiffCombination=\zeta_2\left(\nmXzeroDiffGen\right)=\zeta_2\bigl(\nmXzeroDiffABCD\bigr) ,
\end{longeq}
where
\[
 \begin{aligned}
 A&=d_F^{abcd}d_A^{abcd}/N_R-C_FC_A^3/24,&
 B&=C_FC_A^2(C_A-2C_F),\\
 C&=C_FC_A(C_A-2C_F)^2,& D&=C_F(C_A-2C_F)^3
 \end{aligned}
\]
form the basis of Ref.~\cite{Moch:2026qsw}.

\enlargethispage{\baselineskip}%
In SU($N_c$), the bracket is $\inlf{\nmXzeroDiffSUN}$, or $\nmXzeroDiffSUthree$ in SU(3).
The comparison against eqs.~(13) and (14) of Ref.~\cite{Gehrmann:2026cwg} is given in Sect.~\ref{sec:threefour}.

The first curly bracket in eq.~(3.15) of Ref.~\cite{Moch:2026qsw} is misprinted%
\footnote{The first curly bracket of eq.~(3.15) of Ref.~\cite{Moch:2026qsw} by Moch and Vogt, $\{\nmBracketPrintedOne(1-x)^{-1}-\nmBracketPrintedTwo(1+x)\}$ in both arXiv versions and in the journal version, is twice too large.
It enters the quartic coefficient at $x^0$ as minus the difference of its entries, $\nmPrintedXzeroQuartic$ as printed, where eq.~(\ref{eq:xzerodiff}) requires $-\nmBracketDiff$.
Equation~(3.14) of Ref.~\cite{Moch:2026qsw}, in the same normalization, prints the $x^0$ quartic coefficient $-\nmBracketDiff$, fixing the difference of the bracket entries to $\nmBracketDiff$.
The complete-result $x^2$ quartic coefficient $\nmXkCovQuarticHigh$, extracted from \begin{NoHyper}\url{Pqq3NSp.m}\end{NoHyper} of \mbox{\href{https://arxiv.org/abs/2604.09534v2}{\begin{NoHyper}\url{arXiv:2604.09534v2}\end{NoHyper}}}~\cite{Gehrmann:2026qbl}, fixes the first entry to $\nmBracketOne$ and then the second to $\nmBracketTwo$.
The halved bracket reproduces eq.~(\ref{eq:xzerodiff}) and the coefficients printed by Moch and Vogt in eqs.~(2.10) and (2.14) of Ref.~\cite{Moch:2026pos}.
The other brackets are not affected.}.
The $\ln x$ coefficient of $d_F^{abcd}d_A^{abcd}/N_R$ in eq.~(3.14) of Ref.~\cite{Moch:2026qsw} agrees with the complete result in its $\zeta_2$ term, but its $\zeta_2^2$ and $\zeta_2\zeta_3$ terms have the opposite sign%
\footnote{The expansion in \begin{NoHyper}\url{Pqq3NSpXto0.m}\end{NoHyper} of \mbox{\href{https://arxiv.org/abs/2604.09534v2}{\begin{NoHyper}\url{arXiv:2604.09534v2}\end{NoHyper}}}~\cite{Gehrmann:2026qbl} gives $\nmLnxFourLoop$ for the $\ln x$ term of the $d_F^{abcd}d_A^{abcd}/N_R$ coefficient, while eq.~(3.14) of Ref.~\cite{Moch:2026qsw} prints $\nmLnxPrinted$, with $\nmLnxFourLoopZfour\,\zeta_4=\nmLnxFourLoopZtwoSq\,\zeta_2^2$.
The ancillary files contain the programs that expand the four-loop files \begin{NoHyper}\url{Pqq3NSp.m}\end{NoHyper} and \begin{NoHyper}\url{Pqq3NSpXto0.m}\end{NoHyper} and give these coefficients, including the arithmetic of the bracket misprint and the $\ln x$ sign difference.}.
The solution of the double-logarithmic equation in eq.~(4.7) of Ref.~\cite{Kirschner:1983di} has misprints in its prefactor and Gaussian%
\footnote{In eq.~(4.7) on p.~138 of Ref.~\cite{Kirschner:1983di} the prefactor $\nmMisprintPrefactorPrinted$ should read $\nmMisprintPrefactorCorrected$, and $\nmMisprintGaussPrinted$ should read $\nmMisprintGaussCorrected$, as the substitution $f_i^+=(b_ig^2/c_i)\,u'/u$ in eq.~(4.1) of Ref.~\cite{Kirschner:1983di} shows.}.
The compared coefficients are extracted from the complete results and do not depend on these misprints or the sign difference.

\FloatBarrier
\section{Excluded pole sources and gluons across the softest pair}
\label{app:status}\label{sec:hypotheses}\label{app:assumptions}

Table~\ref{tab:conditions-main} defines the proposed equations.
The further assumptions concern the finite map defined in Sect.~\ref{sec:amplitude} and App.~\ref{sm:amp-map}, the color-degree rule, the small-$N$ and shifted-pole extensions, $d$-invariants in $P^-$, and the extension of the single-source assumption beyond four loops, marked in Table~\ref{tab:conditions-main}, to exclude a new source first acting at five loops or later.

Power counting excludes relative-$\omega$ and relative-$a$ corrections in the negative-signature and adjoint equations from the NNLL poles of $S$, while color algebra excludes the singlet Sudakov term.
The power-counting and color-algebra exclusions use neither $S_3$ nor $S_4$. Crossing-even eikonal terms are proportional to $\Tt$, which vanishes on the singlet, or equal $\psi^+\psi^-=\Delta^2$, already retained. They therefore add no NNLL pole source of $S$ in the positive-signature singlet equation. The color-algebra exclusions are in rows (d) and (h) of Table~N3 of the ancillary notes, and the power count is in Sect.~N6.10 of the ancillary notes.
A new crossing-odd kernel is excluded for $S_3$ through $S_5$ in SU($N_c$) by the argued power count, the two- to four-loop $P^-$ leading logarithms and, in SU($N_c$), the negative-signature adjoint amplitude at one to three loops (assuming the finite-remainder identification beyond one loop), subject to the kernel and row conditions in the subsection ``Scope of the source exclusions'' below. From $S_6$ onward, its absence belongs to the single-source assumption in full.

The complete $S_3$ fixes the combination $y$ of Table~\ref{tab:constants-w4} to zero if the normalization condition holds and no channel-mixing, signature-mixed or singlet-adjoint quark-pair term is present. The complete results also exclude planar color-identity terms first shifting $S_3$ or $S_4$, and the imposed normalization condition excludes the map terms.

The partial calculation of gluons across the softest pair at comparable or larger transverse momentum excludes $\zeta_2ah(F_A^+)^2$ and leaves the $C_FC_A^2$ component of $S_3$ unchanged at leading power in Feynman gauge, with one regulator acting alike on all attachments, a common positive imaginary part in their eikonal denominators, and strictly ordered rung attachments argued. At three loops the quartic component of $S_4$ is reached only by two such gluons with both ends of each attached to external legs and is unchanged at leading power, with the coupling of the two gluons on one energetic line to its external charge and their identification with the iterated term argued. The $C_FC_A^3$ component of $S_4$ is unchanged for one such gluon at leading power, with strictly ordered rung attachments argued. The configuration with one end on an external leg reaches only the $C_F^2C_A$ and $C_F^3$ components of $S_3$ at two loops and can reach components of $S_4$ at three loops, including $C_F^2C_A^2$. Excluding new NNLL sources from it rests on the no-hard-gluon assumption. The $C_FC_A^3$ component beyond one such gluon, together with the $C_F^2C_A^2$ component and the remaining channel-diagonal terms, also rests on the no-hard-gluon assumption. Only the single-source assumption excludes non-planar color-identity equation terms, new sources first acting at five loops or later and unexamined forms, as specified in the subsection ``Scope of the source exclusions'' below. Table~N3 of the ancillary notes gives the remaining source classes and their exclusions.

Table~\ref{tab:constants-w4} lists the constants of the equations and maps. The map terms are $d\,\zeta_2aF_1^+$ and $e\,\zeta_2a\omega$, with map constant $d$ and map coefficient $e$. The relative singlet Born correction is $a\,\zeta_2v$.
The quark-pair class excluded by the no-hard-gluon assumption is \mbox{$\zeta_2aC_F[\eta_1(F_1^+)^2+\eta_2(F_A^+)^2+\eta_{34}F_1^+F_A^+]$}.
Its channel-diagonal part has $k_P=\zeta_2\eta_1C_F$, and the other terms mix color channels.
The singlet-adjoint term $g_{1A}\zeta_2aC_FF_1^+F_A^-$ is an added form for the configuration with one end on an external leg, if that configuration acts in this form.
The signature-mixed term is $g\,\zeta_2\omega^2F_1^+F_1^-$.
The term $\psi^+F_1^+F_1^-$ would give an NLL $\zeta_2$ term absent from $P^{(1)+}$ if it were not suppressed to NNLL.

\begin{table}[!t]
\centering\small
\caption{Constants of the equations and maps. The ``Entry and order'' column gives the first contribution described here. The fitted family with $k_P=d=e=0$ reproduces $S_1$, $S_2$ and the one- to three-loop $\zeta_2$ amplitude coefficients in SU($N_c$), leaving $S_{n\ge3}$ unchanged.}
\label{tab:constants-w4}
\rowsapartmain
\begin{tabular}{>{\raggedright\arraybackslash}p{1.7cm}>{\raggedright\arraybackslash}p{5.3cm}>{\raggedright\arraybackslash}p{8.1cm}}
\hline
Constant & Entry and order & Status\\
\hline
$c_+$ & Relative $\pi^2\omega^2$ map term, starting in $S_1$ & Equal to $\ell_+$ by the normalization condition, fitted within the family of Sect.~N7.1 of the ancillary notes\\
$\ell_+$ & Relative $\pi^2\omega^2$ pair weight, starting in $S_2$ & Fitted within the same family\\
$\alpha$ & Common $\pi^2\omega^2$ signature normalization, one-loop amplitude constant & Fitted within the family, color independent\\
$\beta_e$ & Relative $\pi^2\omega^2$ singlet Born term, starting in $S_1$ & Fitted within the family, color independent\\
$v$ & Relative $a\zeta_2$ singlet Born correction, $S_2$ and one-loop amplitude constant & Fitted, with color degree one. A cutoff calculation would fix it independently\\
$r$ & Scalar amplitude map $1-a\zeta_2r$, one-loop constant & Fitted, with color degree one. A finite-map calculation would fix it independently\\
$k_P$ & Relative $a$ pair correction, starting in $S_3$ & Excluded by the no-hard-gluon assumption, zero through $y=0$ with the normalization condition\\
$d$ & Map constant in $d\zeta_2aF_1^+$, starting in $S_2$ & Set to zero by the normalization condition\\
$e$ & Map term $e\zeta_2a\omega$, starting in $S_1$ & Set to zero by the normalization condition\\
$y$ & $\nmYdef$, the combination entering $S_{n\ge3}$ & Fixed to zero by $S_3$ if the normalization condition holds and no channel-mixing, signature-mixed or singlet-adjoint pair term is present\\
$h$ & $\zeta_2ah(F_A^+)^2$, starting in $S_3$ & Excluded by the no-hard-gluon assumption. Computed absent at leading power, with strictly ordered rung attachments argued. Zero from the complete $S_3,S_4$ at $c_+=\ell_+$ or with $C_F(c_+-\ell_+)$ and $y$ of color degree one. The alternative $h=\nmKappaDegH$, $\kappa=-1$, reproduces $S_3,S_4$ but does not reproduce the $P^-$ leading logarithms at two to four loops\\
$g$ & $g\zeta_2\omega^2F_1^+F_1^-$ at NNLL, shifting $S_3,S_4,S_5$ & Excluded by the no-hard-gluon assumption. Zero from the complete $S_3,S_4$ at $c_+=\ell_+$ or with $C_F(c_+-\ell_+)$ and $y$ of color degree one\\
$g_{1A}$ & $g_{1A}\zeta_2aC_FF_1^+F_A^-$, starting in $S_3$ & Excluded by the no-hard-gluon assumption for the configuration with one external-leg end\\
$\eta_{34}$ & $\eta_{34}\zeta_2aC_FF_1^+F_A^+$, starting in $S_3$ & Excluded by the no-hard-gluon assumption, no $d$-invariant in any $S_n$\\
\hline
\end{tabular}
\end{table}

The leading-power angular average and transverse integral for external-leg attachments are computed with eikonal couplings and massless pair lines as
\begin{equation}
  \left\langle\frac{q\cdot(q+k)}{q^2(q+k)^2}\right\rangle=\frac{\theta(q^2-k^2)}{q^2},\qquad
  \int_{\mu<|q|<\Lambda}d^2q\,\frac{q\cdot(q+k)}{q^2(q+k)^2}
  =\pi\ln\frac{\Lambda^2}{\max(k^2,\mu^2)} .
  \label{eq:hard-transverse-w2}
\end{equation}
Here $\mu$ and $\Lambda$ are the lower and upper transverse cutoffs, with $|k|<\Lambda$.
Transverse-momentum conservation prevents the gluon from being harder than both quarks.
If the attachment is strictly ordered, one quark is soft, its transverse logarithm is removed, and there is no $\zeta_2$ at leading power.
This result is computed for external-leg attachments and argued for rung attachments.
The integral has no constant remainder from comparable transverse momenta.
The principal-value parts cancel between the legs for a gluon with transverse momentum comparable to the softest pair's and both ends attached to external legs. A configuration with one end on an external leg has only one such attachment.
The computed longitudinal integral for the configuration with both ends attached, with $u=q^2$ and $v=(q+k)^2$, is\hfill\mbox{}
\begin{equation}
  \int_0^\infty\frac{dz\,z^{\omega-1}}{(u+z)(v+z)}
  =\frac{\pi}{\sin(\pi\omega)}\,\frac{u^{\omega-1}-v^{\omega-1}}{v-u} .
  \label{eq:hard-longitudinal-w2}
\end{equation}
Its $1/\omega$ residue is one over $uv$.
At this leading pole the longitudinal and cutoff transverse integrals give the soft-gluon term of eq.~(\ref{eq:S}), with a remainder of order $\zeta_2^2a^2\tilde L^0$.
The $O(\omega^0)$ term and the region where the pair's longitudinal invariant is comparable to its squared transverse momentum are counted, not integrated.

The two-loop rung, pair-line and leg attachments are summed algebraically at leading power in Feynman gauge with one regulator acting alike on all attachments for a gluon matching the pair in transverse momentum and rapidity.
The Feynman-gauge gluon two-point function is \mbox{$D_{\mu\nu}(k)=-ig_{\mu\nu}/(k^2+i0)$}, with the same positive imaginary part in all eikonal denominators.

The term proportional to $l^\mu l^\nu$ added in a general covariant gauge to the two-point function of the matching gluon, of momentum $l$, drops out of the $C_A^2$ entry of the two-loop $\zeta_2$ coefficient of the positive-signature color-singlet amplitude divided by its Born term (Sect.~\ref{sec:amplitude}, Table~\ref{tab:amp-nnll}).
The single-graph colors \inlf{$\nmRungRungColor$} and \inlf{$\nmLegRungColor$} cancel beyond the retained soft-gluon term.
The resulting zero $C_A^2$ remainder excludes $\zeta_2ah(F_A^+)^2$, $h=\eta_2C_F$, at leading power, with strictly ordered rung attachments argued.
The $C_A^2$ contribution already in eq.~(\ref{eq:system}) is retained.
At leading power, a gluon that is not the softest parton leaves the $C_FC_A^2$ component of $S_3$ unchanged, with the attachment sum computed and the strictly ordered rung attachments argued.
The configuration with one end on an external leg reaches only the $C_F^2C_A$ and $C_F^3$ components of $S_3$ at two loops, and can reach the $C_F^2C_A^2$ component of $S_4$ at three loops.

\begin{table}[tbp]
\centering\footnotesize\narrowcols
  \caption{Two-loop attachment colors for the routing of eqs.~(\ref{eq:hard-upper-w3}) and (\ref{eq:hard-lower-w3}). The external legs are $L_i$, the soft quark lines are $A,B$, and the upper and lower rungs are $R_1,R_2$. These are individual graph colors. Only the $C_A^2$ part of their subtracted sum is determined.}\label{tab:hard-attachments-w3}
\fracrowsS
\begin{tabular}{p{11cm}p{3cm}}
\hline
Attachment pairs & Color per Born term\\
\hline
$(L_{1},A),\ (L_{1},B),\ (A,L_{2}),\ (A,L_{4}),\ (A,L_{3}),\ (L_{2},B),\ (L_{4},B),\ (B,L_{3})$ & $C_Fb_A$ \\
$(L_{1},L_{2}),\ (L_{1},L_{4}),\ (L_{2},L_{3}),\ (L_{4},L_{3})$ & $b_A^2$ \\
$(L_{1},L_{3}),\ (A,A),\ (A,B),\ (L_{2},L_{4}),\ (B,B)$ & $C_F^2$ \\
$(L_{1},R_{1}),\ (A,R_{2}),\ (B,R_{2}),\ (L_{3},R_{1})$ & $C_F(C_F-b_A)$ \\
$(L_{1},R_{2}),\ (L_{3},R_{2})$ & \inlf{$-\nmLegRungColor$} \\
$(A,R_{1}),\ (L_{2},R_{2}),\ (L_{4},R_{2}),\ (B,R_{1})$ & $-C_F(C_F-b_A)$ \\
$(L_{2},R_{1}),\ (L_{4},R_{1})$ & \inlf{$\nmLegRungColor$} \\
$(R_{1},R_{2})$ & \inlf{$-\nmRungRungColor$} \\
\hline
\end{tabular}

\end{table}

The attachment sum uses $k=\beta p_1+\alpha p_2+k_\perp$ and $s=2p_1\cdot p_2$, with $\alpha_A+\alpha_B+\alpha_l=0$ and $\beta_{qB}=\beta_A+\beta_l$.
All denominators below have the same positive imaginary part before it is taken to zero.
The upper attachments to legs one and three and to the rung, after removal of the common factor $(ig)^3$ and the quark spinors, are
\begin{equation}
 2p_1^\rho\left\{
 \frac{c_1}{(-s\alpha_l+i0)(s\alpha_B+i0)},\quad
 \frac{c_3}{(s\alpha_l+i0)(-s\alpha_A+i0)},\quad
 \frac{c_R}{(-s\alpha_A+i0)(s\alpha_B+i0)}\right\} .
 \label{eq:hard-upper-w3}
\end{equation}
The lower attachments to legs two and four and to the rung are
\begin{equation}
 2p_2^\sigma\left\{
 -\frac{c_2}{(s\beta_l+i0)(s\beta_{qB}+i0)},\quad
 -\frac{c_4}{(-s\beta_l+i0)(s\beta_A+i0)},\quad
 \frac{c'_R}{(s\beta_A+i0)(s\beta_{qB}+i0)}\right\} .
 \label{eq:hard-lower-w3}
\end{equation}

The color $c_R$ is $if^{abc}t^a\otimes t^b$ on the upper pair, with $c'_R$ defined on the lower pair.
In the singlet projection $c_1=c_3=c_L$ and $c_2=c_4=c'_L$.
We sum the upper and lower leg and rung denominators before taking their common positive imaginary part to zero. The identity $1/(z+i0)-1/(z-i0)=-2\pi i\delta(z)$ isolates the delta-function terms. The remaining products of denominators give the second terms below. This algebraic step is computed. The strictly ordered rung contribution and the three-loop identification with the iterated soft term remain argued.
The attachment sums with the colors of Table~\ref{tab:hard-attachments-w3}, divided by their respective $2p_1^\rho$ and $2p_2^\sigma$, for $x=s\alpha_l$, $y=s\alpha_A$, $x'=s\beta_l$ and $y'=s\beta_A$ become
\begin{align}
 U&=c_L\frac{2\pi i\delta(s\alpha_l)}{s\alpha_A-i0}
       +(c_R-c_L)\frac1{(-y+i0)(-x-y+i0)},\nonumber\\
 L&=c'_L\frac{2\pi i\delta(s\beta_l)}{s\beta_A+i0}
       +(c'_R+c'_L)\frac1{(y'+i0)(x'+y'+i0)} .
 \label{eq:hard-sums-w3}
\end{align}
The symbols $x,y,x',y'$ in this calculation are local integration variables, distinct from the momentum fraction $x$ and from the combination $y$ in Table~\ref{tab:constants-w4}.
The external-charge combinations are $-T_A=-T_B$ on the upper pair and $T_A=T_B$ on the lower pair.
The coefficients of $C_A^2$ per Born term in the attachment order $(1,3,R)$ against $(2,4,R')$ are
\begin{equation}
 \frac14\begin{pmatrix}1&1&-1\\1&1&-1\\1&1&-1\end{pmatrix} .
 \label{eq:hard-color-matrix-w3}
\end{equation}
This rank-one matrix multiplies only the product of the two delta-function terms of eq.~(\ref{eq:hard-sums-w3}).
The common-pole algebra is computed with the same $+i0$ prescription on every denominator. The computed common-pole contribution equals the signature-changing term acting on the box.
The remainder after subtraction of that term is $(0,r_{AF},r_{FF})$ in the basis $(C_A^2,C_AC_F,C_F^2)$, with $r_{AF}$ and $r_{FF}$ undetermined by this calculation.
The color projections at three loops were evaluated in SU(2), SU(3), SU(4), SU(5), SU(6), SO(5), SO(7), Sp(4) and G$_2$.
The ancillary file \nolinkurl{three_loop_attachment_colors.txt} lists the placements described in the section ``Data and program availability''.

At three loops such gluons reach the quartic component of $S_4$ only when both ends of each of two gluons attach to external legs.
Their color projections are computed in \countword{\nmGlauberColorGroups} groups and equal $\nmTrcTrcMtwoCoef M_2$ per Born term, the color of $\Trc\Trc$ on the box.
The coupling of two such gluons on one energetic line to its external charge and the identification with the iterated term of eq.~(\ref{eq:system}) are argued.
The quartic component of $S_4$ is unchanged at leading power with these two steps argued.
One such gluon also leaves the $C_FC_A^3$ component unchanged at leading power, with strictly ordered rung attachments argued.
The weight of the configuration with one end on an external leg and the NNLL weights of two gluons that are not the softest parton at three loops are not computed.
The configuration with one end on an external leg requires a rapidity regulator common to its soft, collinear and Glauber regions.
In the Glauber region the gluon's transverse momentum dominates its longitudinal components.
The two-gluon configurations can reach the $C_FC_A^3$ component, so beyond one such gluon this component rests on the no-hard-gluon assumption together with the $C_F^2C_A^2$ component and the remaining channel-diagonal terms.

\subsection*{Scope of the source exclusions}
The factor $\psi^+=O(\zeta_2\omega)$ suppresses relative-$\omega$ and relative-$a$ corrections in the other equations beyond the compared poles. New NNLL poles require another term in the positive-signature singlet equation or its map, or a change of its double-logarithmic adjoint input.
A new crossing-odd kernel changes that input and lies outside the proposed equations by definition.
 Its order is argued by power counting. A term starting at $a^p$ first changes the $P^-$ leading logarithms at $a^{p+1}$ and $S$ at $a^{p+2}$ when its singlet row does not annihilate the Born weights. A nonzero singlet row that annihilates them first changes both at $a^{p+2}$ in SU($N_c$).
The power count and the two-, three- and four-loop $P^-$ leading logarithms protect $S_3,S_4,S_5$, respectively, in SU($N_c$) against a single forward kernel with integer coupling powers, a finite polynomial in the energy logarithm, no boundary terms and all other LL terms fixed, if its singlet row does not annihilate the Born weights. The $P^-$ leading logarithms protect only $S_3,S_4$ this way in SU($N_c$) for a nonzero singlet row that annihilates the Born weights. The one- to three-loop negative-signature adjoint-amplitude logarithms also protect $S_5$ in SU($N_c$) if the adjoint row does not annihilate the Born weights. The negative-signature adjoint-amplitude logarithms protect $S_3$ through $S_5$ in SU($N_c$) for a zero singlet row if the adjoint row does not annihilate the Born weights. Both amplitude arguments assume the finite-remainder identification beyond one loop. If both rows annihilate the Born weights, exclusion rests on the single-source assumption from $S_5$ onward for a nonzero singlet row and from $S_3$ onward for a zero row. From $S_6$ onward absence of a new crossing-odd kernel belongs to the single-source assumption in full (Sect.~N6.10 of the ancillary notes).

The color-degree rule is argued at double-logarithmic accuracy and assumed at NNLL, as stated in Sect.~\ref{sec:threefour}. If $c_+-\ell_+$ and $y$ are arbitrary functions of the group invariants and no singular or additive map term and no channel-mixing, signature-mixed or singlet-adjoint quark-pair term is present, $S_3$ and $S_4$ fix both in each group, leaving no independent test.
Only the single-source assumption excludes non-planar color-identity equation terms, every new source first acting at five loops or later (including planar and color-identity terms), and forms outside the examined list in Table~N3 of the ancillary notes.
 A derivation of the positive-signature singlet equation at NNLL would decide their absence. The complete results exclude planar color-identity terms first shifting $S_3$ or $S_4$. The three- and four-loop comparisons cannot test a source that first acts at five loops or later.
A complete five-loop coefficient would test the absence of a source first acting at five loops jointly with the small-$N$ assumption. Sources first acting at six loops or later remain untested by the five-loop coefficient and affect $S_6$ or later terms of the tower.

The small-$N$ assumption is that eq.~(\ref{eq:relation}) holds at NNLL apart from the retained term. It is checked at $x^0$ through four loops and assumed for the five-loop coefficient and tower. The assumption at $N=-k$, $k\ge1$, covers only $\zeta_2$ because rational NNLL violations are present. Moch and Vogt state eq.~(3.13) of Ref.~\cite{Moch:2026qsw} for $x^0$ and use the method of Vogt et~al.~\cite{Vogt:2012gb} for double logarithms at higher powers of $x$. The $\zeta_2$ form at shifted poles is an assumption made here.
A derivation of the relation at these poles would decide it.

The shifted-pole assumption applies the equations at $N=-k$ with Born weights multiplied by $\lambda_k$. The three- and four-loop $P^+$ even-$k$ and $P^-$ odd-$k$ coefficients check it.
The five-loop $x^2$ coefficients would decide the shifted-pole assumption jointly with the small-$N$ assumption in its $\zeta_2$ form and the absence of a new source first acting at five loops.
A derivation of the shifted equations would decide it.

The assumption on $d$-invariants in $P^-$ excludes such invariants from the missing NNLL contributions to $P^-$ through $a^5$. The four-loop quartic coefficient at $x^0$ and its zero from $x^2$ onward check its $a^4$ part only. A five-loop calculation of $P^-$ would test the $a^5$ part.

The equations use a transverse-momentum cutoff, while the complete results use $\MSbar$. The normalization condition constrains the map between them and is not derived by a factorization-scheme argument. A scheme change multiplying the non-singlet density by $\mathcal Z(a,N)$ changes $P$ by $\beta(a)\,d\ln\mathcal Z/da$. Only $\beta_0$ times the NLL $\zeta_2$ terms of $\ln\mathcal Z$ can reach NNLL if the scheme-change coefficients are polynomial in $n_f$, contain no inverse $\beta_0$, and $\ln\mathcal Z$ is no more singular than $a^kN^{-2k}$ at order $a^k$. Terms involving $\beta_1$ cannot. Since $\beta_0$ contains $n_f$, a nonzero change would then introduce $n_f$ dependence absent from the violations. This is an argument about changes of anomalous dimension, not a derivation of the cutoff-wave map (Sect.~N6.13 of the ancillary notes). The functions $F_2$ and $F_3$ are deep-inelastic structure functions. The physical kernels are $P_{\rm ph}=P+\beta(a)\,d\ln C/da$, with coefficient function $C$ for $F_2$ or the even-$N$ combination of $F_3$ given by Davies, Kom, Moch and Vogt~\cite{Davies:2022ofz}. The NNLL $\zeta_2$ poles of these physical kernels equal those of $P^+$ at three and four loops. The absence of $\pi^2$ terms in physical anomalous dimensions stated by Davies and Vogt~\cite{Davies:2017hyl} concerns explicit even zeta values at integer $N$, whereas the small-$x$ $\zeta_2$ poles arise from the continuation of harmonic sums.

\section{The two-loop constants and the negative-signature sector}
\label{app:twoloop}

Section~\ref{sec:protection-w4} gives the $S_2$ discrepancy and the $C_A$ part of the equations' one-loop singlet residue, $\nmResDiffSingletCA\,\kappa\zeta_2C_A$ per Born term, which is absent from the finite remainder, as shown after eq.~(\ref{eq:sign-mismatch}) in App.~\ref{sec:sign}. The fitted family with $k_P=d=e=0$ of Sect.~\ref{sec:protection-w4} (Sect.~N7.1 of the ancillary notes) fits both without changing $S_{n\ge3}$, with the constants defined in Table~\ref{tab:constants-w4}.

\subsection{The sector of \texorpdfstring{\protect\boldmath$P^-$}{P-}}
\label{sec:lim-pminus}
The leading logarithms agree for $P^-$ at even $k$ and $P^+$ at odd $k$. The quartic coefficient also agrees under the assumption on $d$-invariants.
The complete results minus the equations are $\nmMismatchVthreeSUthree$ and $\nmMismatchVfourSUthree$ for $V_3^-$ and $V_4^-$ at $x^0$ in SU(3), with $\ell_+=\ell_-=c_-=0$.
The differences in this convention at every power $x^k$ for $P^-$ at even $k$ and $P^+$ at odd $k$ contain no quartic invariant and are proportional to $C_F(C_A-2C_F)$, suppressed by $1/N_c^2$ relative to planar terms and outside the large-$N_c$ resummation of $P^-$ beyond LL in Ref.~\cite{Davies:2022ofz}.
The factor also holds at $x^0$ for arbitrary $\ell_+$ with $c_-=\ell_-$ and is lost when they differ.
The numbers $\ell_+$, $\ell_-$ and $c_-$, taken common to each signature, give no general-color solution for both $V_3^-$ and $V_4^-$, since the three-loop difference fixes all three and then misses four loops.
The quartic coefficient $\nmPminusQuarticRho$, with $\rho$ scaling the leading $\psi^+$ in the positive-signature adjoint equation, is unchanged by relative-$\omega$ or relative-$a$ corrections but becomes $\nmPminusQuarticKm$ when $\kappa$ is reversed, and the three five-loop quartic coefficients in eq.~(N13) of the ancillary notes have the same protection.
Nonunique corrections confined to the positive-signature adjoint equation,
with color factors nonzero at $C_A=0$,
 fit the NNLL differences
at $x^0$ without changing $S_3$ through $S_5$
 or LL and NLL $P^-$, while a fitted $a(\nmNLLtermCoef)F_A^+$ term reproduces the NLL coefficients of $P^-$ at three and four loops and leaves the rational and $n_f$-dependent NNLL poles unreproduced.
Neither fit has an identified source or enters a result here.
The missed $P^-$ terms show that the proposed equations are incomplete in the negative-signature singlet equation, its map $c_-$ or the positive-signature adjoint equation, and their source is not identified.
The subleading term of $\psi^-$ enters at the same order as uncomputed $\ell_-$ and $c_-$ terms, negative-signature Born and quark-pair corrections, and adjoint Sudakov-type and $\Tt$ terms, so the mismatch does not single out $\psi^-$.
An NNLL calculation would test the source of the missed $P^-$ terms against $V_3^-$, $V_4^-$ and $P^{(1)-}-P^{(1)+}$.
The three $\Tt$ forms and the convention $\ell_+=\ell_-=c_-=0$, $c_+=\ell_+$ are defined in Sect.~\ref{sec:protection-w4}. Each misses $V_3^-$ and $V_4^-$ separately in general color at first order in $\zeta_2$ and NNLL, with or without the fitted NLL term. Their contributions vanish in the abelian limit for polynomial Casimir coefficients, where the differences $\nmVthreeMinusAbelianDiff$ and $\nmVfourMinusAbelianDiff$ remain. The three forms also leave unreproduced the non-quartic $\lambda_k=2$ differences for $P^-$ at even $k\ge2$ and $P^+$ at odd $k\ge3$ under the shifted-pole assumption. The more general adjoint corrections fitted at $x^0$ also miss the $\lambda_k=2$ differences if the added terms are kept unchanged at the shifted poles. The test does not exclude coefficients singular at $C_A\to0$ or adjoint corrections not proportional to $\Tt$ whose color factors survive at $C_A=0$.

Section~N7.2 of the ancillary notes gives a one-parameter family of signature-dependent numeric constants that fits both abelian $P^-$ differences, with no solution at $\ell_+=\ell_-=c_-$. The abelian $S_3,S_4$ components are already counted in Sect.~\ref{sec:threefour}.
Section~N7.3 of the ancillary notes gives the shift of $v$ equivalent to using only the loop part of $F_A^-$ in the retained term, at $\kappa=+1$ and unit negative-signature adjoint Born term, and the fitted value of $v$.
The relative signature normalization $n_-/n_+=1+\nu\pi^2\omega^2$ multiplies $\psi^+$ and divides $\psi^-$ by this factor, with all other terms unchanged. This uncompensated variation is distinct from the consistent rescaling of App.~\ref{sm:sign-reduction}.
It leaves $S_3$ and $S_4$ unchanged for every $\nu$.
In general color, it permits a one-parameter family of numeric, color-independent constants in $(\nu,\ell_+,\ell_-,c_-)$ matching $V_3^-$ alone, but no such constants match $V_3^-$ and $V_4^-$ together.
 Replacing the full $F_A^-$ in the retained term by its loop part changes $S_2$ and other amplitude entries at one to three loops, but leaves every $S_n$ with $n\ge3$, the two-loop $C_A^2$ entry and the three-loop $C_A^3$ and $C_A^2C_F$ entries unchanged.
 The equations do not choose between the two forms, which one-loop cutoff amplitudes would distinguish.

\subsection{Leading logarithms of \texorpdfstring{\protect\boldmath$P^-$}{P-} and the quartic coefficients}
\label{sec:Pminus}
In negative signature the signature-changing term is double-logarithmic and fixes the non-planar
part of the leading logarithms of $P^-$.
The system gives $\nmLLminusTwo$, $\nmLLminusThree$ and $\nmLLminusFour$ for the full coefficients of $a^nN^{1-2n}$ in $P^-$ at $n=2,3,4$, planar part included. These are $\nmLLminusTwoSUthree$, $\nmLLminusThreeSUthree$ and $\nmLLminusFourSUthree$ in SU(3).
They agree with eqs.~(4.6) and (4.7) of Ref.~\cite{Moch:2004pa} at two loops, with eq.~(4.17) of Ref.~\cite{Moch:2004pa} at three loops
and with the four-loop results of Ref.~\cite{Gehrmann:2026qbl}.
These leading logarithms follow from the
double-logarithmic equations of Refs.~\cite{Kirschner:1983di,Bartels:1995iu}, and the equations with $\kappa=-1$ do not reproduce
them.

Section~N3.5 of the ancillary notes gives the corresponding one-loop amplitude residues. The four-loop leading logarithm of $P^-$ is printed as $\nmPthreeMinusLnSixPrinted$ in quantum chromodynamics (QCD) in eq.~(3.16) of Ref.~\cite{Moch:2026qsw}. The equations give the $x^0\ln^6x$ coefficient $\nmPthreeMinusLnSixQCD$ of $P^{(3)-}$ in SU(3), equal to this value to the printed precision.
The non-planar leading logarithms of $P^-$ confirm the negative-signature sign printed by Kirschner and Lipatov~\cite{Kirschner:1983di} and test the double-logarithmic equations of Ref.~\cite{Kirschner:1983di} and of Bartels, Ermolaev and Ryskin~\cite{Bartels:1995iu}, without testing the NNLL term. The positive-signature sign is derived in App.~\ref{sec:sign}.
Only the quartic-invariant component is reproduced in the $\zeta_2\,x^0\ln^4x$ coefficient of $P^{(3)-}$.
The equations give it in proportion to the scale of $\psi^+$ in the positive-signature adjoint equation, and for the unscaled $\psi^+$ it equals the value $\nmPminusQuartic$ of the complete four-loop results.
The other four components are not reproduced (App.~\ref{sec:lim-pminus}).
The same term predicts the quartic-invariant part of the $\zeta_2\,x^0\ln^6x$ coefficient of $P^{(4)-}$, proportional to the four-loop coefficient (eq.~(N13) in Sect.~N11.1 of the ancillary notes).
This prediction rests on the assumption on $d$-invariants in $P^-$ through $a^5$ and on the absence of a new five-loop source with its own quartic color factor, which the four-loop coefficient cannot exclude.

\section{The powers \texorpdfstring{\protect\boldmath$x^k$}{xk}}\label{app:xk}
\label{sec:xk}
The coefficients beyond $x^0$ come from our expansion of the complete three- and four-loop functions. A second expansion in a different computer-algebra system agrees for every $k\le\nmXkSecondKmax$ and every logarithmic power.
The zeros of $\xi^\pm(-k)$ do not account for the non-quartic mismatch in $P^-$ at even $k$ and $P^+$ at odd $k$.

\label{sm:shifted}

Splitting functions have poles at $N=-k$ for every $k\ge0$.
Equation~(3.15) of Ref.~\cite{Moch:2026qsw},
with the halved bracket discussed in App.~\ref{app:core-data},
 gives the $\zeta_2\,x^k\ln^4x$ differences from the NNLL predictions for $P^+$ at even $k$ and $P^-$ at odd $k$, the $x^k$ counterparts of eq.~(\ref{eq:xzerodiff}).

The equations act at $N=-k$ in the variable \mbox{$\omega=N+k$}, with Born weights $\lambda_kb_R$, if the shifted-pole assumption holds (App.~\ref{app:assumptions}).
Here $\lambda_k$ is the residue of $P^{(0)}$ at $N=-k$ divided by its residue at $N=0$.
The residues $\nmPzeroResidues$ for $k=0,\dots,\nmXkKmax$ give $\lambda_k=\nmLambdaVec$.
We write $\lambda=\lambda_k$ in this appendix. The ratio $\lambda_k$ is the $x^k$ coefficient of $\inlf{\nmLambdaGenFun}$ in general, equal to two for $k\ge2$.
The values of $k$ thus form two classes, $k=0$ and $k=1$ with $\lambda=1$, and $k\ge2$ with $\lambda=2$.
The complete results for $P^+$ at even $k$ and $P^-$ at odd $k$ test these two values of $\lambda$.

Since $e^{i\pi k}=(-1)^k$, the signature factor obeys $\xi^\tau(\omega-k)=(-1)^k\,\xi^{(-1)^k\tau}(\omega)$.
The sign cancels in $\phi^\pm$ and $\psi^\pm$, so odd $k$ interchanges the signature systems.
The positive-signature singlet equation thus describes $P^+$ at even $k$ and $P^-$ at odd $k$.
Keeping the soft-gluon and quark-pair terms unchanged at $N=-k$ is an extrapolation, which we do not derive and on which the results beyond $x^0$ rest.
The equations depend on $k$ only through $\lambda_k$ and the parity of $k$, so this dependence extends the results below to every $k$.
We solved the equations for $k\le\nmXkEquationKmax$.

The equations at $N=-k$ use $\phi^\pm$ and $\psi^\pm$ evaluated at $N=\omega-k$, and the counterpart of $S$ in eq.~(\ref{eq:S}) is
\begin{equation}
  S=P(P-\omega)+2a\lambda_kC_F ,
  \label{eq:xk-S}
\end{equation}
with $\omega=N+k$, for $P=P^+$ or $P=P^-$.
We denote by $V_n$, or $V_n^\pm$ for the two signatures, the $\zeta_2$ part of the $a^n\omega^{4-2n}$ coefficient in $S$ of eq.~(\ref{eq:xk-S}), so that $V_n=S_n$ for $P^+$ at $k=0$.
The three- and four-loop $V_n$ agree with the complete results in all \countword{\nmXkComponents} color structures, \countword{\nmXkComponentsThree} at three loops and \countword{\nmXkComponentsFour} at four loops, for $P^+$ at every even $k$ and $P^-$ at every odd $k$ using the proposed equations and the shifted-pole assumption at $N=-k$.
Only the leading logarithms and the $d_F^{abcd}d_A^{abcd}/N_R$ coefficient in the $\zeta_2\,x^k\ln^4x$ part of $P^{(3)}$ agree for $P^-$ at even $k$ and $P^+$ at odd $k$. The quartic agreement holds with the assumption on $d$-invariants in $P^-$ and, at $x^0$, also the leading $\psi^+$ in the positive-signature adjoint equation (App.~\ref{sec:lim-pminus}).
The complete three- and four-loop $\zeta_2$ NNLL coefficients depend on $k$ only through parity and $\lambda_k$, as the equations do. This shared dependence tests the shifted-pole assumption. The distinction between confirmed inputs, relations valid for every variant and independent tests in Sect.~\ref{app:count} also applies here. Section~N6.4 of the ancillary notes counts the tests with further constants admitted.

The rational NLL and NNLL terms depend on $k$ beyond parity and lie outside the equations at every $k$, as at $x^0$.
For example, the constant of $P^{(0)}$ at $N=-k$ is $\nmPzeroConstKzero$, $\nmPzeroConstKone$, $\nmPzeroConstKtwo$, \inlp{$\nmPzeroConstKthree$} and \inlp{$\nmPzeroConstKfour$} for $k=0,\dots,\nmXkKmax$.
The complete NLL poles contain no $\zeta_2$, so the rational NLL and NNLL terms do not reach the compared NNLL $\zeta_2$ coefficients $V_n$. The residues of $P^{(0)}$ fix only the Born weights. The agreement tests the assumed form of the other terms at $N=-k$.

The quantity $S$ already has LL poles for $P^-$ at even $k$ and $P^+$ at odd $k$. There we compare the $V_n$ directly with coefficients of the complete results.
The equations give the $d_F^{abcd}d_A^{abcd}/N_R$ coefficient in the $\zeta_2\,x^k\ln^4x$ part of $P^{(3)}$ for $P^-$ at even $k$ and $P^+$ at odd $k$ as
\begin{equation}
  \nmXkUncovQuarticFormula ,
  \label{eq:uncov}
\end{equation}
and for $P^+$ at even $k$ and $P^-$ at odd $k$ the same coefficient is $\nmXkCovQuarticFormula$.
The quartic coefficient for $P^-$ at even $k$ and $P^+$ at odd $k$ is $\nmXkUncovQuarticLow$ at $k=0$ and $k=1$ and $\nmXkUncovQuarticHigh$ at $k\ge2$, eq.~(\ref{eq:uncov}).
At $k=0$, it is the coefficient $\nmPminusQuartic$ of $P^{(3)-}$, subject to the assumption on $d$-invariants in $P^-$ (App.~\ref{app:assumptions}).
From $x^2$ on, its zero checks consistency with that assumption.
At $k\ge2$, the quartic coefficients in both pairings allow only $\lambda=2$, independently of $P^{(0)}$.
The other $\zeta_2$ NNLL components for $P^-$ at even $k$ and $P^+$ at odd $k$ are not reproduced (App.~\ref{sec:lim-pminus}).

Table~N7 of the ancillary notes gives the complete comparison, including the non-quartic differences.
Its $P^+$ even-$k$ and $P^-$ odd-$k$ rows at $\lambda=2$ independently check the shifted-pole assumption.
 The quartic part of the other pairing also selects $\lambda=2$, with the assumption on $d$-invariants in $P^-$. Only harmonic polylogarithms with all indices zero contribute at two to four loops to the LL and $\zeta_2$ NNLL coefficients. Their rational prefactors have the form $A+Bx+(C+Dx)/(1-x^2)$, generating the coefficients at $k=0$, $k=1$, and even and odd $k\ge2$ separately.
The proposed equations give the five-loop $x^2$ coefficient in eq.~(N5) of the ancillary notes with the shifted-pole assumption, the small-$N$ assumption in its $\zeta_2$ form, and no new source first acting at five loops. A source confined to the shifted positive-signature singlet equation would change it while leaving the $x^0$ comparisons unchanged.

\section{The forward four-quark amplitude}\label{app:checks}\label{app:amplitude}
The coefficients $\mathcal H^{[1]}$ and $\mathcal H^{[2]}$ multiply the color tensors of eq.~(3.2) of Ref.~\cite{Caola:2021rqz}. Their projection is
\begin{equation}
  A_1=N_c\mathcal{H}^{[1]}+\mathcal{H}^{[2]},\qquad A_A=\mathcal{H}^{[2]} .
  \label{eq:amp-proj}
\end{equation}
The first tensor is $N_c$ times the singlet projector, and the second contains both channels, so $\mathcal H^{[1]}$ alone is not the singlet amplitude. The forward limit sends the invariant of legs one and four to zero in the conventions of Ref.~\cite{Caola:2021rqz}. It is our momentum transfer $t$, carried by legs one and three in our labeling. The processes of eqs.~(7.1) and (7.2) of Ref.~\cite{Caola:2021rqz} give the regions $s_r>0$ and $s_c>0$, respectively. Half their sum and half their difference at fixed color tensors and equal $\tilde L$ give positive and negative signature. We compare helicities that do not flip along the forward direction. The compared $\zeta_2a^L\tilde L^{2L-2}$ coefficients divided by the Born term contain no $n_f$ and are unchanged if $\mu^2$ is multiplied by a constant. Two symbolic forward expansions of the ancillary finite remainders of Ref.~\cite{Caola:2021rqz} agree in every coefficient of both processes through three loops at $\mu^2=|t|$.

\begin{table}[htb]
  \centering
  \caption{Coefficient of $\zeta_2\,a^L\tilde L^{2L-2}$ in the positive-signature singlet amplitude divided by its
  Born term, $A_1^+/A_1^{(1)}$, at $\mu^2=|t|$, in the basis $(C_A^L,C_A^{L-1}C_F,\dots,C_F^L)$ of SU($N_c$). At one
  loop the entry is the $\zeta_2$ constant in the basis $(C_A,C_F)$. The third column gives eq.~(\ref{eq:system}) with the constants reaching these coefficients. The relative correction of the
  singlet Born term is $a\,\zeta_2(v_AC_A+v_FC_F)$. The scalar factor of the map from the cutoff to dimensional
  regularization differs from one by $-a\,\zeta_2(r_AC_A+r_FC_F)$. The correction of the quark-pair term is
  $a\,\zeta_2(p_AC_A+p_FC_F)(F_1^+)^2$.
  These $p_A,p_F$ are the components of $k_P$ in Table~\ref{tab:constants-w4}, distinct from the free coefficients of each $\Tt$ correction tested in Sect.~\ref{sec:protection-w4}.
  Here $\ell_+$ is the $\pi^2\omega^2$ part of the weight of the positive-signature
  quark-pair term. The constants $\alpha$ and $\beta_e$ are the $\pi^2\omega^2$ parts of a normalization factor common to the signature factors
  and of the singlet Born term. The constants of relative order $a$ have color factors of
  degree one and those of relative order $a^0$ none. The last column gives, for $\kappa=-1$, the entries that the constants do not reach, the $C_A^2$ entry at two loops and the $C_A^3$ and $C_A^2C_F$ entries at three loops.}\label{tab:amp-nnll}
  \small
  \nmfracsmall\fracrowsSupp
  \renewcommand{\,}{\mskip\thinmuskip\allowbreak}
  \makeatletter
  \binoppenalty=\@M
  \relpenalty=\@M
  \makeatother
  \begin{tabular}{p{1.8cm}p{2.7cm}p{8.6cm}p{1.7cm}}
  \hline
  Order & \raggedright Finite remainders\newline of Ref.~\cite{Caola:2021rqz} & Eq.~(\ref{eq:system}), $\kappa=\nmKappa$ & $\kappa$ reversed \\
  \hline
  one loop & $\nmAmpDataOneVec$ & {\raggedright $\nmAmpEqOneVec$\par} & \\
  two loops & $\nmAmpDataTwoVec$ & {\raggedright $\nmAmpEqTwoVec$\par} & $\nmAmpKmTwoLoopCA$ \\
  three loops & $\nmAmpDataThreeVec$ & {\raggedright $\nmAmpEqThreeVec$\par} & \raggedright $\nmAmpKmThreeLoopCA$, $\nmAmpKmThreeLoopCACF$ \tabularnewline
  \hline
  \end{tabular}
\end{table}

\subsection{The finite map}\label{sm:amp-map}
The minimally subtracted remainder identified with the cutoff amplitude in Sect.~\ref{sec:amplitude} is $H=Z^{-1}M$ at $\mu^2=|t|$. A cutoff amplitude formed from the dipole terms $\tfrac12\ln^2((-s_{ij}-i0)/\mu^2)$ and the planar quark-pair box $b_R\ell_r^2$ agrees with the finite remainder in the four one-loop leading-logarithmic coefficients. All four one-loop leading logarithms differ without the box. The one-loop remainder is
\begin{equation}
  \begin{aligned}
  H^{(1)}=\Big[&\sum_{i<j}\mathbf{T}_i\cdot\mathbf{T}_j\big(\ell_{ij}^2-\zeta_2\big)+B\big(\ell_r^2+\pi^2\big)\\
  &+\bigl(-\beta_0+\nmHoneLrCoefCF\,C_F+\Tt\bigr)\ell_r+\nmHoneRational\Big]H^{(0)} ,
  \end{aligned}
  \label{eq:amp-H1}
\end{equation}
where $\ell_{ij}=\ln(-s_{ij}-i0)$ in units of $\mu^2$, so the logarithms of $|t|$ vanish, and $B=C_F-\Tt/2$. The coefficient of $\zeta_2$ in the $\tilde L$-independent part of the positive-signature singlet remainder, per Born term at $\mu^2=|t|$, is $\inlp{\nmAmpOneLoopZetaConst}$ and the corresponding adjoint $\zeta_2$ coefficient is $\inlp{\nmOneLoopAdjZetaConst}$. Their cutoff counterparts depend on the finite dipole and box terms. A sharp dipole transverse cutoff with the box kept as $B(\ell_r^2+\pi^2)$ is one examined prescription. Its box constant is assumed, not computed. Every examined prescription gives no $C_A$ part of the map on the positive-signature singlet (Sect.~N9.2 of the ancillary notes). The equations give a $C_A$ part $\nmResDiffSingletCA\,\kappa\zeta_2C_A$ per Born term in the one-loop singlet residue, absent from the finite remainder (App.~\ref{sec:sign}, after eq.~(\ref{eq:sign-mismatch})). The $C_A$ part $v_A$ of the fitted Born correction removes it in the family of Sect.~\ref{sec:protection-w4} and Sect.~N7.1 of the ancillary notes. Table~\ref{tab:amp-nnll} gives the coefficients. Changing the assumed box constant moves only a $C_F$ singlet term and cannot reach the three entries free of constants.

The finite-map conditions concern maximal weight, counting $\tilde L$ and $i\pi$ as weight one and $\zeta_2$ as weight two. The assumed map has the polynomial-in-logarithm dipole form and three-loop quadrupole correction defined in Sect.~\ref{sec:amplitude}, even under exchange of energy channels, and without a logarithm-dependent color-identity part. The finite map between the cutoff and minimal subtraction is not derived. The dipole form of the infrared poles through two loops found by Aybat, Dixon and Sterman~\cite{Aybat:2006mz} motivates this class but does not fix the finite map. Every allowed term that changes a compared maximal-weight coefficient also changes a leading logarithm. Requiring the leading-logarithm agreement therefore leaves those coefficients unchanged within this class. The scan of $\nmAmpStructCount$ structures through orders $a$ to $a^3$ uses the operators $1$, $\Tt$, $\Trc$ and their commutators (Sect.~N9.1 of the ancillary notes). Terms of lower weight cannot reach the compared $\zeta_2a^L\tilde L^{2L-2}$ coefficients per Born term.

The color-degree rule of Sect.~\ref{sec:threefour} restricts the undetermined constants' contributions to the positive-signature singlet amplitude to $C_F^L$ or a degree-one factor times $C_F^{L-1}$ at $L$ loops, as used in Sect.~\ref{sec:amplitude}. A $C_A^2/C_F$ part in the scalar-map, relative Born or channel-diagonal pair constant would change the two-loop $C_A^2$ and three-loop $C_A^2C_F$ entries. The color-degree rule excludes it. Equations~(\ref{eq:amp-rel}) and (\ref{eq:amp-rel-full-w3}) then test the common wave and map, not the source, after $S_3,S_4$ are fixed.

The helicity-flip remainders vanish at leading power through two loops. Their three-loop constant has negative signature and cannot reach the compared coefficients. These statements and the finite-map scans concern the finite remainder. The one-loop flip remainder has forward terms of order $\epsilon$ that combine with infrared poles at higher orders.

\section{The tower}\label{sec:predictions}\label{sec:tower}\label{app:tower}
Equation~(\ref{eq:adjoint-general-w3}) of App.~\ref{app:core-color} becomes the Riccati equation at $\kappa=+1$ in SU($N_c$),
\begin{align}
  2aN_c\,\frac{dF_A^-}{d\omega}&=\omega F_A^--\frac{2aN_c}{\omega}F_A^--\big(F_A^-\big)^2+Q ,\nonumber\\
  Q&=\frac{a}{N_c}+\frac{4a}{N_c\,\omega}F_1^++\frac{2a(N_c^2-4)}{N_c\,\omega}F_A^+ .
  \label{eq:tower-Q-w2}
\end{align}
\makeatletter%
\protected@edef\@currentlabel{\theequation}\label{eq:tower-Q}%
\makeatother%

\subsection{The negative-signature adjoint equation}
\label{sm:tower-wave}

For SU($N_c$), the substitution $\nmLinearSub$ turns the Riccati equation for the double-logarithmic wave $F_A^-$ into
\begin{equation}
  \nmLinearEq ,
  \label{eq:linear}
\end{equation}
whose coefficients contain $F_1^+$ from eq.~(\ref{eq:LL}) and the positive-signature adjoint wave $F_A^+$, both known in closed form from Kirschner and Lipatov~\cite{Kirschner:1983di}, Ref.~\cite{Bartels:1995iu}, and Kiyo, Kodaira and Tochimura~\cite{Kiyo:1997ct}.
The linear equation for $F_A^+$ is the parabolic-cylinder equation.
Equation~(\ref{eq:linear}) instead has coefficients containing $F_1^+$ and $F_A^+$.
We have found no closed form for its solution at finite $N_c$.

One solution is selected by the requirement that $F_A^-$ vanish at large $\omega$. This solution follows the asymptotic series of $F_A^-$ in $a/\omega^2$.

The coefficients $d_n$ of $F_A^-=\sum_{n\ge1}d_n\,a^n\omega^{1-2n}$ follow from the double-logarithmic eq.~(\ref{eq:system}) order by order in $a/\omega^2$.
The inputs are the positive-signature waves $F_1^+=\sum_{n\ge1}f_n\,a^n\omega^{1-2n}$ and \mbox{$F_A^+=\sum_{n\ge1}c^+_n\,a^n\omega^{1-2n}$}.
The recursions in SU($N_c$), with $\nmCatalanF$ and Catalan numbers $\mathrm{Cat}_n$, are
\begin{align}
  c^+_n&=-2N_c(2n-3)\,c^+_{n-1}+\sum_{k=1}^{n-1}c^+_k\,c^+_{n-k} ,\nonumber\\
  d_n&=-4N_c(n-2)\,d_{n-1}+\sum_{k=1}^{n-1}d_k\,d_{n-k}-\frac{4}{N_c}\,f_{n-1}
       -\frac{2(N_c^2-4)}{N_c}\,c^+_{n-1} ,
  \label{eq:tower-recursion}
\end{align}
for $n\ge2$, starting from $c^+_1=d_1=2b_A=-1/N_c$.

In each line, the first term is the Sudakov contribution and the sum is the quark-pair contribution.
The last two terms of the second line come from signature change, using $[\Trc]_{A1}=-1/N_c$ and $[\Trc]_{AA}=2/N_c-N_c/2$ in eq.~(\ref{eq:TrcSUN}).

The coefficients $d_1$ to $d_3$ were obtained in Sect.~\ref{sec:threefour} and give $d_1=-1/N_c$, \inlf{$d_2=\nmDtwoSUN$} and \inlf{$d_3=\nmDthreeSUN$} in SU($N_c$).
The coefficients $d_2$ and $d_3$ reproduce eq.~(\ref{eq:S34SUN}) by eq.~(\ref{eq:Sn-dn}), while the Born term gives the non-pole $S_2=\nmStwoOwn$.

The proposed equations give the NNLL poles of $S$ at fixed coupling and first order in $\zeta_2$ to every order in $a$, with no new source beyond four loops. These poles form the non-planar $\zeta_2$ tower.
Kirschner and Lipatov~\cite{Kirschner:1983di} solve the positive-signature equations and the negative-signature singlet equation, but give only the $\omega\to0$ behavior for the negative-signature adjoint wave $F_A^-$, which we solve for here.
The $x$-space series of the negative-signature adjoint wave is an entire function of $\sqrt{a}\,\ln(1/x)$, and the LL branch point $\omega_0$ of eq.~(\ref{eq:LL}) is the rightmost singularity of $S$ for every $N_c\ge2$. Both statements hold at fixed coupling and first order in $\zeta_2$ (Sect.~N10.1 of the ancillary notes).

The coefficients obey the bound in eq.~(N12) of the ancillary notes. Their factorial growth is compensated on Mellin inversion by $(2n-2)!$, so the $x$-space series converges. The leading late-order term of $d_n$ is absent in SU(2), where a second sector dominates (Sect.~N10.3 of the ancillary notes).

\section{Five and six loops}
\label{sec:fiveloop}
\label{app:five}

The proposed equations give the five-loop pole, with no new source first acting at five loops,
\begin{longeq}\label{eq:S5}
  S_5=\nmSfiveGen ,
\end{longeq}
which for SU($N_c$) becomes
\begin{equation}
  S_5=\nmSfiveSUN ,
  \label{eq:S5SUN}
\end{equation}
and gives $\nmSfiveSUthree$ in SU(3).

At six loops the reduction of $M_4$ brings in $d_F^{abcd}d_A^{abef}d_A^{cdef}/N_R$ through the four-rung adjoint ladder graph (eq.~(\ref{eq:color-M34}) in App.~\ref{app:core-color}).
Solutions of the equations in the $\nmRepsExtCount$ representations of App.~\ref{app:core-color} agree with the general-color $S_5$ of eq.~(\ref{eq:S5}) and with $S_6$ of eq.~(\ref{eq:S6}).

\begin{sloppypar}
Equation~(\ref{eq:color-d45}) with the moments $M_3$ and $M_4$ of eq.~(\ref{eq:color-M34}) gives eq.~(\ref{eq:S5}) and, with no new source first acting at five or six loops,
\begin{longeq}\label{eq:S6}
  S_6=\nmSsixGen ,
\end{longeq}
which in SU($N_c$) is
\begin{equation*}
  S_6=\nmSsixSUN ,
\end{equation*}
or $\nmSsixSUthree$ in SU(3).

The invariant $d_F^{abcd}d_A^{abef}d_A^{cdef}/N_R$ enters $S_6$ only through $M_4$.
The eight invariants of eq.~(\ref{eq:S5}) satisfy the following linear relation for the defining representations of SU($N_c$), SO($N$) and Sp($N$)
\begin{equation}
  \nmInvRelationClassical ,
  \label{eq:fiveloop-relation}
\end{equation}
which fails, among the $\nmRepsExtCount$ representations, for every representation of $\mathrm{G}_2$, $\mathrm{F}_4$, $\mathrm{E}_6$, $\mathrm{E}_7$ and $\mathrm{E}_8$.
It also fails for every other classical-group representation considered except the three-dimensional SU(2) representation, the SO(3) vector.
Results for these defining representations alone therefore cannot separate the $C_FC_A^4$, $C_A\,d_F^{abcd}d_A^{abcd}/N_R$ and $C_F\,d_A^{abcd}d_A^{abcd}/N_A$ coefficients in eq.~(\ref{eq:fivegen-w2}).
Nor can they separate the three quartic coefficients of eq.~(N13) of the ancillary notes.
In Mellin space, these are $\nmPfourMinusPoleVec$ for $\zeta_2\,a^5N^{-7}$ in the basis $(C_A\,d_F^{abcd}d_A^{abcd}/N_R,\,C_F\,d_F^{abcd}d_A^{abcd}/N_R,\,C_F\,d_A^{abcd}d_A^{abcd}/N_A)$.

A new non-planar NNLL pole source in the positive-signature singlet equation starting at $a^5$ could add to $S_5$ any combination of the eight invariants whose $N_c^5$ coefficient in SU($N_c$) vanishes.
This requirement is one linear condition, with coefficients $\nmLargeNcCondVec$ on the invariants of eq.~(\ref{eq:S5}).
Equation~(\ref{eq:S5}) itself satisfies it.
\end{sloppypar}

\subsection{New sources at five loops}
\label{sm:five-sources}

A change $\delta S_5$, for example from a new pole source starting at five loops, shifts the coefficient by
\begin{equation}
  \delta\Big[P^{(4)+}\Big]_{\zeta_2x^0\ln^6x}=-\frac{\delta S_5}{\nmFiveShiftFactor\,\zeta_2} ,
  \label{eq:five-shift}
\end{equation}
and in SU(3) the part from $S_5$ is this expression with $\delta S_5=\nmSfiveSUthree$.
A source \mbox{$q\,\zeta_2\,C_F(C_A-2C_F)C_A^3$} in $S_5$ shifts the SU(3) coefficient $\nmFiveLoopXzeroSUthree$, given after eq.~(\ref{eq:fivegen-w2}), by \inlp{$\nmQsourceShiftSUthree\,q$}.
At $x^2$ the same shift is $-\delta V_5/(\nmFiveShiftFactor\,\zeta_2)$ (eq.~(N5) in Sect.~N8.1 of the ancillary notes).

A change $\delta S$ in $S$ shifts $P^+$ by $-\delta S\,(\omega^2-8aC_F)^{-1/2}$ at first order in $\zeta_2$ (App.~\ref{sec:reduction}).
The $\zeta_2$ part of the $a^m x^0\ln^{2m-4}(1/x)/(2m-4)!$ coefficient in $P^+$ is therefore $-T_m$, where
\begin{equation}
  \nmTmMap ,
  \label{eq:Tm}
\end{equation}
and $S_1=\nmSoneData$ and $S_2=\nmStwoData$ come from the complete results.
Through four loops, this map reproduces the complete general-color $\zeta_2\,x^0\ln^{2m-4}x$ coefficients.
Table~N8 of the ancillary notes gives the general-color coefficient of eq.~(\ref{eq:fivegen-w2}), which in SU($N_c$) becomes
\begin{equation}
  \Big[P^{(4)+}\Big]_{\zeta_2x^0\ln^6x}=\nmFiveLoopXzeroSUN ,
  \label{eq:fiveSUN}
\end{equation}
and reduces in SU(3) to $\nmFiveLoopXzeroSUthree$, the value after eq.~(\ref{eq:fivegen-w2}), with corresponding $\zeta_2N^{-7}$ coefficient $\nmFiveLoopNsevenSUthree$.

A new non-planar five-loop source adds \countword{\nmSourceFiveFreeParams} parameters in general color or \countword{\nmSourceFiveFreeParamsSUthree} in SU(3). Such a source leaves only the large-$N_c$ limit fixed. The large-$N_c$ limit $\nmFiveLoopLargeN$ of the five-loop $\zeta_2x^0\ln^6x$ coefficient in eq.~(\ref{eq:fivegen-w2}) equals the large-$N_c$ limit of eq.~(3.4) of Ref.~\cite{Davies:2022ofz} and follows from eq.~(15) of Ref.~\cite{Herzog:2018kwj}. Section~N11.4 of the ancillary notes gives the parameter count and the exclusions of added pair terms by $S_3,S_4$.

\FloatBarrier
{\footnotesize\raggedright
}

\end{document}